\documentclass[aps,prd,superscriptaddress,showkeys,nofootinbib]{revtex4}

\usepackage[T1]{fontenc}
\usepackage{aas_macros}
\usepackage{lmodern}
\usepackage{ifthen}
\usepackage{textcomp}   
\usepackage{gensymb}
\usepackage[english]{babel}
\usepackage{graphicx}
\usepackage{xcolor}
\usepackage{soul}
\usepackage{rotating}
\usepackage{amsmath,amsthm}
\usepackage{dsfont}\let\mathbb\mathds
\usepackage{tabularx}
\usepackage[caption=false]{subfig}
\usepackage{multirow}
\usepackage{siunitx}
\usepackage{booktabs} 

\usepackage[breaklinks=true,colorlinks=false,hidelinks,hyperfootnotes=false]{hyperref}

\hypersetup{pdfauthor={P.~Schmidt-Wellenburg}}
\graphicspath{{./Images/}}

\newlength{\bildtitel}
\newcommand\REVIEW[1]{\message{LaTeX Warning: \noexpand untreated nEDM-REVIEW command in \jobname .tex: l\the\inputlineno}}
\newcommand{\diff}[1]{\operatorname{d}\ifthenelse{\equal{#1}{}}{\,}{\!#1}}

\newcommand{\Part}[2]{\displaystyle\frac{\partial #1}{\partial #2}}

\newcommand{\MeVc}{\ensuremath{\mathrm{MeV}\!/\!{\it c}}}

\newcommand{\ecm}{\ensuremath{\si{\elementarycharge}\!\cdot\!\cm}}

\newcommand{\cm}{\ensuremath{\mathrm{cm}}}

\newcommand{\mum}{\ensuremath{\micro \mathrm{m} }}

\newcommand{\EDM}{{\mathchoice{}{}{\scriptscriptstyle}{}\text{EDM}}}

\begin{document}

\title{A compact frozen-spin trap for the search for the electric dipole moment of the muon}

\author{A.~Adelmann}
\affiliation{ETH Zürich, 8093 Zürich, Switzerland}
\affiliation{PSI Center for Scientific Computing, Theory, and Data, 5232 Villigen PSI,
Switzerland} 
\author{A.R.~Bainbridge}
\affiliation{ASTeC, STFC Daresbury Laboratory, Sci-Tech Daresbury, Warrington, WA4 4AD, UK}
\affiliation{Cockcroft Institute, Sci-Tech Daresbury, Warrington, WA4 4AD, UK}
%
\author{I.~Bailey}
\affiliation{Cockcroft Institute, Sci-Tech Daresbury, Warrington, WA4 4AD, UK}
\affiliation{Lancaster University, Lancaster, UK}
\author{A.~Baldini}
\affiliation{University of Pisa and INFN, Pisa, Italy}
\author{S.~Basnet}
\affiliation{Tsung-Dao Lee Institute, Shanghai Jiao Tong University, Shanghai, China}
\affiliation{School of Physics and Astronomy, Shanghai Jiao Tong University, Shanghai, China}
\author{N.~Berger}
\affiliation{PRISMA+ Cluster of Excellence and Institute of Nuclear Physics, Johannes Gutenberg University Mainz, Mainz, Germany}
%
\author{C.~Calzolaio}
\affiliation{PSI Center for Accelerator Science and Engineering, 5232 Villigen PSI, Switzerland}
\author{L.~Caminada}
\affiliation{Physik-Institut, Universität Zürich, Winterthurerstrasse 190, CH–8057 Zürich, Switzerland}
\affiliation{PSI Center for Neutron and Muon Sciences, 5232 Villigen PSI, Switzerland}
\author{G.~Cavoto}
\affiliation{Sapienza Universit\`a di Roma, Dip.\ di Fisica, P.le A.\ Moro 2, 00185 Roma, Italy}
\affiliation{Instituto Nazionale di Fisica Nucleare, Sez.~di Roma, P.\,le A.~Moro 2, 00185 Roma, Italy}
\author{F.~Cei}
\affiliation{University of Pisa and INFN, Pisa, Italy}
\author{R.~Chakraborty}
\affiliation{PSI Center for Neutron and Muon Sciences, 5232 Villigen PSI, Switzerland}
\author{C.~Chavez~Barajas}
\affiliation{University of Liverpool, Liverpool, UK}
%
%
\author{M.~Chiappini}
\affiliation{University of Pisa and INFN, Pisa, Italy}
%
%
\author{A.~Crivellin}
\affiliation{Physik-Institut, Universität Zürich, Winterthurerstrasse 190, CH–8057 Zürich, Switzerland}
\author{C.~Dutsov}
\affiliation{PSI Center for Neutron and Muon Sciences, 5232 Villigen PSI, Switzerland}
\author{A.~Ebrahimi}
\affiliation{PSI Center for Neutron and Muon Sciences, 5232 Villigen PSI, Switzerland}
\author{M.~Francesconi}
\affiliation{University of Pisa and INFN, Pisa, Italy}
\author{L.~Galli}
\affiliation{University of Pisa and INFN, Pisa, Italy}
\author{G.~Gallucci}
\affiliation{University of Pisa and INFN, Pisa, Italy}
\author{M.~Giovannozzi}
\affiliation{CERN, Beams Department, 1211 Geneva, Switzerland}
\author{H.~Goyal}
\affiliation{PSI Center for Neutron and Muon Sciences, 5232 Villigen PSI, Switzerland}
\author{M.~Grassi}
\affiliation{University of Pisa and INFN, Pisa, Italy}
\author{A.~Gurgone}
\affiliation{University of Pisa and INFN, Pisa, Italy}
\author{M.~Hildebrandt}
\affiliation{PSI Center for Neutron and Muon Sciences, 5232 Villigen PSI, Switzerland} 
\author{M.~Hoferichter}
\affiliation{Albert Einstein Center for Fundamental Physics, Institute for Theoretical Physics, University of Bern, Sidlerstrasse 5, 3012 Bern, Switzerland}
\author{D.~Höhl}
\affiliation{PSI Center for Neutron and Muon Sciences, 5232 Villigen PSI, Switzerland} 
\author{T.~Hu}
\affiliation{Tsung-Dao Lee Institute, Shanghai Jiao Tong University, Shanghai, China}
\affiliation{School of Physics and Astronomy, Shanghai Jiao Tong University, Shanghai, China}
%
%
\author{T.~Hume}
\affiliation{PSI Center for Neutron and Muon Sciences, 5232 Villigen PSI, Switzerland}
\author{J.A.~Jaeger}
\affiliation{PSI Center for Neutron and Muon Sciences, 5232 Villigen PSI, Switzerland}
\author{P.~Juknevicius}
\affiliation{PSI Center for Neutron and Muon Sciences, 5232 Villigen PSI, Switzerland}
\author{H.C.~Kästli}
\affiliation{PSI Center for Neutron and Muon Sciences, 5232 Villigen PSI, Switzerland}
\author{A.~Keshavarzi}
\affiliation{Department of Physics and Astronomy, University of Manchester, Manchester, UK}
\author{K.S.~Khaw}
\affiliation{Tsung-Dao Lee Institute, Shanghai Jiao Tong University, Shanghai, China}
\affiliation{School of Physics and Astronomy, Shanghai Jiao Tong University, Shanghai, China}
\author{K.~Kirch}
\affiliation{ETH Zürich, 8093 Zürich, Switzerland}
\affiliation{PSI Center for Neutron and Muon Sciences, 5232 Villigen PSI, Switzerland}
\author{A.~Kozlinskiy}
\affiliation{PRISMA+ Cluster of Excellence and Institute of Nuclear Physics, Johannes Gutenberg University Mainz, Mainz, Germany}
\author{M.~Lancaster}
\affiliation{Department of Physics and Astronomy, University of Manchester, Manchester, UK}
\author{B.~M\"arkisch}
\affiliation{School of Natural Sciences, Technical University of Munich, Garching, Germany }
\author{L.~Morvaj}
\affiliation{PSI Center for Neutron and Muon Sciences, 5232 Villigen PSI, Switzerland}
%
\author{A.~Papa}
\affiliation{University of Pisa and INFN, Pisa, Italy}
\affiliation{PSI Center for Neutron and Muon Sciences, 5232 Villigen PSI, Switzerland}
\author{M.~Paraliev}
\affiliation{PSI Center for Accelerator Science and Engineering, 5232 Villigen PSI, Switzerland}
\author{D.~Pasciuto}
\affiliation{Instituto Nazionale di Fisica Nucleare, Sez.~di Roma, P.\,le A.~Moro 2, 00185 Roma, Italy}
\author{J.~Price}
\affiliation{University of Liverpool, Liverpool, UK}
\author{F.~Renga}
\affiliation{Instituto Nazionale di Fisica Nucleare, Sez.~di Roma, P.\,le A.~Moro 2, 00185 Roma, Italy}
\author{M.~Sakurai}
\affiliation{University College London, London, UK}
\author{D.~Sanz-Becerra}
\affiliation{PSI Center for Neutron and Muon Sciences, 5232 Villigen PSI, Switzerland} 
\author{\underline{P.~Schmidt-Wellenburg}}
\email{philipp.schmidt-wellenburg@psi.ch} 
\affiliation{PSI Center for Neutron and Muon Sciences, 5232 Villigen PSI, Switzerland} 
%
%
\author{Y.\,Z.~Shang}
\affiliation{Tsung-Dao Lee Institute, Shanghai Jiao Tong University, Shanghai, China}
\affiliation{School of Physics and Astronomy, Shanghai Jiao Tong University, Shanghai, China}
\author{Y.~Takeuchi}
\affiliation{Tsung-Dao Lee Institute, Shanghai Jiao Tong University, Shanghai, China}
\affiliation{School of Physics and Astronomy, Shanghai Jiao Tong University, Shanghai, China}
\author{M.\,E.~Tegano}
\affiliation{University of Pisa and INFN, Pisa, Italy}
\author{T.~Teubner}
\affiliation{University of Liverpool, Liverpool, UK}
\author{F.~Trillaud}
\affiliation{Instituto de Ingeniería, Universidad Nacional Autónoma de México, Cuidad de México, México}
\author{D.~Uglietti}
\affiliation{Swiss Plasma Center, École Polytechnique Fédérale de Lausanne, Villigen, Switzerland}
\author{D.~Vasilkova}
\affiliation{University of Liverpool, Liverpool, UK}
\author{A.~Venturini}
\affiliation{University of Pisa and INFN, Pisa, Italy}
\author{B.~Vitali}
\affiliation{University of Pisa and INFN, Pisa, Italy}
\author{C.~Voena}
\affiliation{Instituto Nazionale di Fisica Nucleare, Sez.~di Roma, P.\,le A.~Moro 2, 00185 Roma, Italy}
\author{J.~Vossebeld}
\affiliation{University of Liverpool, Liverpool, UK}
\author{F.~Wauters}
\affiliation{PRISMA+ Cluster of Excellence and Institute of Nuclear Physics, Johannes Gutenberg University Mainz, Mainz, Germany}
\author{G.\,M.~Wong}
\affiliation{Tsung-Dao Lee Institute, Shanghai Jiao Tong University, Shanghai, China}
\affiliation{School of Physics and Astronomy, Shanghai Jiao Tong University, Shanghai, China}
\author{Y.~Zeng}
\affiliation{Tsung-Dao Lee Institute, Shanghai Jiao Tong University, Shanghai, China}
\affiliation{School of Physics and Astronomy, Shanghai Jiao Tong University, Shanghai, China}
\begin{abstract}
The electric dipole moments~(EDM) of fundamental particles inherently violate parity~(P) and time-reversal~(T) symmetries. 
By virtue of the CPT theorem in quantum field theory, the latter also implies the violation of the combined charge-conjugation and parity~(CP) symmetry. 
We aim to measure the EDM of the muon using the frozen-spin technique within a compact storage trap. 
This method exploits the high effective electric field, $E\approx\SI{165}{MV/m}$, experienced in the rest frame of the muon with a momentum of about $\SI{23}{MeV/}c$ when it passes through a solenoidal magnetic field of $|\vec{B}|=\SI{2.5}{T}$.
In this paper, we outline the fundamental considerations for a muon EDM search and present a conceptual design for a demonstration experiment to be conducted at secondary muon beamlines of the Paul Scherrer Institute in Switzerland. 
In Phase~I, with an anticipated data acquisition period of 200 days, the expected sensitivity to a muon EDM is $\sigma(d)\leq\SI{4E-21}{\ecm}$. 
In a subsequent phase, Phase~II, we propose to improve the sensitivity to $\sigma(d)\leq\SI{6E-23}{\ecm}$ using a dedicated instrument installed on a different beamline that produces muons of momentum \SI{125}{\MeVc}.
\end{abstract}

\maketitle
\tableofcontents

\section{Introduction}
\label{Sec:IntroductionMotivation}
\subsection{Motivation}

Electric dipole moments (EDMs) of elementary particles violate time-reversal symmetry. According to the CPT theorem~\cite{Lueders1957}, this also implies the violation of combined charge-conjugation and parity-inversion (CP) symmetry, making EDMs powerful tools for probing physics beyond the current Standard Model (SM) of particle physics.
There are two possible sources of CP violation~(CPV) in the SM.
Best known is the CP-violating phase of the Cabibbo-Kobayashi-Maskawa (CKM) matrix~\cite{Kobayashi:1973fv}, parametrized in a basis-independent manner by the Jarlskog invariant~\cite{Jarlskog1985PhysRevLett}. 
Although this phase is close to maximal~\cite{ParticleDataGroup:2024cfk,Hocker:2001xe,UTfit:2005ras}, the resulting EDM values that can be accessed in an experiment are far too small~\cite{Pospelov:2013sca,Seng:2014lea} for detection anytime soon.
The second possible source of CPV in the SM is $\theta$, the coefficient of the topological term of the QCD Lagrangian~\cite{hooft1976}. It is limited to be smaller than $\simeq\num{1e-10}$ by the search result for a neutron EDM~\cite{Abel2020} assuming chiral estimates for the matrix element~\cite{Crewther:1979pi}, although a robust calculation in lattice QCD has so far remained elusive~\cite{Dragos:2019oxn,Alexandrou:2020mds,Bhattacharya:2021lol}.
Contrary to hadronic EDM searches, the muon EDM is not directly sensitive to a possible contribution from the CPV $\theta$-term of QCD, but due to high-order loop effects the present limit on $\theta$ still allows a contribution larger than the one via the CKM mechanism~\cite{Ghosh2018}.

Intriguingly, these two sources of CPV in the SM are known to be insufficient to explain the observed matter-antimatter asymmetry~\cite{Sakharov1967}, requiring additional sources of CPV beyond the SM (BSM).
Therefore, CP-violating observables are very sensitive probes of BSM physics. More than 60 years ago, E.M.~Purcell, N.F.~Ramsey, and their student J.H.~Smith~\cite{Smith1957} published the first search for an EDM of the neutron. Only a year later, the first search for a muon EDM was completed, resulting in an upper limit of \SI{2.9e-15}{\ecm}~(95\% C.L.)\,\cite{Berley1958PRL}. Since then, many searches around the world have been concluded with increasing sensitivity on neutrons, atoms, and molecules~\cite{Jungmann2013,Chupp2019RMP,Cairncross2019NatRevPhys}; however, so far all have found only null results. 

In BSM scenarios assuming minimal flavor violation~(MFV) in the lepton sector~\cite{Cirigliano2005}, as often implemented within the Minimal Supersymmetric Standard Model~(MSSM), a simple scaling by the ratio $m_{\mu}/m_e$ is predicted for the muon vs.\ the electron EDM~\cite{Chivukula:1987fw,Hall:1990ac,Buras:2000dm,DAmbrosio:2002vsn}. 
In this case, the electron EDM, $d_e \leq\SI{4.1e-30}{\ecm}$~(CL 90\%)~\cite{Roussy2023Science}, places severe limits on the muon. 
However, MFV is, to some extent, an ad hoc symmetry, mainly invented to allow light particle spectra within the MSSM to reduce the degree of fine-tuning in the Higgs sector, while respecting at the same time flavor constraints.

\begin{figure}[t]%
	\includegraphics[width=0.66\textwidth]{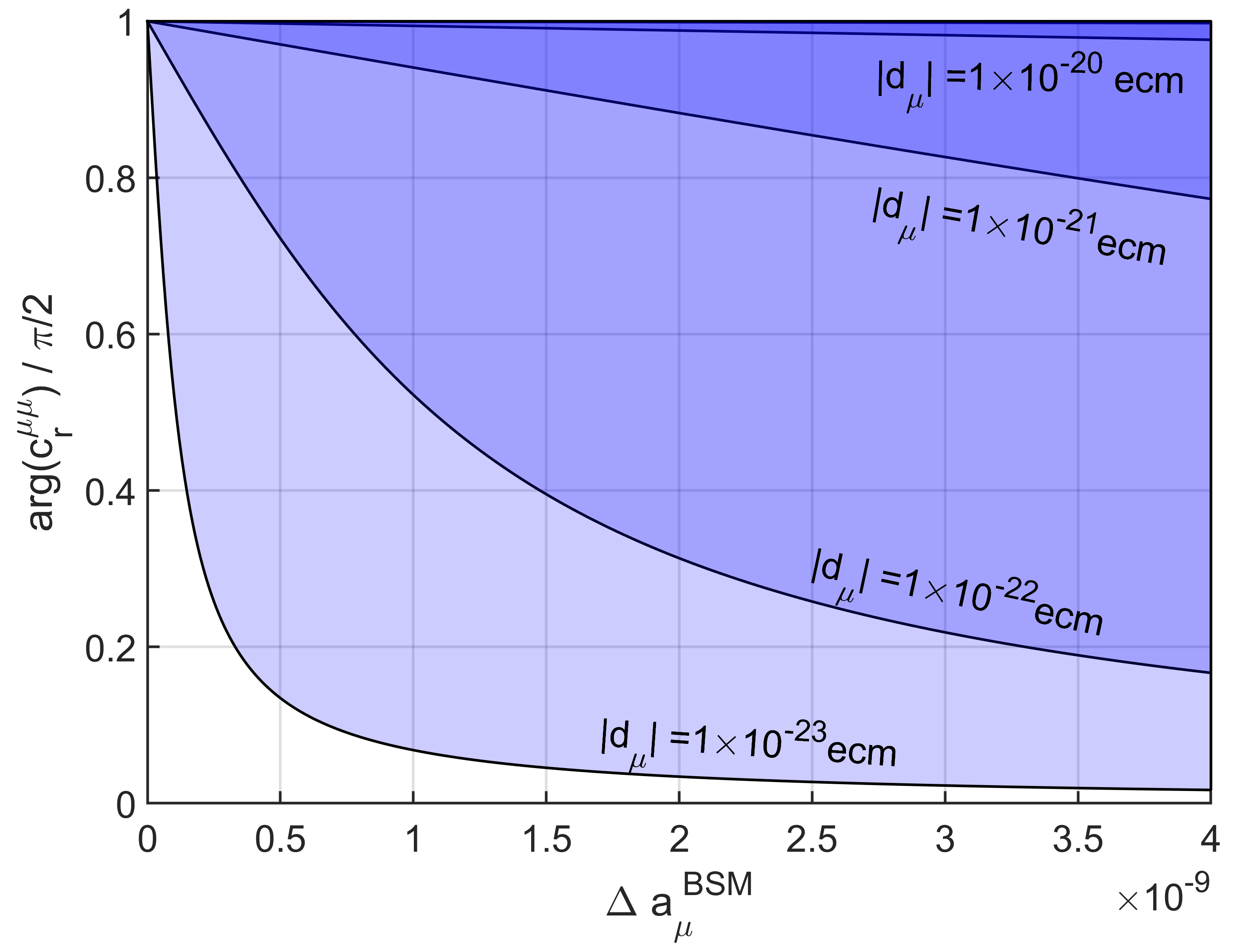}%
	\caption[Contour plot of AMM phase of the associated Wilson coefficient.]{Contours of $d_\mu$ as a function of a potential BSM effect in the AMM $\Delta a^{\rm BSM}_\mu$ vs.\ the phase of the associated Wilson coefficient~\cite{Crivellin2018}.}
	\label{fig:muEDMvsamu}%
\end{figure}

In theories with a flavor structure beyond the MFV paradigm, the simple scaling can be circumvented allowing for the possibility of a sizable muon EDM\@. 
Within a general model-independent effective-field-theory~(EFT) approach, the imaginary part of the Wilson coefficient of the relevant dipole operator corresponds to the muon EDM, while the real part is associated with a potential BSM effect in the anomalous magnetic moment~(AMM) of the muon~\cite{Muong-2:2023cdq,Muong-2:2024hpx,Aoyama:2020ynm,Colangelo:2022jxc}, leading to an interesting interplay between the two.
From this perspective, constraints on the muon EDM can be converted to the phase of the Wilson coefficient, see Fig.~\ref{fig:muEDMvsamu}, which shows that a phase of order one would be within reach of a dedicated muon EDM experiment for large parts of the parameter space~\cite{Feng2001,Crivellin2018}.

Another key point to make a sizable muon EDM possible is that $\mu\to e\gamma$ transitions must be avoided. This is possible as long as the electron couplings are sufficiently small, or can by achieved by disentangling the muon from the electron sector via a symmetry, such as an $L_\mu-L_\tau$ symmetry~\cite{He1991PhysRevD,Foot1991ModPhysLettA,He1991PhysRevDa}, which, even after breaking, protects the EDM of the electron and $g-2$ from large BSM contributions~\cite{Altmannshofer2016PhysRevD}. 
Therefore, one can, in general, protect the electron EDM and $\mu\to e\gamma$ from dangerously large BSM contributions while having a muon EDM within the measurable range. However, in a simplified or UV complete model approach, this requires the mechanism of chiral enhancement~\cite{Crivellin2018}. This means that the chirality flip, inherent in a dipole operator, is not generated by the muon mass, but by a generic coupling to the SM Higgs that can be larger and in general complex. 

Furthermore, it is possible to obtain a significant EDM without incurring significant fine-tuning related to the muon Yukawa coupling~\cite{Bigaran2022PhysRevD}, while the observable consequences of scenarios with large EDMs in $h\to\mu\mu$ and $Z\to\mu\mu$ could be investigated in future colliders~\cite{Crivellin2021JHEP}. Finally, from an EFT perspective~\cite{Pruna2017,Crivellin2018,Crivellin2019}, it is clear that the muon EDM is not constrained by other observables and that its measurement is the only way to determine the imaginary part of the associated Wilson coefficient.

In fact, the direct limit of the muon EDM $d_\mu<\SI{1.8e-19}{\ecm}$ (CL 95\%)~\cite{Bennett2009PRD} is currently the only result directly probing the EDM of a fundamental particle, since limits on the electron EDM~\cite{Roussy2023Science,Andreev2018} are derived from atomic systems and constraints on the $\tau$ EDM via $e^+e^-\to\tau^+\tau^-$~\cite{Belle:2021ybo,Bernabeu:2006wf,Crivellin:2021spu}. 
Indirect limits derived from $d_e$ via CP-odd insertions in three-loop diagrams~\cite{Grozin2009PhysAtomNucl,Crivellin2018,Ema2022PhysRevLett} slightly improve the current direct limit and provide another target for a dedicated search, but could be evaded if other sources of CPV were present. 
Here we describe the concept of a direct search for an EDM of muons employing the frozen-spin technique~\cite{Khriplovich1998PLB,Farley2004PRL,Adelmann2010JPG} in a compact muon trap at PSI.

\newpage
\subsection{Overview of the experiment}
\label{sec:ExperimentOverview}
\begin{figure}%
\centering
			\includegraphics[width=0.75\textwidth]{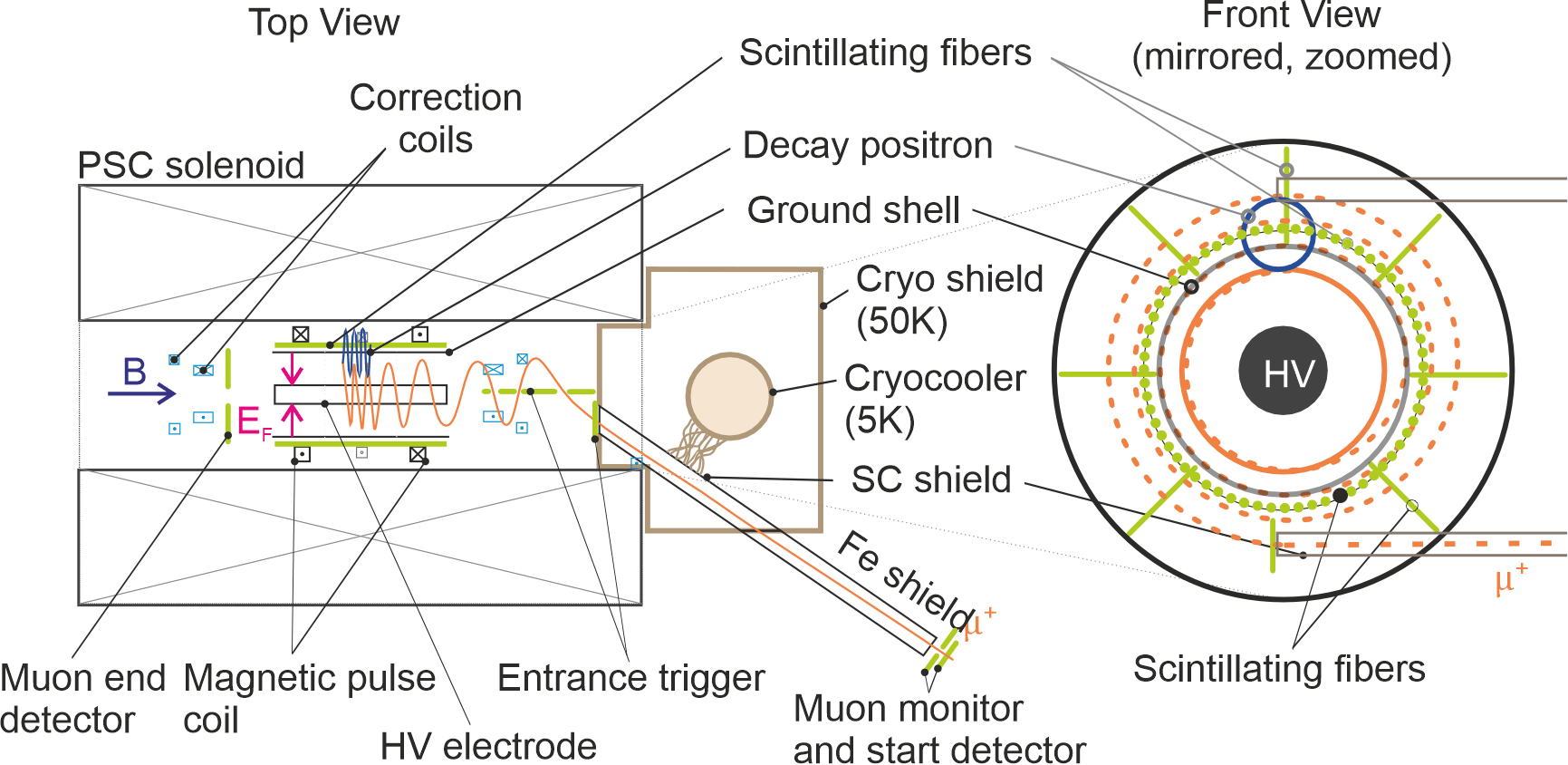}
	\caption{Sketch, not to scale, of the compact superconducting solenoid and the experimental setup for the search for the EDM of the muon. The warm bore of the solenoid has an inner diameter of \SI{200}{mm} and an outer diameter of about \SI{1000}{mm}. Muons are injected along two magnetically shielded parallel injection paths, each made up of a superconducting (SC) shield and an iron shield, which are spaced by about \SI{90}{mm} vertically as shown on the front view.}%
	\label{fig:PhaseISetup}%
\end{figure}

At PSI, we plan to search for the muon EDM in two phases. 
\begin{description}
		\item{\bf Phase~I}: In this exploratory phase, we will set up an experiment to demonstrate the frozen-spin method and search for a muon EDM using an existing solenoid. The instrument will be connected to a surface-muon beamline at PSI, delivering about \SI{4e6}{s^{-1}} muons at a momentum of $p=\SI{28}{\MeVc}$ in a transverse phase space of $\epsilon_{xx'}= \SI{192}{\pi mm\,mrad}$ and $\epsilon_{yy'}= \SI{171}{\pi mm\,mrad}$. Although the sensitivity to a muon EDM will be sufficient to improve the current best measurement~\cite{Bennett2009PRD}, the main purpose is to establish all necessary techniques and methods for a measurement with the highest possible sensitivity. Figure~\ref{fig:PhaseISetup} shows an overview sketch of the experimental setup for Phase I.
		\item{\bf Phase~II}: The future instrument will use a dedicated magnet with minimal field gradient between injection and storage region to increase the acceptance phase space and integrate all lessons learned from Phase~I. In addition, it will benefit from being coupled to the highest-intensity muon beam at PSI, with more than \SI{1e8}{s^{-1}} muons at a momentum of $p=\SI{125}{\MeVc}$ in a transverse phase space of $\epsilon_{xx'}= \SI{920}{\pi mm\,mrad}$ and $\epsilon_{yy'}= \SI{213}{\pi mm\,mrad}$.
\end{description}

Muons enter the experiment through one of two parallel injection paths shielded from the magnetic field of the solenoid by a combination of magnetic steel tube and a SC shield~\cite{Barna2017}, see Sec.\,\ref{sec:Injection}. By changing the height of the experiment and reversing all magnetic fields, we can change from a clockwise~(CW) to a counter clockwise~(CCW) injection, canceling major systematic effects (see Sec.\,\ref{Sec:SystematicEffects}). 

A set of detectors are used at the entrance and exit of the injection paths to align the experiment with the beam direction and measure the time of flight~(ToF) of each injected muon (see Sec.\,\ref{sec:MuDetection}). 
The second ToF scintillating detector is part of the muon entrance trigger (see Sec.\,\ref{sec:MuInjTrigger}), creating a trigger signal only for muons that are within the storage phase space of the trap, the tagged muons, and to start the data acquisition~(DAQ). 

While a tagged muon spirals along the solenoid field, the trigger signal launches a high-current pulse, which is injected into the split pulse coil in the center of the solenoid (see Sec.\,\ref{sec:B_FieldPulse}), accurately timed to stop the longitudinal motion of the muon. 
The resulting magnetic pulse traps the muon permanently in the weakly-focusing field at the center of the solenoid.
Section~\ref{sec:MagneticField} describes the magnetic field including correction and weakly-focusing coils required for a high injection efficiency. 

Muons that are not trapped pass through and are detected in the muon end-detector. 
The electric field to establish the frozen-spin condition is applied between the charged cylindrical electrodes and the ground cylinder is described in Sec.\,\ref{sec:electrodes}.

A tracker based on scintillating fibers (see Sec.\,\ref{sec:PositronTracker}) detects the positron from the muon decay. 
By measuring the time evolution of the decay asymmetry between positrons ejected along and opposite to the magnetic-field direction, we will deduce the electric dipole moment of the muon.  

In Phase II, the primary experimental features remain largely unchanged. However, a dedicated solenoid magnet will be designed to better accommodate the higher momenta and enhance injection efficiency. 

\section{The frozen-spin technique}
  \label{Sec:FrozenSpinTechnique}
An EDM of a fermion exposed to an electric field results in a two-level system with energy eigenstates
\begin{equation}
		V_\pm = \pm dE = \pm\frac{\hbar\omega_e}{2},
\label{eq:eFreq}
\end{equation}
where $\omega_e$ is the precession frequency proportional to the transition energy between the two states.
%

Most experiments searching for an EDM~\cite{Graner2016PRL,Andreev2018,Abel2020,Ayres2021} implement Ramsey's method of separated oscillating fields~\cite{Ramsey1950PR} in combination with alternating electric fields to measure the EDM frequency $\omega_e$. 
However, exposing muons to a static electric field between two electrodes will result in an acceleration, moving the muon quickly into a region without field, e.g.\ the surface of one of the electrodes. 
Instead of using a static electric field, we will use a static magnetic field, $\vec{B}$, in which we store muons with a velocity, $\vec{v}=\vec{\beta}c$, where $c$ is the speed of light in vacuum and $\vec{\beta}=\vec{p}/E$ is the relativistic factor, the ratio between momentum and total energy, in a closed orbit, and $\gamma=\left(1-\beta^2\right)^{-1/2}$. 
In the rest frame of the muon, this results in an electric field $|\vec{E}^\ast|=\gamma c |\vec{\beta}\times\vec{B}| \approx\SI{1}{GV/m}$ for $B=\SI{3}{T}$ and $\beta\approx0.75$ in Phase~II\@.

\subsection{The muon spin precession with an EDM in electromagnetic fields}
The spin precession relative to the momentum vector of a muon with a magnetic, $\vec{\mu}=gq/(2m)\vec{s}$, and electric,  $\vec{d}=\eta q/(2mc)\vec{s}$, dipole moment in a magnetic, $\vec{B}$, and electric field, $\vec{E}$, with charge $q=\pm e$, is given by the Thomas-BMT equation~\cite{Thomas1927,Bargmann1959}

\begin{widetext}
    \begin{align}
    	\vec{\Omega}=&-\frac{q}{m}\left[a\vec{B}-\frac{a\gamma}{\left(\gamma+1\right)}\left(\vec{\beta}\cdot\vec{B}\right)\vec{\beta}-\left(a+\frac{1}{1-\gamma^2}\right)
    	\frac{\vec{\beta}\times\vec{E}}{c}\right]  \nonumber \\
    	&-\frac{\eta q}{2mc}\left[c\vec{\beta}\times\vec{B}+\vec{E}-\frac{\gamma\left(\vec{\beta}\cdot\vec{E}\right)\vec{\beta}}{(\gamma+1)}\right],
        \label{eq:omegaMuWithEDM}
    \end{align}
\end{widetext}
where $a=\left(g-2\right)/2$ is the anomalous magnetic moment~(AMM), extended by a second term for the EDM contribution.

The first line of Eq.~\eqref{eq:omegaMuWithEDM} is the anomalous precession frequency, $\omega_{\rm a}$, the difference between the Larmor precession and the cyclotron precession. 
The second line is the precession $\omega_{\rm e}$ due to the EDM coupling to the electric field in the rest frame of the muon moving through the magnetic field $\vec{B}$, oriented perpendicular to $\vec{B}$. 

If momentum, magnetic and electric fields form an orthogonal basis, the scalar products of momentum with the fields, $\vec{\beta}\cdot\vec{B}=\vec{\beta}\cdot\vec{E}=0$. 

\begin{figure}%
	\centering
	\subfloat[]{
	\includegraphics[width=0.47\textwidth]{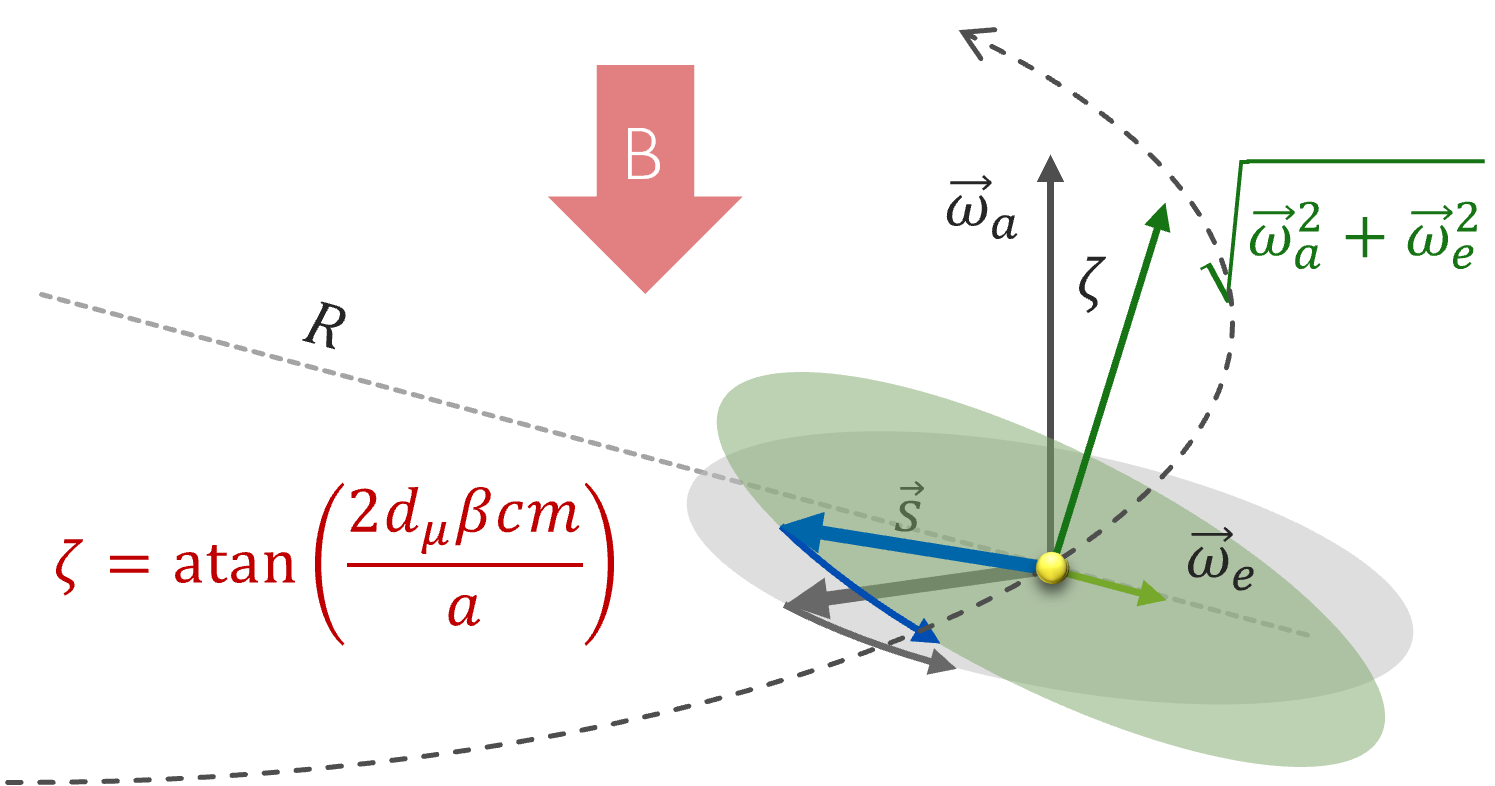}
	\label{subfig:SpinPrecessionMuon}}%
	\hfill
	\subfloat[]{
	\includegraphics[width=0.47\textwidth]{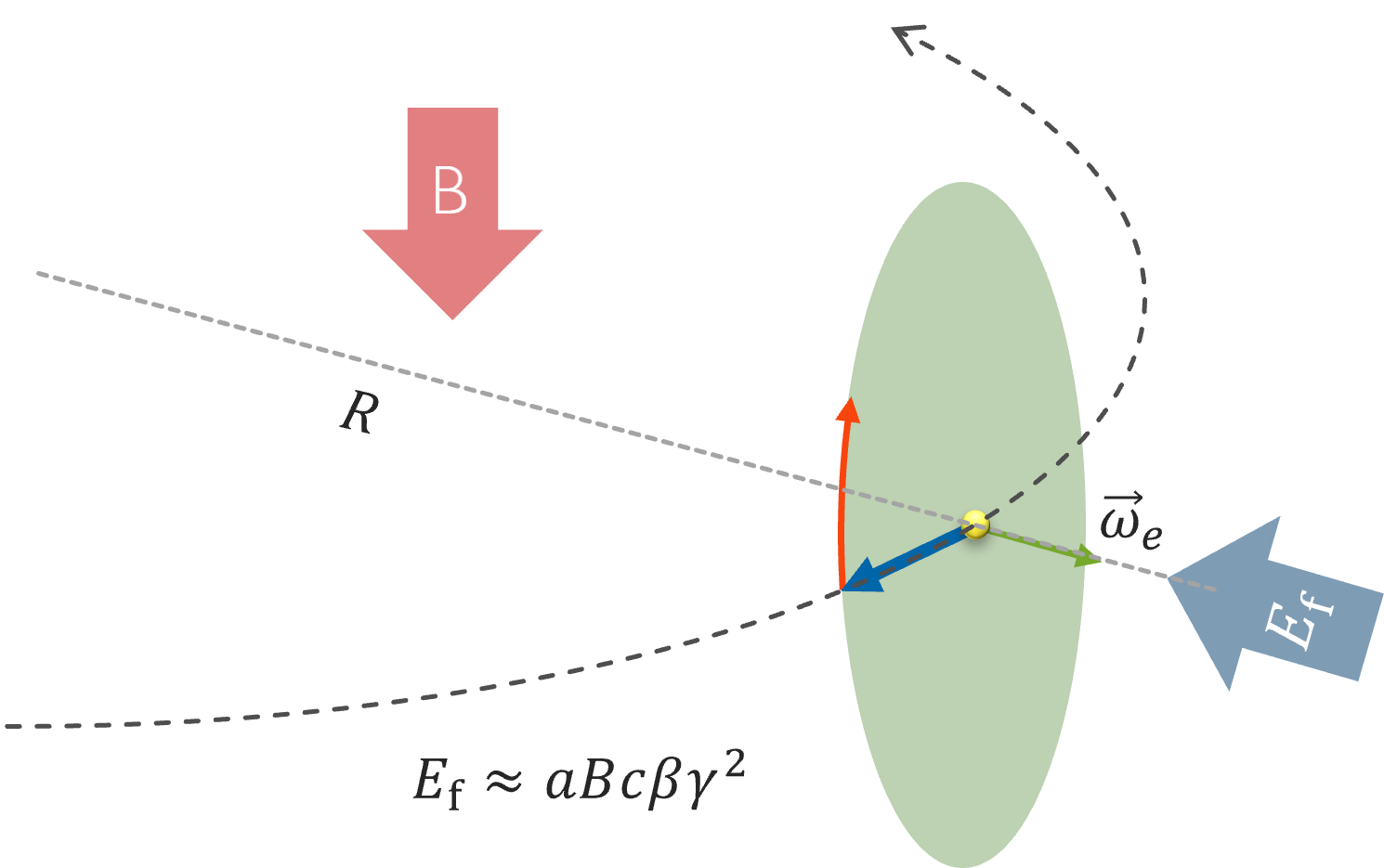}
	\label{subfig:FrozenSpinPrecessionMuon}}%
\caption{Spin precession of a muon stored on a circular orbit of radius $R$ using a magnetic field $\vec{B}$. (a)~Combined precession due to AMM and EDM, illustrated by the green tilted disc, in a configuration optimized for the search for the $(g-2)$ of the muon~\cite{Abe2019,Abi2021PRL}. (b)~Precession due to the EDM only, in a storage trap configuration using the frozen-spin technique to zero the AMM precession~\cite{Farley2004PRL,Adelmann2010JPG,LOI2021} by applying a radial electric field along the entire orbit.}%
\label{fig:SpinPrecessionRing}%
\end{figure}

In the presence of a muon EDM, the precession plane, shown as a green disc in Fig.~\ref{subfig:SpinPrecessionMuon}, will be tilted out of the orbital plane defined by the muon movement by an angle $\zeta =\arctan\left(2d_\mu\beta m c/a\right)$.
As a consequence, a longitudinal oscillation becomes observable shifted by $\pi/2$ in phase, relative to the transverse oscillation, characteristic for the AMM and the frequency changes to $\Omega=\sqrt{\omega_a^2 + \omega_e^2}$.
The amplitude of the longitudinal oscillation is proportional to $\zeta$, and hence the EDM\@. 
Note that attributing the observed precession frequently uniquely to the AMM would cause an overestimate of the AMM, as $\Omega>\omega_a$.  

%
%
%
If the electric field is adjusted to fulfill the condition 
\begin{equation}
		a\vec{B} = \left(a-\frac{1}{\gamma^2-1}\right)\frac{\vec{\beta}\times\vec{E}}{c},
\label{eq:FrozenSpinCondition}
\end{equation}
approximately by setting $E=E_{\rm f}=aBc\beta\gamma^2$ as proposed in Refs.~\cite{Khriplovich1998PLB,Farley2004PRL,Adelmann2010JPG}, no precession occurs in the absence of an EDM and the spin is ``frozen'' relative to the momentum vector.
%
In the presence of an EDM the spin will start to precess around the electric field $\vec{E}^\ast$, in the rest frame of the muon, as illustrated in Fig.~\ref{subfig:FrozenSpinPrecessionMuon}. 
By tracking muon decay positrons along and opposite to the magnetic-field direction, we measure the change in asymmetry over time.
Being proportional to the mean polarization,
%
%
\begin{equation}
	P(t) = P_0\sin\left(\omega_e t\right)	 \approx P_0\omega_e t = 2 P_0 \frac{d_\mu}{\hbar}\beta c B t,
	\label{eq:polarizationEDM}
\end{equation}
we deduce the EDM, where $P_0$ is the mean initial polarization after injection and we neglect the decoherence during storage. 
From the slope $\diff{P(t)}/\diff{d_\mu}$
%
%
multiplied by the mean spin analysis power in the weak decay, $\tilde\alpha$, we calculate the sensitivity as
\begin{equation}
		\sigma(d_\mu)=\frac{\hbar}{2P_0 \beta c B\sqrt{N} \gamma\tau_\mu \tilde\alpha},
\label{eq:EDMsensitivity}
\end{equation}
for a search for muon EDM by replacing $t$ with the mean free laboratory lifetime of the muon in the trap,
 $\gamma\tau_\mu$, and scaling by $1/\sqrt{N}$ for the Poisson statistics of $N$ observed positrons from muon decays. The detailed derivation can be found in Appendix~\ref{app:StatSens}.

\subsection{General experimental considerations}
A single measurement consists of the storage of a single muon on an orbit in a strong magnetic field in combination with a radial electric field adjusted to fulfill the frozen-spin condition.
This will be repeated, as often as the stability of all experimental nuisance parameters remain sub-leading, about 400 times per second for several hours. 
Then the magnetic field and the propagation orientation of the muon will be reversed and the single measurements continued until the next reversal, and so forth.

\begin{figure}%
\centering
\includegraphics[width=0.77\textwidth]{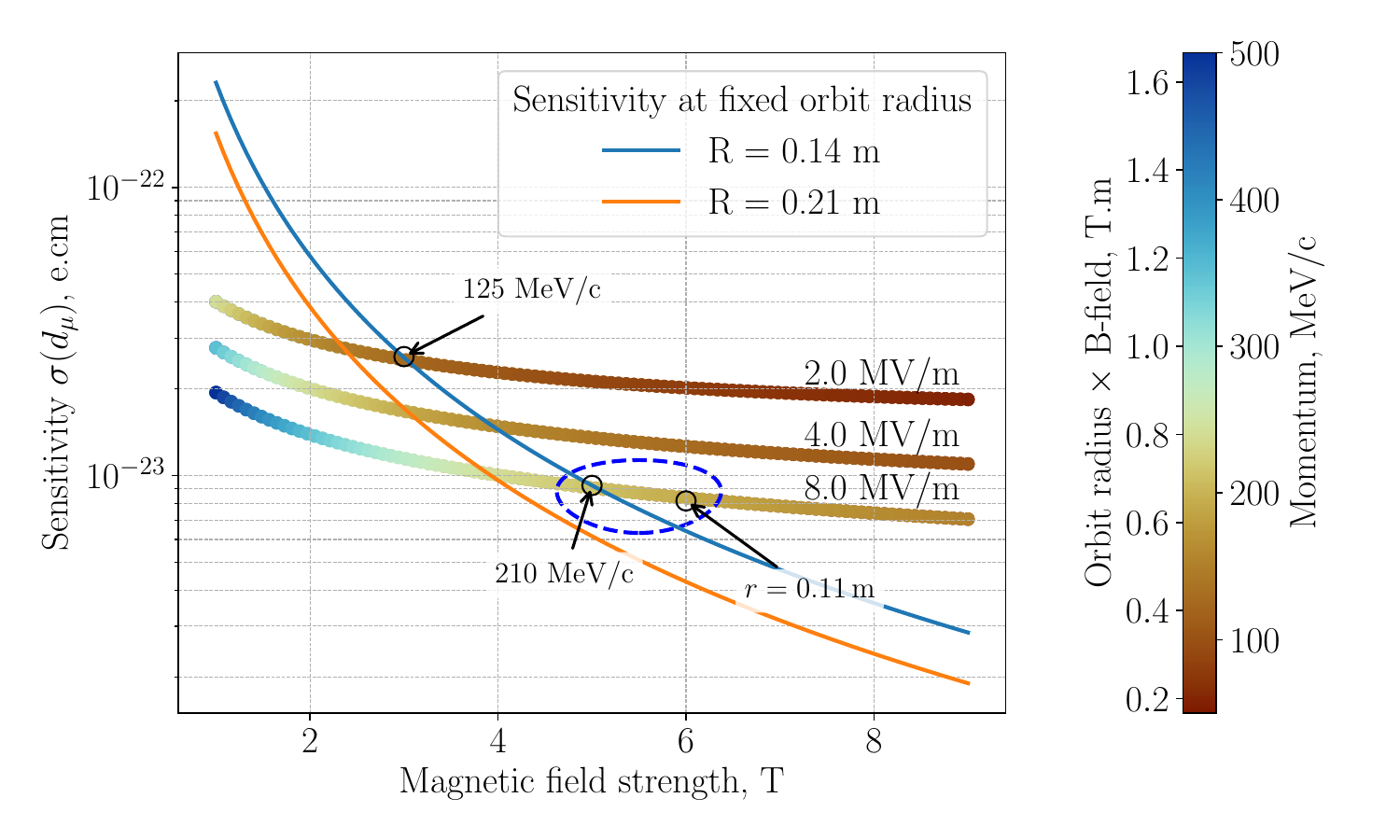}
\label{subfig:muEDMSensSingle}%
\caption{Statistical sensitivity for muon EDM searches over 200 days of data taking, assuming a single-muon ``on request.'' Sensitivity increases with magnetic field strength keeping the radius constant. Electric fields above \SI{8}{MV/m} are very difficult to obtain for large, stripped electrodes with extremely thin material thickness, required for positron transmission. The region within the circle indicates the possible parameter space for a muon EDM search using muons with momentum up to \SI{210}{\MeVc}.} %
\label{fig:muEDMsens}%
\end{figure}

Figure~\ref{fig:muEDMsens} shows the sensitivity landscape for single-muon storage ``on request'' for EDM searches using the frozen-spin method assuming 200 days of data collection.
``On request'' indicates the assumption that as soon as the previous muon decay was detected the next muon can enter the trap without additional delay.

The statistical sensitivity, c.f.\ Eq.~\eqref{eq:EDMsensitivity}, is inversely proportional to $\sigma\propto\left( B\sqrt{\gamma^2-1} \right)$ and experimentally limited by the achievable static electric field, $E_{\rm f}$, in vacuum. 
The exact value of $E_{\rm f}$ for which Eq.~\eqref{eq:FrozenSpinCondition} is fulfilled will be adjusted by measuring $\Omega(E_\rho)$ as a function of the applied radial electric field, $E_\rho$. 
Although static electric fields of $E=\SI{10}{MV/m}$ were reported for neutron beam experiments using sturdy massive electrodes separated by a gap of \SI{1}{cm}~\cite{Dress1977}, we assume that the strength of the electric field will be limited to $E=\SI{8}{MV/m}$, as fragile electrodes made of thin foils will not allow conditioning. 
 
The sensitivity increases with increasing magnetic field, $B$, for a constant orbit radius. 
Two examples are shown in Fig.~\ref{fig:muEDMsens} for $r=\SI{0.21}{m}$ and $r=\SI{0.14}{m}$, as an orange and blue line. 
The magnetic field can only be increased when the electric field increases at the same time, shown as three nearly horizontal colorful lines for $E=$~\SIlist{2;4;8}{MV/m}. 
The color scale of the equi-electric-field lines indicates the muon momentum matching the frozen-spin condition, as well as the orbit radius at this momentum multiplied by the B-field strength. 

Another constraint for the design of the optimal frozen-spin storage trap is the requirement that the positron detector, tracking decay positrons, must be placed within the muon orbit for muon momentum above \SI{100}{\MeVc}, as most positrons will be emitted inward. 
Taking into account all constraints, we highlight three possible design points of the Phase II instrument in Fig.~\ref{fig:muEDMsens}, indicated with ``\SI{125}{\MeVc}'', ``\SI{210}{\MeVc}'', and ``$r=\SI{0.11}{m}$''. The first two assume the same orbit radius of \SI{0.14}{m} on the blue upper curve, and only the magnetic and electric field are adjusted to the higher muon momentum, while the third option would also include a more compact positron detector for an orbit radius of only $\SI{11}{cm}$. 
In addition to the sensitivity for a single muon, it is also important to take the properties of the muon beam correctly into account, as only a small fraction of the beam phase space matches the acceptance phase space for trapping. 
For muon beams at $p=\SI{28}{\MeVc}$ and $p=\SI{125}{\MeVc}$ these have been studied. For the $p=\SI{210}{\MeVc}$ scenario, using muons from backward decays of pions with $p=\SI{370}{\MeVc}$, requires a modification of the existing beam line, a future option. 
Therefore, for Phase II we envisage an instrument that provides a magnetic field of up to $B=\SI{5}{T}$, an electric field of up to $E=\SI{8}{MV/m}$, and a detector system fitting into a muon orbit of $r=\SI{14}{cm}$.

Phase I measurement will use an existing solenoid with a maximum magnetic-field strength of \SI{5}{T}, connected to a surface-muon beam tuned to a momentum of $p=\SI{28}{\MeVc}$. As the muons pass a short distance through air, several plastic scintillators, and vacuum windows, the momentum of trapped muons will be in the range of \SIrange{22}{23}{\MeVc}.
The magnetic field then needs to be tuned to about $B=\SI{2.5}{T}$ to store the muon on an orbit with radius $r=\SI{30}{mm}$ resulting in a cyclotron period of about $\SI{3}{ns}$.
For such a small radius, a lateral injection, as in the (g-2)-experiment at Fermilab~\cite{Abi2021PRL} is not possible. 
Hence, we will adapt the longitudinal injection pioneered for the J-PARC g-2/EDM experiment~\cite{Iinuma2016NIMA,Abe2019} to inject muons into the solenoid. 
After a mean lifetime of $\gamma\tau=\SI{2.24}{\micro\second}$, the muon decays into a positron and two neutrinos. 
Detecting the change in the spatial distribution of positron tracks using a detector made of scintillating fibers, gives us access to the muon EDM\@.

\begin{figure}
	\centering
	\subfloat[]{\includegraphics[width=0.47\textwidth]{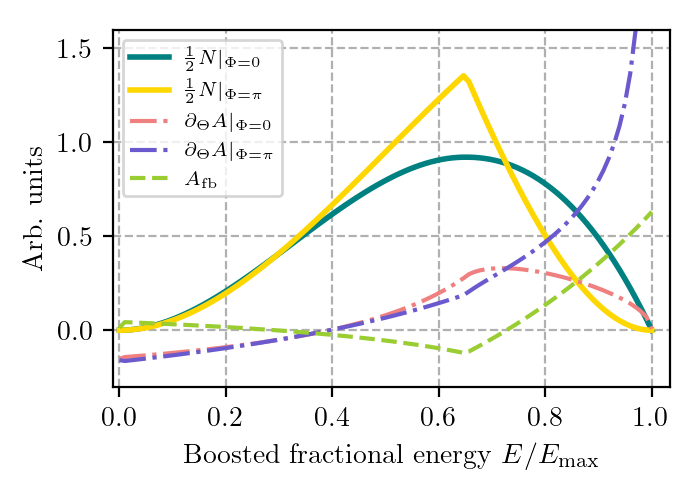}}
	\label{subfig:n_dot_a_23}
	\hfill
	\subfloat[]{\includegraphics[width=0.47\textwidth]{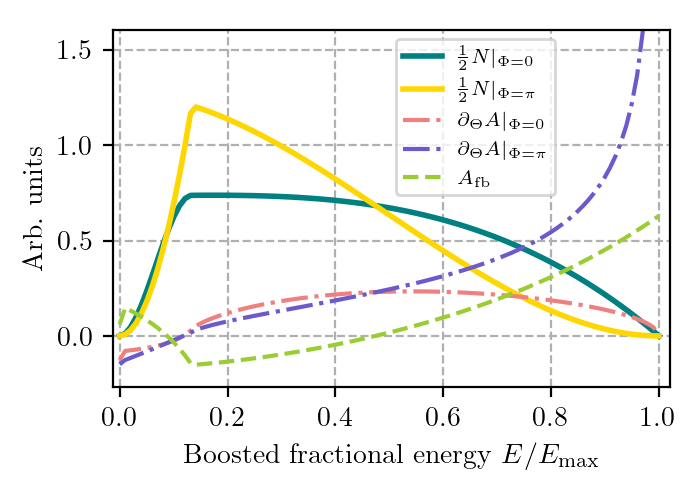}}
	\label{subfig:n_dot_a_125}
	\caption{Energy spectrum of decay positrons $N$, rate of change of asymmetry $\partial_\Theta A$ evaluated for the spin parallel, $\phi=0$, and anti-parallel, ($\phi=\pi$), respectively for (a)~Phase~I, with $p = \SI{23}{\MeVc}$ and (b)~Phase~II with $p = \SI{125}{\MeVc}$. The green dashed line shows the forward-backward asymmetry $A_\mathrm{FB}$.}
	\label{fig:n_dot_a}
\end{figure}

Due to parity violation of the weak decay, the direction of the ejected positron is correlated to the spin orientation.
Figure~\ref{fig:n_dot_a} shows the change in the decay positron energy distribution for two extreme spin orientations, along~(dark green) and opposite to (yellow) the muon momentum. 
Without an applied electric field, the measured positron energy distribution will oscillate between these two extremes as a function of decay time, allowing the extraction of the anomalous precession frequency $\Omega$.
The frozen-spin condition can then be found by measuring $\Omega$ as a function of the high voltage, $U$, and interpolating to $\Omega(U)=0$. 

Once the frozen-spin condition is met and in the case of a non-zero EDM, the muon spin will precess out of the orbit plane, c.f.\ Eq.~\eqref{eq:omegaMuWithEDM}, and start to point along or opposite to the magnetic field direction. 
By detecting the decay asymmetry, proportional to the longitudinal polarization $P(t)\propto(N_a-N_c)/(N_a+N_c)$, where $N_a$ and $N_c$ are the numbers of positrons tracked along or opposite to the $B$-field direction, we can detect a non-zero EDM\@. 
The directional decay asymmetry, $A_{\rm FB}$, as a function of the positron momentum is depicted in dashed light green in Fig.~\ref{fig:n_dot_a}.
In Phase I, our design goal for the tracking detectors is to be sensitive to positron momenta above \SI{23}{\MeVc}, which results in a mean asymmetry of about $\tilde\alpha\approx32\%$.

In summary, using Eq.~\eqref{eq:EDMsensitivity} with $\gamma=1.02$, $P_0=0.95$, $E_{\rm f}=\SI{2}{kV/cm}$ we expect a sensitivity of $\sigma(d_\mu)<\SI{3E-16}{\ecm}$ per muon, which translates to $\sigma(d_\mu)<\SI{4E-21}{\ecm}$ in a year of data collection, assuming $N=\SI{400}{\per\second}$ positrons detected, in Phase I\@. The characteristic parameters (see Sec.~\ref{sec:kinematics}) for detection ratio and mean decay asymmetry $\tilde\alpha$, and the expected sensitivity of Phase I\&II are listed in Table~\ref{tab:StatSens}.

\begin{widetext}
\begin{table}%
\def\arraystretch{1.2}
\centering
\caption{Annual statistical sensitivity of the muon EDM measurement of Phase~I and II\@. In Phase~I, we inject surface muons with $p=\SI{28}{\MeVc}$ into the solenoid. As the muons have to pass thin scintillators and air from the exit of the secondary beam line to the center of the solenoid, the final momentum will be $p=\SI{23}{\MeVc}$, corresponding to $\gamma=1.02$. The scenario ``\SI{210}{\MeVc}'' assumes a replacement of several elements in the existing beam line. Channel transmission and injection efficiencies were deduced in MC simulations. Note, only in the case that multiple muons may be injected and stored at once a sensitivity below \SI{1E-23}{\ecm} is possible. Values for $\tilde \alpha$ and the ratio between muon storage and positron detection rate are taken from Tab.\,\ref{tab:AnalysisMethods}.}\vspace{6pt}
\begin{tabular}{|l|c|c|c|}
\hline
& Phase~I & Phase~II &\\[-12pt]
& \SI{28}{\MeVc} & \SI{125}{\MeVc} &\SI{210}{\MeVc}\\ \hline
Muon flux /$\mu^+\si{\per\second}$ 				&	  \num{4e6}	& \num{1.2e8} & $\sim$\num{2.5e8} \\ \hline 
Channel transmission					&	 0.03 			& 0.005				& 0.01 		\\[-12pt]
\quad(\scriptsize{ratio: exit channel/exit beam})					&	  			&     & 		\\	\hline
Injection efficiency 				&	 0.004			& 0.17  			& 0.17			\\[-12pt]
\quad(\scriptsize{ratio: stored muons/exit channel})				&	 			&   			& 		\\	\hline
Muon storage rate /\si{\per\second} 		 	&	 500			  &	\num{100e3} &	\num{200e3}\\ \hline
Gamma factor $\gamma$					&	1.02				& 1.56 				& 2.24 \\ \hline
$e^+$ detection rate /\si{\per\second}	&	 400					& \num{90e3}  & \num{150e3}\\ \hline\hline
{\bf Detections per 200 days}	&	\num{5.8e9}&  \num{1.5E12} &  \num{3E12}\\ \hline \hline

Mean decay asymmetry $\tilde\alpha$			&	0.32			  & 0.30 &  0.28			\\	\hline
Initial polarization $P_0$	  & 0.95				&	0.95 &	0.95				\\	\hline \hline
Magnetic field $B$ 	/T				& 2.5           & 3    &  5    				\\	\hline

{\bf Sensitivity in one year} /$\ecm$&  <\num{4e-21} & < \num{6E-23} & < \num{1E-23} \\ \hline

\end{tabular}
\label{tab:StatSens}
\end{table}
\end{widetext}	

\section{Design considerations}
\label{sec:DesignConsiderations}
	\subsection{Analysis methods and figure of merit calculations}
    
\begin{figure}
    \centering
    \includegraphics[width=0.5\linewidth]{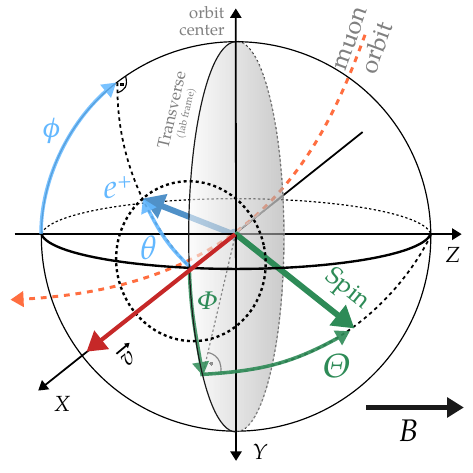}
    \caption{Schematic of the kinematic variables used to derive the muon decay asymmetry. Uppercase $\Phi, \Theta$ relate to the direction of the spin at the time of decay, and lowercase $\phi, \theta$ describe the direction of emission of the decay positron. The angle between the muon spin and the positron momentum in the muon rest frame is denoted with $y$ (not shown here).}
    \label{fig:MuDecayKinematics}
\end{figure}

If the muon spin has a component in the longitudinal direction, then the
probability for positron emission along $p_\uparrow$ or opposite $p_\downarrow$ to the direction of the main magnetic field will differ and an asymmetry,
\begin{equation}
	A = p_\uparrow - p_\downarrow,
	\label{eq:asym_theory}
\end{equation}
proportional to the longitudinal polarization, $P(t)$, will be observed.
The probabilities $p_{\uparrow,\downarrow}$ are calculated from the kinematics of the Michel decay~\cite{Michel1950,Bouchiat1957,Kinoshita1957} and depend on the spin direction at the time of muon decay, $\Phi$ and $\Theta$, and on the direction of the positron emission, $\phi$ and $\theta$ (see Fig.\,\ref{fig:MuDecayKinematics}).
The reference frame is chosen such that $\Theta = \pi / 2$ coincides with the magnetic field direction and $\Theta = 0$ is in the plane of the muon orbit.

In the rest frame of the muon, neglecting radiative corrections and the positron mass, the differential distribution of Michel decay positron is
\begin{equation}
	W(x, \cos y)dx~ d(\cos y) = x^2 \left((3-2x) + (2x - 1)\cos y\right)dx~d(\cos y),
\end{equation}
where $x = E/E_\mathrm{max}$ is the relative positron energy, and $y$ is the angle between the momenta of the muon and decay positron.

The experimental observable for an EDM signal in the laboratory reference frame is the time derivative of the asymmetry
\begin{equation}
	\frac{\partial A}{\partial t} = \frac{\partial
		A}{\partial \Theta}\frac{\partial \Theta}{\partial t}  = \partial_\Theta A~ \omega_e = \frac{2}{(p_\uparrow + p_\downarrow)^2} \left(
	p_\downarrow \frac{\partial p_\uparrow}{\partial \Theta} -
	p_\uparrow\frac{\partial p_\downarrow}{\partial \Theta}
	\right) \dot \Theta ,
	\label{eq:asym_theory_5}
\end{equation}
with $p_{\uparrow,\downarrow} = p_{\uparrow,\downarrow}(\Theta(t))$, $\omega_e =
	\dot\Theta$, and $\partial_\Theta A \equiv \partial A / \partial \Theta$.
Note that in the laboratory reference frame, the energy of the positron is $E' = \sqrt{(E'_\parallel)^2 + (E'_\perp)^2}$ where $ E'_\parallel = \gamma(\cos y+\beta)E$, $E'_\perp = E\sin y$.
This is valid under the assumption that the rest mass of the positron is much smaller than its kinetic energy.
Here, $\parallel$ and $\perp$ denote the components parallel or perpendicular to the muon velocity, $\vec{v}$. 
For arbitrary spin directions, we make the substitution $\cos y \rightarrow \cos \theta \sin(\Phi-\phi)\sin\Theta + \sin \theta \cos\Theta$.

We define the Lorentz boosted variables $u$ and $v$:
\begin{equation}
	u = \frac{E'}{E'_\mathrm{max}} = x\frac{\sqrt{\gamma^2 (\cos y+\beta)^2 + 1 - \cos^2 y}}{\gamma(1+\beta)},\quad
	v = \cos\left(\arctan\frac{E'_\parallel}{E'_\perp}\right) = \frac{1}{\sqrt{(E'_\parallel/E'_\perp)^2 + 1}}.
\end{equation}
The boosted angular and energy distribution is then:
\begin{equation}
	\int_0^1\int_{v_\mathrm{lim}}^{1} W'(u,v)\frac{\partial x}{\partial u} \frac{\partial y}{\partial v} dvdu=\int_0^1\int_{-1}^1 W(x,\cos y)d(\cos y)dx,\text{ where }
	v_\mathrm{lim} = \frac{u + \beta - 1}{u\beta}.
\end{equation}
Integrating the distribution over all kinematically allowed angles we obtain the energy spectrum of decay positrons
\begin{equation}
	N(u) = \int_{v_\mathrm{lim}}^{1} W'(u,v)\frac{\partial x}{\partial u} \frac{\partial y}{\partial v} dv,
\end{equation}
which implicitly depends on the spin direction $(\Phi, \Theta)$.

As the sensitivity to an EDM is inversely proportional to $\tilde\alpha\sqrt{N}$ (see Eq.~\eqref{eq:EDMsensitivity}), we maximize the figure of merit~(FoM) given by
\begin{equation}
	F = \tilde\alpha \sqrt{\frac{N_{e^+}}{N_{\mu^+}}},
	\label{eq:figure_of_merit}
\end{equation}
in the experimental design to minimize the statistical uncertainty of the measurement. 
Here $N_{e^+}$ is the number of detected decay positrons and 
\begin{equation}
	\tilde\alpha = \frac{1}{N_{\mu^+}(u_0)} \int_{u_0}^1 \partial_\Theta A(u) N(u) \diff u,
	\label{eq:tilde_alpha}
\end{equation}
is the weighted average of the rate of change of the parity violating decay asymmetry, taken above energy threshold $u_0$.
The normalization factor $N_{\mu^+}(u_0)$ is the total number of muons decaying to positrons with energy above the threshold $u_0$. 
The energy spectrum $N(u)$ of the decay positrons and the derivative of the asymmetry with respect to spin $\partial_\Theta A(u)$, evaluated at first order, are shown in Fig.~\ref{fig:n_dot_a} for Phases~I and II, respectively. The effect of higher order correction~\cite{KinoshitaPR1959} as calculated in Ref.~\cite{BanerjeeSciPost2023} are considered for analysis.

Neglecting constraints from solid angle limitations and energy thresholds of the positron detection scheme, the simplest form of analysis is taking the asymmetry of all positrons emitted along and counter to the magnetic-field direction. 
This corresponds to setting $u_0 = 0$ in Eq.~\eqref{eq:tilde_alpha}.
In this case, the FoM is
$F = \tilde\alpha = 0.1655$, for both phases, close to the
value $1/6$ expected from the Michel decay asymmetry without taking into
account radiative corrections. 
In the following, we show that an improvement is possible by, ({\it i\,}) setting an energy threshold (T-method), $u_0$, above which all positrons are used in the analysis, or ({\it ii\,}), a binned analysis (W-method), taking into account the emission angles $(\phi, \theta$) and the energy $u$ of each positron. 
%

%

%
For the T-method, we evaluate Eq.~\eqref{eq:tilde_alpha} as a function of the low-energy positron threshold $u_0$. The maximum figure of merit $F = 0.221$ at $u_0 = 0.626$ and $F = 0.173$ at $u_0 = 0.183$, corresponding to thresholds of $E_\mathrm{thr} = \SI{41.4}{MeV}$ and
$\SI{25.2}{MeV}$ in Phase~I and~II, respectively.

\begin{figure}
	\centering
	\subfloat[]{\includegraphics[width=0.49\columnwidth]{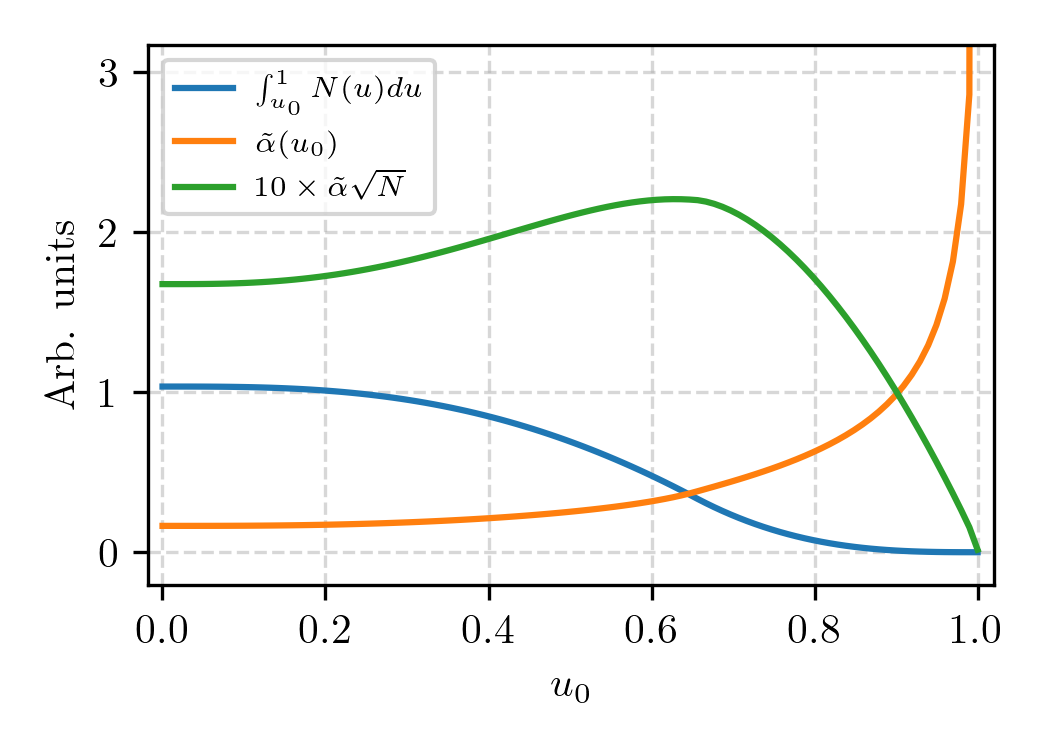}}
	\subfloat[]{\includegraphics[width=0.49\columnwidth]{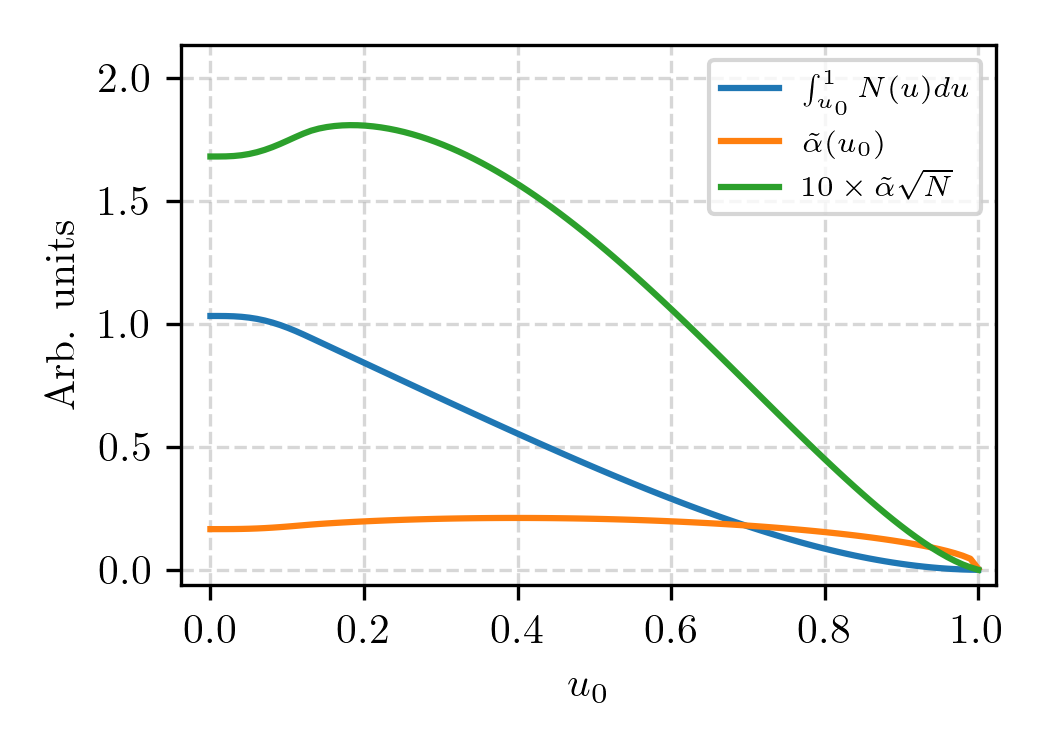}}
	\caption{Integral FoM considering the T-method as a function of the threshold $u_0$.
		(a)~Phase~I for $p = \SI{23}{\MeVc}$ with antiparallel muon momentum and spin and (b)~ Phase~II, $p=\SI{125}{\MeVc}$, parallel aligned muon momentum and spin. The figure
		of merit, $F = \tilde{\alpha}\sqrt{N}$, is multiplied by a factor 10 in order to keep a similar scale to
		the other curves.}
	\label{fig:integral_fom}
\end{figure}
%
The decrease of the FoM for Phase~II is the consequence of using right-handed muons with $p=\SI{125}{\MeVc}$ from backward decaying pions with $p_\pi=\SI{240}{\MeVc}$. Note that the usage of muons from forward decay is not possible as the the rates are much lower, the polarization worse, and the beam is contaminated with pions and positrons.
As the asymmetry changes sign with $u$, see Fig.~\ref{fig:n_dot_a}, a further increase in sensitivity is possible when the angular distribution of the decay positrons is taken into account.

We distinguish three variants of the W-method: ({\it i\,})~binning only in $u$,
({\it ii\,})~binning in $u$ and $\theta$ and ({\it iii\,})~binning in $u$, $\theta$ and $\phi$.
For each individual bin $(u, \theta, \phi)$ we define
\begin{equation}
	W_{ijk} = \alpha_{ijk} \sqrt{N_{ijk}},
	\label{eq:bin_fom}
\end{equation}
where
\begin{equation}
	\alpha_{ijk} = \frac{1}{N_{ijk}} \int_{u_{i-1}}^{u_i}
	\int_{\theta_{j-1}}^{\theta_j} \int_{\phi_{k-1}}^{\phi_k} \partial_\Theta A(u,
	\theta, \phi) N(u, \theta, \phi) \sin \theta~\diff u\,\diff\theta\,\diff\phi,
	\label{eq:bin_alpha}
\end{equation}
and
\begin{equation}
	N_{ijk} = \int_{u_{i-1}}^{u_i} \int_{\theta_{j-1}}^{\theta_j} \int_{\phi_{k-1}}^{\phi_k} N(u, \theta, \phi) \sin \theta~\diff u\,\diff\theta\,\diff\phi.
\end{equation}
The total FoM for the W-method is then
\begin{equation}
	\mathcal{W} =  \sqrt{\sum_{ijk} W_{ijk}^2},
	\label{eq:bin_fom_total}
\end{equation}
and the weighted mean rate of change of asymmetry,
\begin{equation}
    \tilde a = \frac{\sum_{ijk} \alpha_{ijk} N_{ijk}}{\sum_{ijk} N_{ijk}}.
    \label{eq:tilde_a}
\end{equation}
%

\begin{figure}
	\centering
	\subfloat[]{
	\includegraphics[width=0.49\columnwidth]{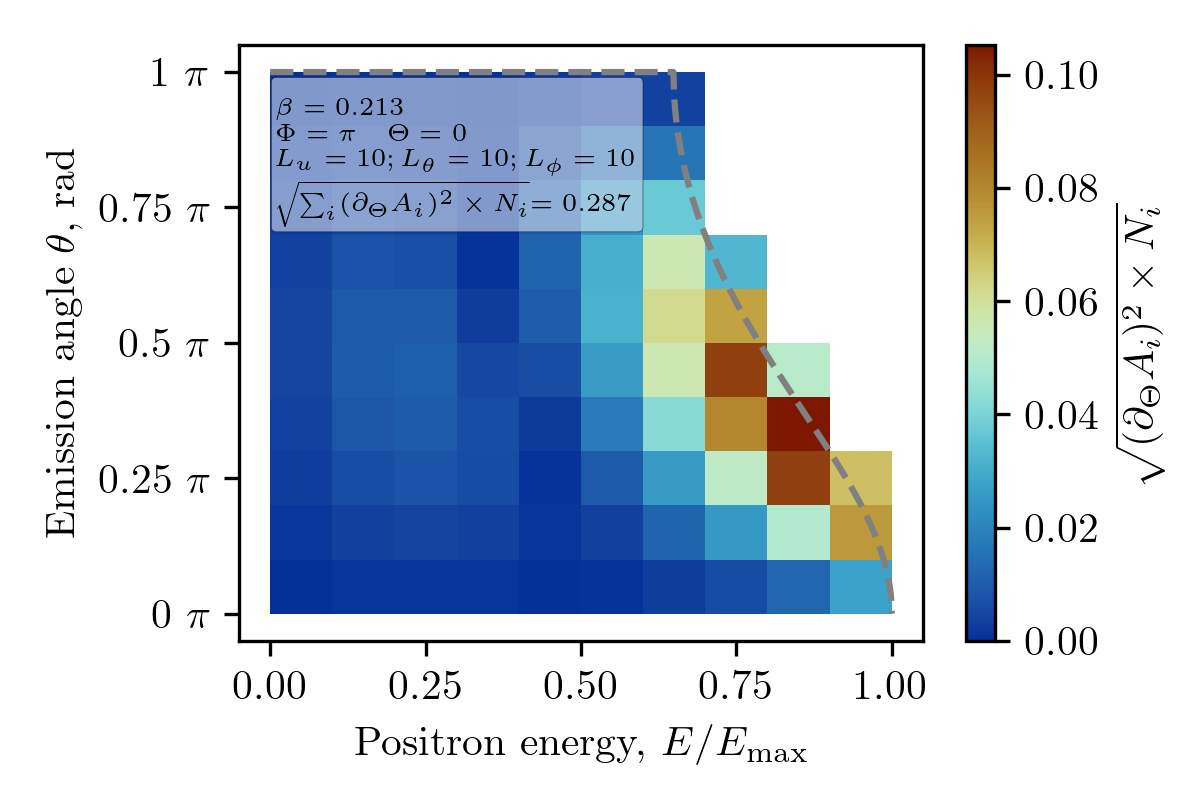}}
	\hfill
	\subfloat[]{\includegraphics[width=0.49\columnwidth]{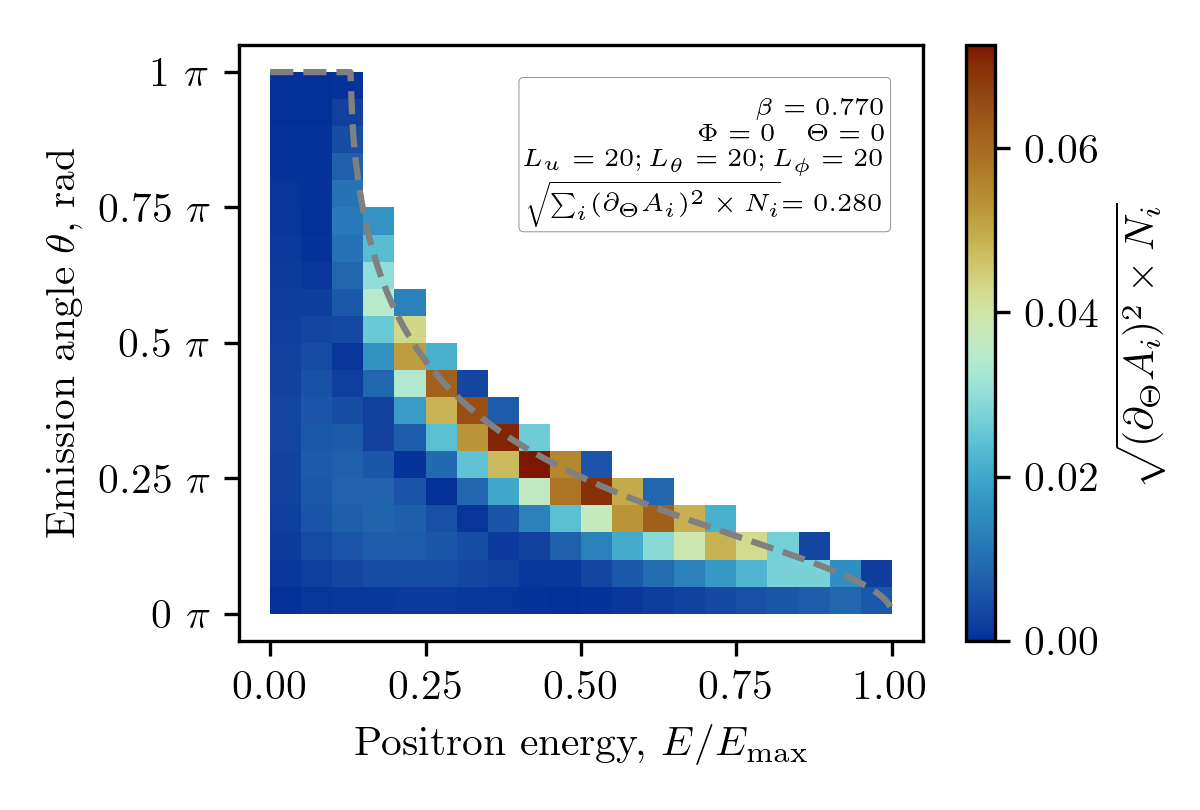}}
	\caption{Binned analysis using the W-method for the (a)~Phase~I experiment at $p=\SI{23}{\MeVc}$ and (b)~Phase~II at $p=\SI{125}{\MeVc}$. The dashed black line shows the
		kinematic constraint on the emission angle $\theta$. The distribution is shown only
		for $u$ and $\theta$ summed quadratically over $\phi$.}
	\label{fig:binned_fom_phaseI}
\end{figure}


Performing a binned analysis, the resulting FoM in bin $ijk$ and the deduced rate of change in asymmetry are
\begin{equation}W^2_{ijk} = \frac{(N_\uparrow - N_\downarrow)^2}{N_\uparrow + N_\downarrow},\text{ and }  \dot A_{ijk} = \frac{d}{dt} \left( \frac{N_\uparrow - N_\downarrow}{N_\uparrow + N_\downarrow} \right).
\end{equation}
The variance of $\dot A_{ijk}$ is $\sigma^2_{ijk} = W_{ijk}^{-2}$ and its weighted average
\begin{equation}
    \dot A_\mathrm{m} = \frac{\sum_{ijk} \dot{A}_{ijk} W_{ijk}^2}{\sum_{ijk} W_{ijk}^2}.
\end{equation}
%
The measured EDM is then
\begin{equation}
    d_\mu = \frac{\hbar}{2c} \frac{\dot A_\mathrm{m}}{\beta B P_0 }.
\end{equation}

In summary, $F = 0.1655$ is the baseline for Phase~I
and Phase~II when simply taking the asymmetry between positrons emitted along and opposite to the magnetic-field direction. 
Note, that any detection systems would result in a lower detection threshold and effectively a suboptimal T-method. 
By measuring the momentum of the emitted positron we can optimize the T-method improving  $F$ to $0.217$ for Phase~I and $0.178$ for Phase~II. 
Using a positron tracking detector, see Sec.~\ref{sec:PositronTracker}, we will measure also the emission angle and use the W-method for analysis, approaching the ideal value of $\mathcal{W}_{\mathrm{max}} \simeq 0.295$.
A summary of the parameters of interest is shown in Table~\ref{tab:AnalysisMethods}.

Hence, in Phase~I the positron tracker must be optimized for positrons with energies greater than \SI{27.6}{MeV} with emission angle $\theta$ between $\pi/6$ to $\pi/2$ with respect to the muon momentum.
By applying an energy threshold, any possible stochastic uncertainties resulting from the emission of positrons at a small angle and small momentum becomes negligible. In Phase~II emission angles up to $3\pi/4$ and energies above \SI{21}{MeV} are more sensitive to an EDM signal.

\begin{table}
\centering
\caption{Summary of analysis methods. The ratio $N_{e^+}/N_{\mu^+}$ is the fraction of detected positrons with respect to the total number of injected muons and $\tilde \alpha$ is the mean change in asymmetry above a threshold, relative to the maximum positron energies. The presented values are the theoretical maximum and do not include effects of multiple scattering of positrons or limited detector acceptance.}
\begin{tabular}{@{}p{2.2cm}p{2cm}ccc@{\hspace{10pt}}p{2cm}ccc@{}}
\toprule
\multicolumn{1}{c}{\multirow{2}{*}{Method}} & \multicolumn{4}{c}{Phase I @ $\SI{23}{\MeVc}$} & \multicolumn{4}{c}{Phase II @ \SI{125}{\MeVc}} \\ \cmidrule(l{2pt}r{10pt}){2-5} \cmidrule(l{2pt}r{2pt}){6-9} 
\multicolumn{1}{c}{} & Threshold \scriptsize$\times \SI{68.9}{MeV}$ & \multicolumn{1}{c}{$\tilde \alpha$} & \multicolumn{1}{c}{$N_{e^+}/N_{\mu^+}$} & \multicolumn{1}{l}{$F$} & Threshold \scriptsize$\times \SI{140.2}{MeV}$ & \multicolumn{1}{c}{$\tilde \alpha$} & \multicolumn{1}{c}{$N_{e^+}/N_{\mu^+}$} & \multicolumn{1}{c}{$F$} \\ \midrule
Simple & None & 0.166 & 1.0 & 0.166 & None & 0.166 & 1.0 & 0.166 \\ \midrule
T-method & 0.626 & 0.345 & 0.397 & 0.217 & 0.183 & 0.195 & 0.835 & 0.178 \\ \midrule
W-method  & None & 0.251 & 1.0 & 0.251 & None & 0.183 & 1.0 & 0.183 \\[-2pt]
\scriptsize(20 energy bins) & & & & & & & & \\
W-method & 0.4 & 0.280 & 0.800 & 0.250 & 0.15 & 0.194 & 0.876 & 0.183 \\[-2pt]
 \scriptsize(20 energy bins) & & & & & & & & \\ \midrule
W-method \scriptsize (20x20x20 bins) & None & 0.292 & 1.0 & 0.292 & None & 0.280 & 1.0 & 0.280 \\[-1pt]
W-method \scriptsize (20x20x20 bins) & 0.4 & 0.326 & 0.800 & 0.291 & 0.15 & 0.299 & 0.876 & 0.280 \\ \bottomrule
\end{tabular}
\label{tab:AnalysisMethods}
\end{table}

	\label{sec:kinematics}
 %
  \subsection{Overview of systematic effects}
    \label{Sec:SystematicEffects}
The principal goal of the frozen-spin technique is to increase the sensitivity to an EDM of a charged particle in a electromagnetic storage trap by adjusting an electric field to cancel any spin precession due to the anomalous magnetic moment~(AMM).
However, in a storage trap made up of magnetic and electric fields, this cancellation may not be perfect, and an EDM-like signal might appear from a coupling of the magnetic dipole moment~(MDM) to the electromagnetic~(EM) fields of the experimental setup. In the following we categorize all systematic effects arising from MDM coupling as real spin systematic effects.
However, effects that cause a change in positron detection asymmetry over time, but are not due to spin precession, are classified as apparent spin precession effects, such as drifts in the detector system's sensitivity or changes of acceptance during the characteristic time scale of the measurement. 

To evaluate systematic effects related to the electromagnetic fields in the experiment, it is necessary to study the relativistic spin motion in electric $\vec{E}$ and magnetic $\vec{B}$ fields described by the Thomas-BMT Eq.~\eqref{eq:omegaMuWithEDM}, and assuming that $\vec \beta \cdot
	\vec E = 0$, $\vec \beta \cdot \vec B = 0$, and the angular velocity of the spin due to a non-zero EDM is
\begin{equation}
	\omega_e =  \frac{2d_\mu}{\hbar}\left(\beta c B +E_\mathrm{f}\right).
	\label{eq:edm_angle_vs_de}
\end{equation}

For Phase~I and Phase~II of the experiment $\beta = 0.21$ and $0.77$, $B = \SI{2.5}{T}$ and \SI{3}{T}, respectively. The angular velocity corresponding to an EDM at the limit of our expected sensitivity for the two phases is then:
\begin{align}
	\omega_e^\mathrm{I}   = \SI{19.8}{\micro\radian/\micro\second} & \text{\quad for \quad}  \sigma(d_\mu) =
	\SI{4e-21}{\ecm},
	\label{eq:limit_on_angular_velocity_prec}                                                          \\
	\omega_e^\mathrm{II}  = \SI{0.84}{\micro\radian/\micro\second}  & \text{\quad for \quad }  \sigma(d_\mu)  = \SI{4e-23}{\ecm}.
	\label{eq:limit_on_angular_velocity_final}
\end{align}

Changes in the longitudinal polarization mimicking an EDM signal could be caused by:
$(i)$ a change with time of the mean radial $B$-field $\langle B_\rho \rangle \neq 0$ on the time scale of the muon lifetime. 
In this case, the dominant effect is a change in the detected asymmetry due to $\dot B_\rho$  from a change in muon momentum such that the orbit moves in a direction parallel to the field of the main solenoid until a new equilibrium position is reached. 
If this change of $\langle B_\rho \rangle \neq 0$ is correlated with the injection trigger, the mean of the longitudinal component of the momentum vector $\langle p_z \rangle$ is different from zero and changes over time resulting in an EDM-like signature. 
In the uncorrelated case, this effect reduces the sensitivity.
Note that a static $\langle B_\rho\rangle$ would merely shift the position of the equilibrium within the weakly focusing field but would not introduce a systematic effect. 
To minimize this systematic effect, we limit the radial $B$-field with frequencies in the range of \SI{100}{kHz} to \SI{1}{MHz}. 
This $B$-field could be due to residual current flowing through the kicker coils of the injection setup. 

Similarly, muons could experience $(ii)$ an azimuthal $B$-field $B_\theta$, which would result in a precession of the spin around the momentum on a cone when momentum and spin were not perfectly aligned, thus mimicking an EDM signal. 
The average over the orbit of this field component is non-zero only if there is electrical current flowing through the muon orbit, a net current of less than \SI{10}{mA} corresponds to a systematic effect of less than \SI{1E-23}{\ecm}.

A crucial component of the experiment is the radial electric field needed to freeze the spin motion. 
A systematic effect occurs if there is $(iii)$ a net longitudinal component of the field $\langle E_z \rangle \neq 0$. 
The consequences are twofold: the orbit would be displaced until a new equilibrium point is reached such that in the muon rest frame there is an effective radial $B$-field $\langle B_\rho^\ast \rangle = -\langle \gamma\beta E_z /c\rangle$, which also results in spin precession out of plane, generating a false EDM\@. 
The overall effect of a longitudinal $E$-field leads to 
\begin{equation}
    \Omega^{E_z}_\rho  =
	-\frac{ea}{mc} \left(1 - \frac{1}{a(\gamma^2 - 1)} -
	\frac{1}{\beta^2}\right) \beta \langle E_z \rangle.
 \label{eq:longitudinal_efield}
\end{equation}
The factor in brackets is of the order of \num{e3} for both experimental phases where $\gamma \approx 1$. This significantly amplifies the systematic effects due to $E_z$ compared to those due to a radial $B$-field. 
As other systematic effects related to the anomalous magnetic moment, this one is strongly suppressed by alternating clockwise (CW) and counter-clockwise (CCW) injection of muons to perform measurements in both configurations. This motivates the periodic vertical displacement of the entire apparatus by about \SI{90}{mm} and a simultaneous reversal of the direction of the magnetic field.
The measured asymmetry, defined by Eq.~\eqref{eq:asym_theory}, is based on positrons moving along or opposite to the main $B$-field. By combining the measurements made using both orientations of the magnetic field, the true EDM is effectively doubled, while the false signal is canceled. This corresponds to the method of reversing the electric field relative to the magnetic field in established EDM experiments~\cite{Chupp2019RMP}. In our case, it is the motional magnetic field, $v  E_z/c^2$, relative to the motional electric field, $v B_z$, in the muon rest frame which changes. 
As in foundational EDM searches~\cite{Graner2016PRL,Abel2020,Andreev2018}, a successful suppression of the false EDM depends on tightly controlling the stability of key parameters when reversing the relative field orientations.
 
These are: $(iv)$ the difference between the mean muon momentum $\Delta p$ averaged over all injected muons; $(v)$ the difference in the spin polarization and initial spin orientation; $(vi)$ differences in the residual $(g-2)$ spin-precession $\omega_z$ due to differences in the $E$- or $B$-field; $(vii)$ differences in the mean $E_z$, i.e., $\langle E_z^\mathrm{cw} \rangle \neq \langle E_z^\mathrm{ccw}\rangle$, which may occur if the muons are not traversing the same physical volume in the two experimental modes.

If there is a displacement between the mean center of the muon orbit and the axis of symmetry of the weakly focusing field, a magnetron oscillation will occur, causing the orbit to nutate about the axis of symmetry with $\omega_\mathrm{m} = \omega_\mathrm{b}/2\omega_\mathrm{c}$, similar to the behavior seen in Penning traps~\cite{Brown1982}. 
As the magnetron oscillation is proportional to the cyclotron frequency, it will change sign between magnetic-field configurations. 
As a result, muons in the CW and CCW configurations may experience different EM-fields. 
Therefore, it is crucial to control the displacement between the center of the muon orbit and the axis of symmetry of the weakly focusing field, along the gradients of the longitudinal $E$-field in the transverse plane, specifically $\partial_x E_z$ and $\partial_y E_z$.

Misalignment of or non-uniformity in the $E$- and $B$-field causes the spin to undergo small periodic movements around different axes. 
For example, the center of the muon orbit can shift from the central axis of the radial $E$-field, and the central axis of the $E$-field can be tilted relative to that of the main solenoid. 
Such a misalignment results in oscillations in the $(g-2)$ motion due to periodically lower and higher $E_\rho$ and oscillations in $E_z$ proportional to the tilt of the electrodes. 
Since rotations do not commute, the combined effect of these motions generates a net rotation around the radial axis, leading to a false EDM signal. 
This is an example of a geometric phase~\cite{Pendlebury2004}, also known as the Berry phase~\cite{Berry1984} causing a systematic effect, $(viii)$ . 

Another class of systematic effects involves the detection of decay positrons, which are used to deduce the spin phase. 
An example is $(ix)$ early to late variations in the EDM detection system such as gain, acceptance, and noise thresholds. 
If these changes differ between the detectors that monitor positrons that move along the $B$-field and those that move in the opposite direction, a false EDM could be generated. 
However, if the same set of detectors is affected in the same way for the CW and CCW configuration, the systematic effect would change sign, while the true EDM signal remains unaffected. 

In~\cite{Cavoto2024EPJC}, we derive analytical equations describing the precession due to AMM in the EM field, which were validated by \textsc{Geant4} Monte Carlo simulations using realistic field maps.
The agreement between the analytical model and the simulation results is excellent, demonstrating that the equations accurately describe the spin motion on short, medium, and long timescales.
Table~\ref{tab:systematics} provides a summary of all the significant systematic effects considered in this analysis.

\begin{table}
\caption{Summary of systematic effects. When known, both the limit value and the expected value of a given parameter are shown. The ``Limit value'' corresponds to an effect resulting in an false EDM signal of the size of the expected statistical sensitivity.}
\def\arraystretch{1.05}
\begin{tabular}{@{}lcccc@{}}
\toprule
    \multicolumn{1}{c}{\multirow{2}{*}{\textbf{Systematic effect}}} & \multicolumn{2}{c|}{\textbf{Phase I}} & \multicolumn{2}{c}{\textbf{Phase II}} \\ \cmidrule(l){2-5} 
    \multicolumn{1}{c}{} & \multicolumn{1}{c}{\textbf{\begin{tabular}[c]{@{}c@{}}Expected value\\[-6pt] (Limit value)\end{tabular}}} & \multicolumn{1}{c}{\textbf{\begin{tabular}[c]{@{}c@{}}Syst.\\[-6pt] \scriptsize \SI{e-21}{\ecm}\end{tabular}}} & \multicolumn{1}{c}{\textbf{\begin{tabular}[c]{@{}c@{}}Expected value\\[-6pt] (Limit value)\end{tabular}}} & \multicolumn{1}{c}{\textbf{\begin{tabular}[c]{@{}c@{}}Syst.\\[-6pt]  \scriptsize \SI{e-23}{\ecm}\end{tabular}}} \\ \midrule
Radial $B$-field $(i)$                & \SI{5}{\micro T}       & \multirow{2}{*}[3pt]{0.03}       & \SI{20}{\micro T}     & \multirow{2}{*}[3pt]{0.75} \\[-6pt]
@\SI{100}{kHz}                        &  (\SI{140}{\micro T})  &                             & (\SI{40}{\micro T})   &   \\ \hline 
Azimuthal B-field $(ii)$              & $< \SI{10}{mA}$        & \multirow{2}{*}[3pt]{\num{1E-2}} & $< \SI{10}{mA}$       & \multirow{2}{*}[3pt]{0.3} \\[-6pt]
 Current flowing through orbit        & (\SI{250}{mA})         &                             & (\SI{40}{mA})         &   \\ \hline
Longitudinal $E$-field $E_z$, $(iii)$ & \multirow{2}{*}[3pt]{$<\num{e-4}E_\mathrm{f}$} & \multirow{2}{*}{--}& \multirow{2}{*}[3pt]{$<\num{1.5e-5}E_\mathrm{f}$} & \multirow{2}{*}{--} \\[-6pt]
without CW/CCW storage                &                        &                             &                             &  \\ \cline{1-5}
\quad CW/CCW mean momentum            & 0.2\%                  & --                          & \multirow{2}{*}[3pt]{(0.1)\%} & -- \\[-6pt]
\quad difference $\Delta p$, $(iv)$   & (0.5)\%                & --                          &                          & -- \\ \cline{2-5}
\quad CW/CCW difference in initial    & \multirow{2}{*}[3pt]{\SI{25}{mrad}} & \multirow{2}{*}[3pt]{--} & \multirow{2}{*}[3pt]{\SI{5}{mrad}} & \multirow{2}{*}[3pt]{--} \\[-1pt]
\quad polarization, $(v)$             &                        &                             &                          &  \\ \cline{2-5}
\quad Radial $E$-field adjustment, $(vi)$ & 0.1\%              & --                          & 0.01\%                   & -- \\ \cline{2-5}
\quad Main $B$-field adjustment, $(vi)$   & 0.01\%             & --                          & 0.001\%                  & -- \\ \cline{2-5}
\quad CW/CCW orbit displacement, $(vii)$  & \SI{1}{mm}         & --                          & \SI{1}{mm}               & -- \\ \cline{2-5}
\quad $\partial_x E_z$, $\partial_y E_z$, $(vii)$  & (\SI{0.56}{mV/mm/mm})& --                          & (\SI{0.15}{mV/mm/mm})      & -- \\ \cline{1-5}
{\bf Total} $E$-field related systematics & -- & 0.75 & -- & 1.5 \\ \midrule
Resonant geometrical                & Pitch < \SI{1}{mrad}    & \multirow{2}{*}[3pt]{\num{2e-2}} & Pitch < 1 mrad       &  \multirow{2}{*}[3pt]{0.15} \\[-1pt]
phase accumulation $(viii)$           & Offset < \SI{2}{mm}     &                              & Offset < \SI{2}{mm} &  \\ \hline\hline
TOTAL &  & 0.75  &  & 1.70 \\ \bottomrule
\end{tabular}
\label{tab:systematics}
\end{table}

Systematic effects not only mimic an EDM signal but, when occurring stochastically, can reduce the overall sensitivity of the measurement. 
The stored muons in the trap exhibit a finite-momentum distribution, and the frozen-spin condition is met only on average.
This results in gradual de-phasing over time, diminishing the average polarization, which is directly proportional to the EDM signal (see Eq.~\eqref{eq:EDMsensitivity}).

\begin{figure}
    \centering
    \subfloat[]{\includegraphics[width=0.495\linewidth]{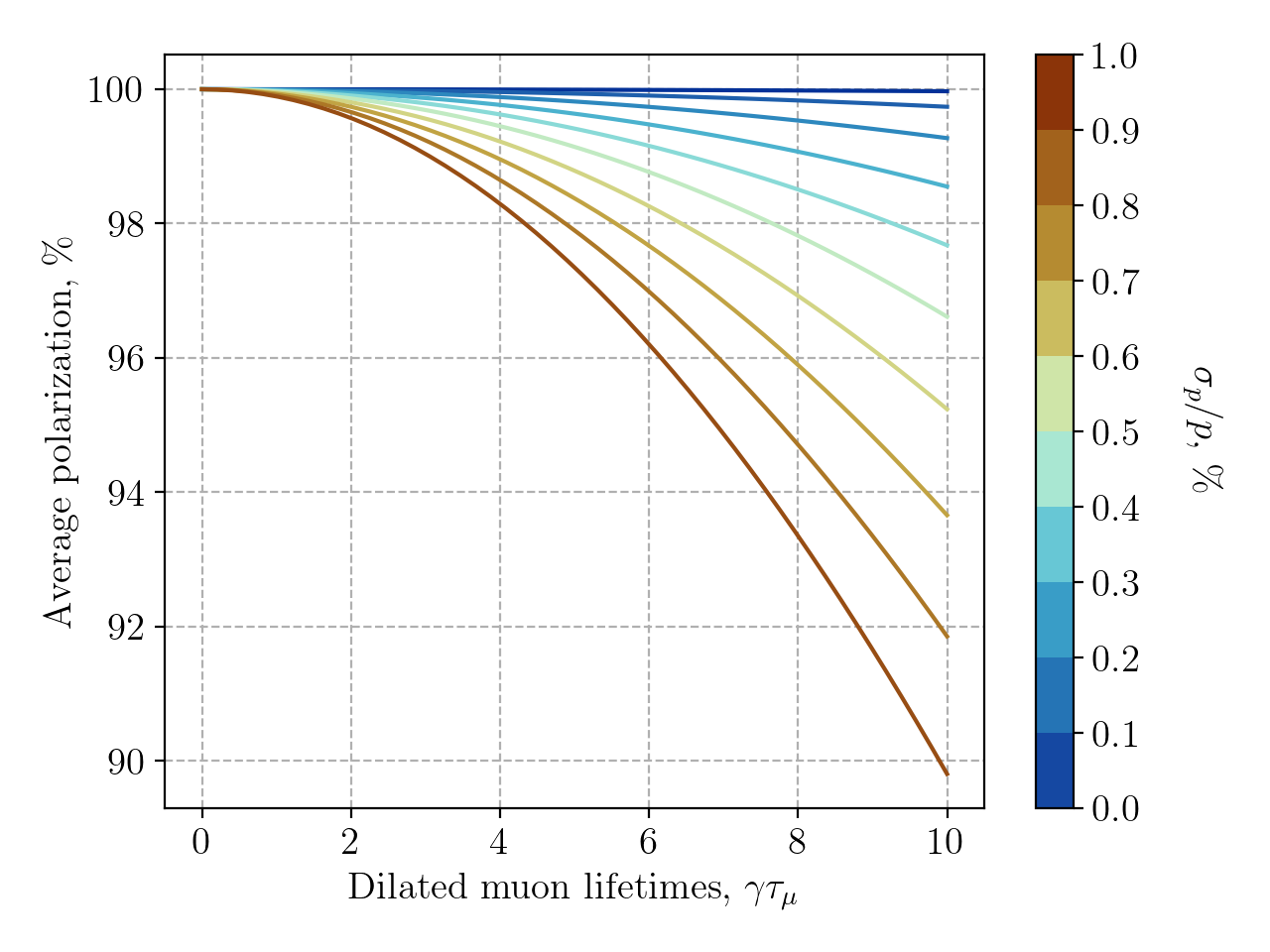}}
    \subfloat[]{\includegraphics[width=0.495\linewidth]{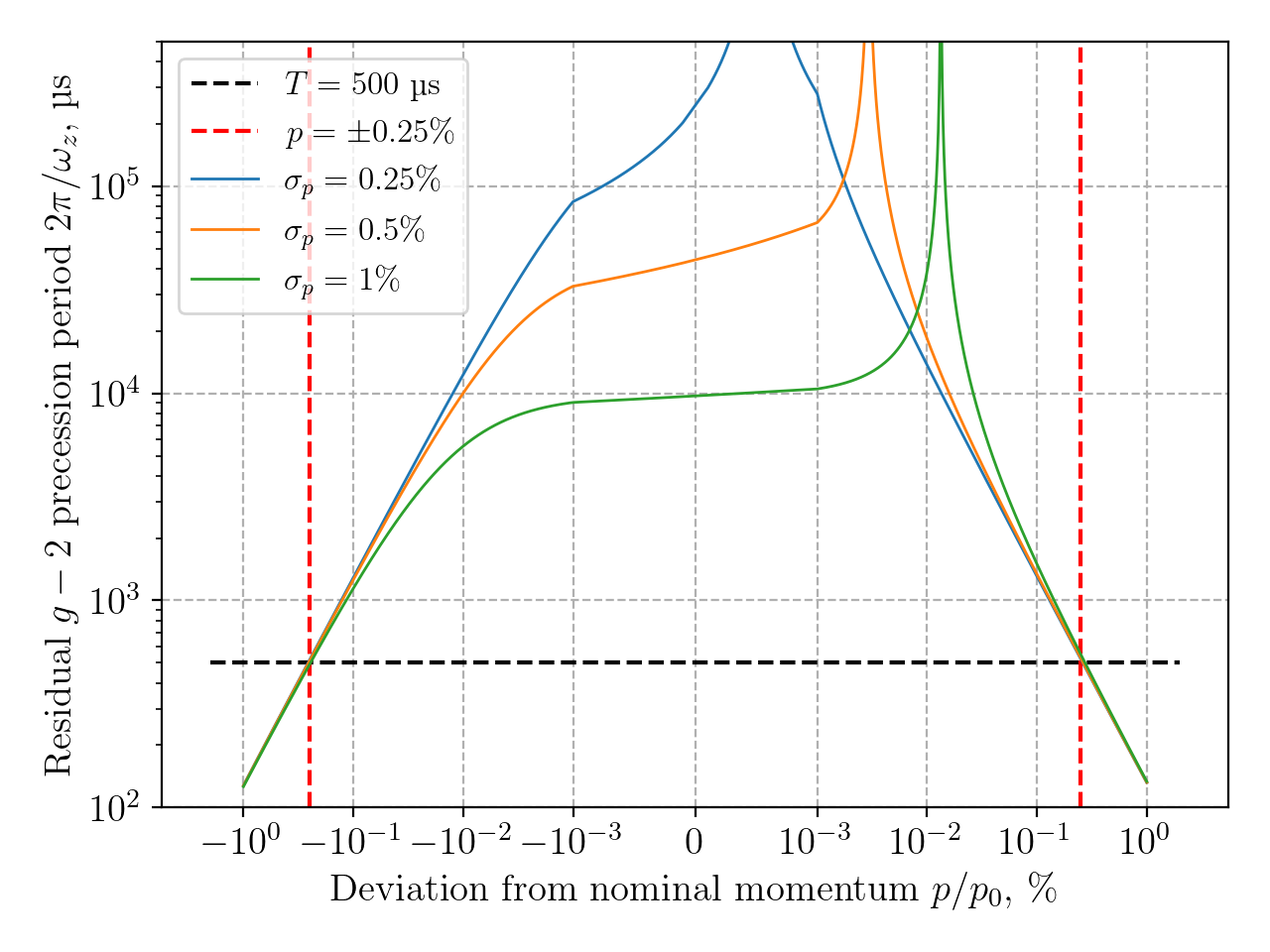}}
    \caption{Polarization, averaged over all detected muon decays, as a function of dilated muon lifetimes~(a) and residual $g-2$ precession period~(b), as a function of the momentum spread or deviation of the momentum in \%, assuming a Gaussian distribution with mean $p_0$ and standard deviation $\sigma_p$. }
    \label{fig:decoherence}
\end{figure}

Figure~\ref{fig:decoherence}~(a) shows the depolarization as a function of dilated muon lifetimes, $\gamma\tau_\mu$, for a Gaussian momentum spread ranging from $\sigma_p / p_0$ from 0\% to 1\%, in a \SI{3}{T} magnetic field.
The depolarization depends solely on the magnetic-field strength and the relative momentum spread, remaining unaffected by the nominal momentum or the distance between the electrodes providing the frozen-spin $E$-field.

Small changes in the magnetic field in the secondary beam line that transports the muons can result in a change of the central momentum $p$. 
This affects the frozen-spin condition and results in a residual g-2 precession, illustrated in 
Fig.~\ref{fig:decoherence} for different values of $\sigma_p$, where $\omega_z = -\frac{e}{m} (aB -E_f/(c\beta\gamma^2))$.
We observe that the standard deviation of the momentum distribution has no significant effect on the mean $g-2$ precession period. 
Considerations related to systematic effects, discussed above, limit the momentum difference between CW and CCW injections to 0.5\%, shown with a dashed red line.
The horizontal black dashed line indicates $T_{(g-2)} = 2\pi/\omega_z = \SI{500}{\micro\second}$,
approximately 40 times the measurement time of six dilated muon lifetimes $\gamma\tau_\mu$, $\gamma = 1.02$ for Phase~I\@, corresponding to 99.75\% of all decay positrons or 95\% of the statistical signal.

In conclusion, a 0.5\% relative standard deviation of the muon momentum results in less than 2\% loss of polarization for a \SI{20}{\micro\second} wide measurement window. 
The observed $g-2$ precession period is related to the deviation of the mean muon momentum from its nominal value.
As we record the momentum of all trapped muons using ToF, we can correct for any momentum deviation between the two injection modes, by muon selection.

  \newpage
\section{Features of the experimental setup}
The experimental setup is being optimized using Monte Carlo simulations, \textsc{musrSIM}~\cite{Sedlak2012PhPro}, \textsc{Geant4}~\cite{Agostinelli2003}, and \textsc{G4Beamline}~\cite{Roberts2011}, for diverse interfacing but independent studies. A genetic algorithm in combination with two surrogate models, which were based on a Monte Carlo~(MC) particle track simulation using \textsc{G4Beamline} was used to optimize the parameters relevant for a successful injection, resulting in a stored muon.

The injection efficiency of the system depends on various initial parameters, including the injection angles, the position of the injection channel, the position, dimensions, and current of the correction coils, as well as the strength and timing of the kicker pulse and the weakly-focusing field.
The geometry of and currents in the correction coils affect not only injection efficiency but also heat dissipation. 
The mean polarization, defined as the average spin orientation relative to the momentum vector, depends on the specific magnetic field experienced by each muon during injection and the stored momentum distribution.
A multi-objective optimization algorithm was implemented to find solutions characterized by the best compromise, also called Pareto fronts~\cite{Deb2002IEEE}.
This surrogate optimization combined with results from iterative optimization of specific features, e.g.\ magnetic pulse characteristics simulated using \textsc{ANSYS}, results in an injection efficiency of 0.45\% using the set of parameters listed in Table~\ref{tab:OptimalParameters}, which will be used as a baseline, subject to further refinement, for the setup of the muEDM Phase~I experiment.

\begin{table}%
\centering
\caption{Design parameters deduced in optimization of injection efficiency, mean muon polarization and minimal heat dissipation resulting in a spiral injection efficiency within the solenoid of 0.45\%.} 
\begin{tabular}{|r|l|r|}
		\hline 
		System 	& Parameter & Value \\ \hline\hline
		\multirow{3}{*}{Injection tube}  	& shielding factor SF & 100 \\ \cline{2-3}	
        & exit $(x,y,z)$  /mm	& $(0,\pm47,-455)$ \\ \cline{2-3}
& orientation $(\theta,\phi)$ /\degree   					& $(-42.3,10.6)$ \\ \hline
\multirow{2}{*}{Correction coil 1}	& (z-position, inner radius, outer radius, length) /mm  & ($285, 40, 70, 30$) \\ \cline{2-3} 
& current / $\si{A/mm^2}$	&1.8    \\ \hline
\multirow{2}{*}{Correction coil 2}	& (z-position, inner radius, outer radius, length) /mm  & ($235, 40, 60, 30$) \\ \cline{2-3} 
																				& current / $\si{A/mm^2}$																			&  0.744   \\ \hline
		\multirow{2}{*}{Weakly-focusing coil 2}	& (z-position, inner radius, outer radius, length) /mm  & ($0, 50, 60, 10$) \\ \cline{2-3} 
																				& current / $\si{A/mm^2}$																			&   1.5 \\ \hline	
		\multirow{2}{*}{Magnetic pulse coil}& (z-position, inner radius, outer radius, length) /mm  & ($0, 41,  42, 90$)\\ \cline{2-3}
																				& current per quadrant  / A	 																			&  192  \\ \hline
                                                                                Timing	& (delay, FWHM) /\si{\micro\second}										& (\SI{100}{ns}, \SI{35}{ns}) \\ \hline 
\end{tabular}
\label{tab:OptimalParameters}
\end{table}

\begin{figure}%
\centering 
\includegraphics[width=\columnwidth]{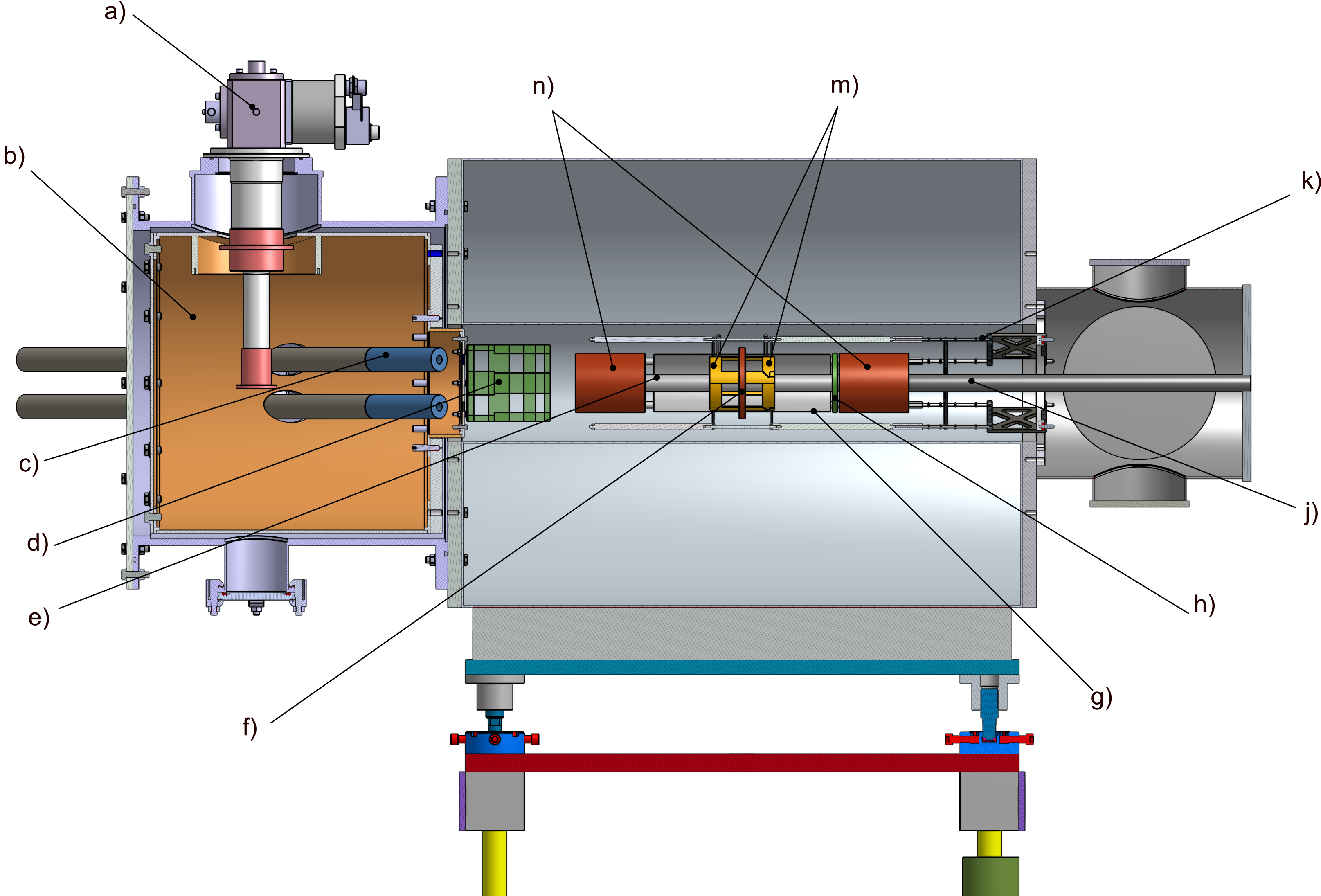}%
\caption{Computer rendered overview of the experimental setup for Phase I\@. See text for a description of all components.}%
\label{fig:CADOverview}%
\end{figure}

Figure~\ref{fig:CADOverview} shows an computer rendered overview of the experimental setup. For discussion of the experimental features, we follow the trajectory of the muon from the exit of the beam line until it decays into a positron. 
For injection, we will use a superconducting shield~(c) cooled to about \SI{4}{K} by a cryo cooler~(a), within a thermal shield~(b) at \SI{60}{K}. 
The gray tubes are made of normal steel and are used to shield the muon path in the magnetic field fringe region below \SI{0.5}{T}. 
As the muons leave the injection tubes, they pass through the entrance trigger unit~(d) creating a TTL signal if the muon matches the acceptance phase space for trapping. 
The magnetic field inside the SC-magnet is adjusted using a set of water-cooled correction coils~(n) and the weakly focusing coil~(f) to create the trapping potential. 
As the muon passes through the central region, the TTL signal will have launched a current pulse creating a short radial magnetic pulse using the pulse coil~(m). 
If the muon is not trapped it will be detected in the muon end detector~(h), otherwise it will circulate in the weakly-focusing field region until it decays into neutrinos and a positron. 
A meticulously adjusted radial electric field using the HV electrode~(e) connected to the HV feed-through~(j) and the ground shell~(g) establishes the frozen-spin condition. 
The positron detection system inside the HV electrode and outside the ground shell made of scintillating fibers is not shown for clarity. 
To minimize the multiple scattering of the positrons, the holding structure~(k) and all features are designed to minimize the material budget.

		\subsection{Properties of the surface muon beam}
				\label{sec:piE1}
				The experiment of Phase~I will be mounted on the $\pi$E1 beam line\footnote{\url{https://www.psi.ch/en/sbl/pie1-beamline}, last visited \today.} at PSI, a high-intensity pion and muon beam line with beam momenta ranging from \SIrange{10}{500}{MeV/c} and a momentum spread of better than \SI{0.8}{\%}. 
The beam is extracted from a thick carbon target in the forward direction and passes three dipoles to select the momentum. 
A horizontal slit system in the center of the second dipole, where the dispersion function is the largest, is used to select the momentum spread. 
The beam then passes through a Wien filter, also known as a separator or spin-rotator, which will be used to select a muon beam with very low contamination from positrons or pions. 
The perpendicular combination of a magnetic and an electric field changes the relative spin orientation of the incident muons as a function of field strengths, as the name ``spin-rotator'' suggests. 
By accurately adjusting the magnetic and electric fields of the separator, we will tune the initial muon spin orientation relative to the momentum.
Further downstream a fourth dipole deflects the muons at an angle of \SI{38}{\degree} into two quadrupole triplets to transport the beam to the experimental area. 

The beam was characterized, measuring the beam rate and the transverse phase space, by mounting a scintillating fiber~(SciFi) beam monitoring detector~\cite{Papa2015NIMA} in the focus of the last quadrupole triplet. 

\begin{figure}%
	\centering
	\subfloat[]{
			\includegraphics[width=0.48\columnwidth]{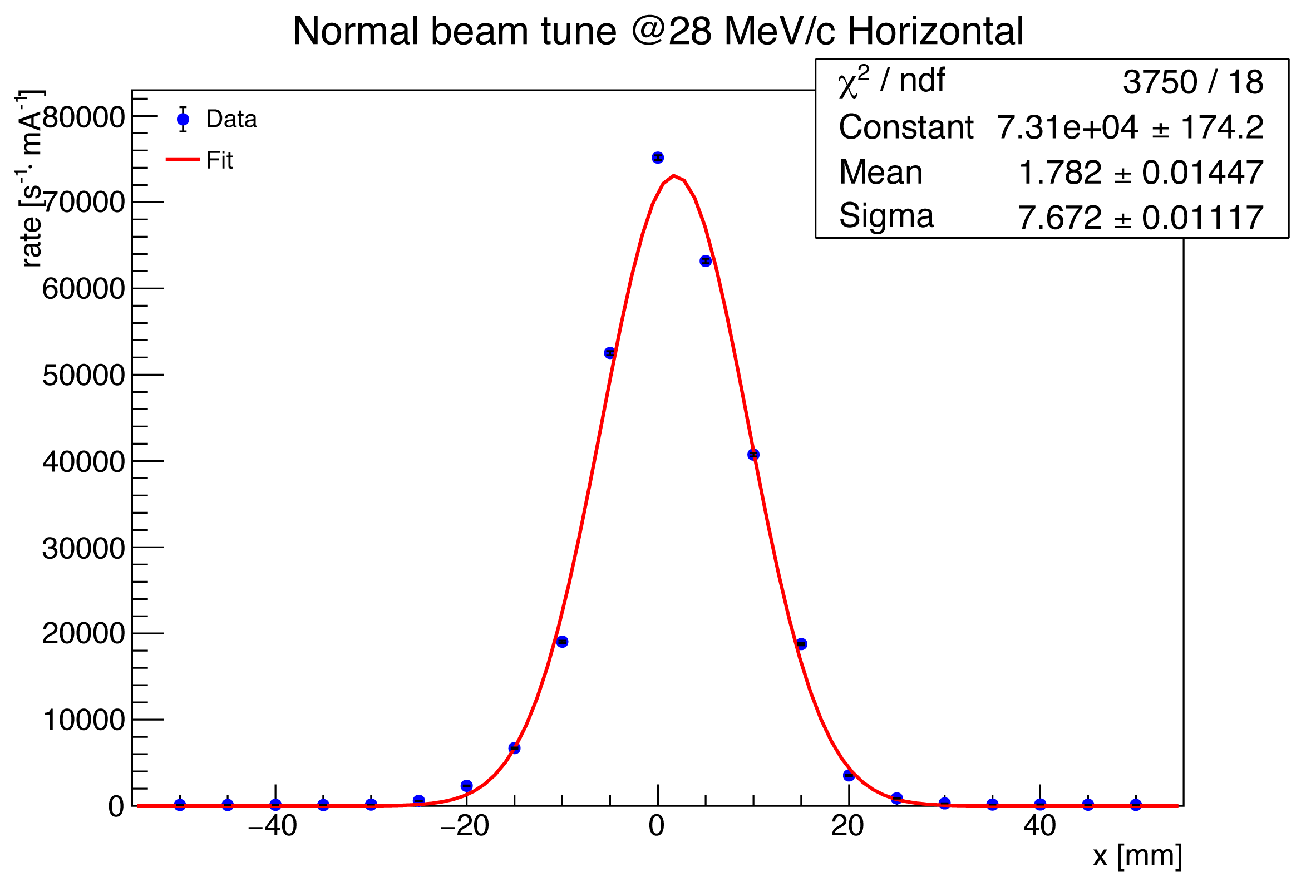}}%
	\hfill
	\subfloat[]{
			\includegraphics[width=0.48\columnwidth]{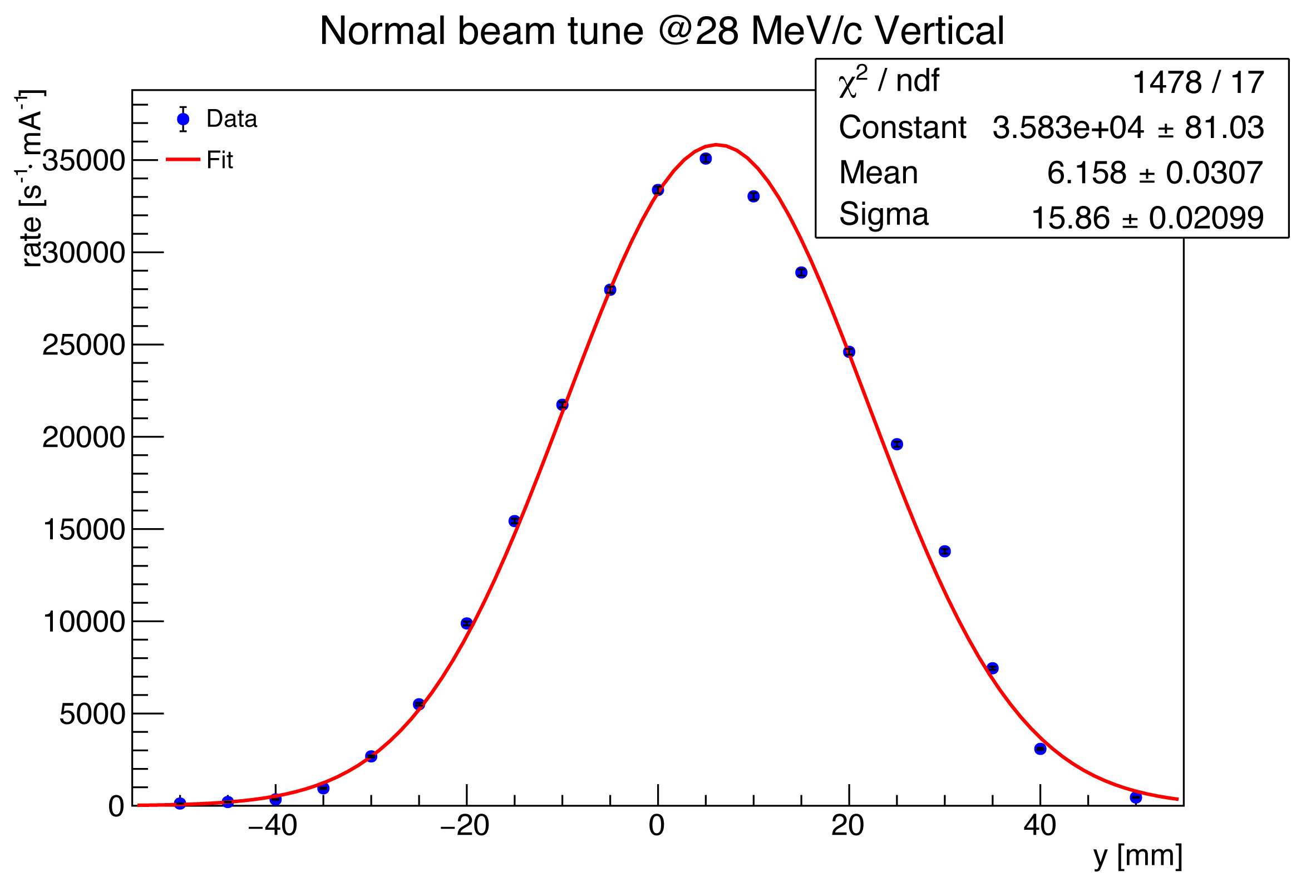}}\\%
	\subfloat{
		\includegraphics[width=0.49\columnwidth]{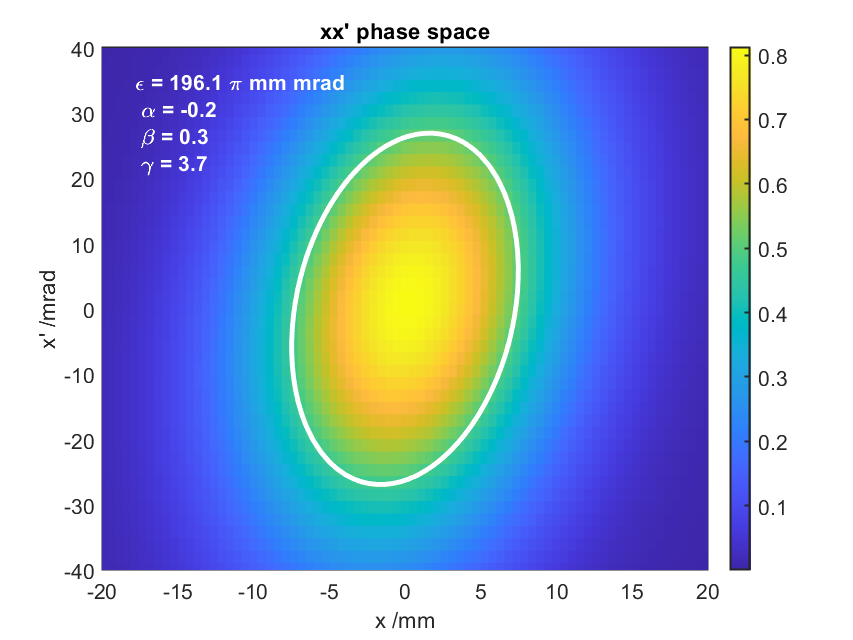}}
	\hfill
	\subfloat{
		\includegraphics[width=0.49\columnwidth]{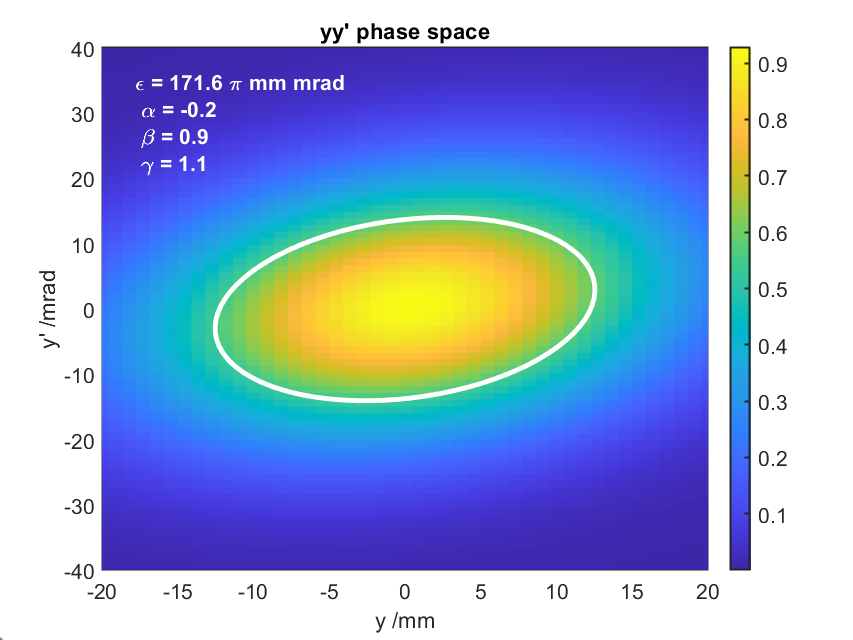}}
	\caption{(a)~Horizontal and (b)~vertical muon beam profile for muons with a momentum of \SI{28}{\MeVc} at $\pi$E1. Using the Twiss parameters summarized in Table~\ref{tab:piE1Twiss} we generated (c)~$x$ and (d)~$y$ phase-space distributions with \num{5e6} events as input for simulations of the transmission through the injection tube. }%
	\label{fig:BeamProperties}%
\end{figure}

The transverse phase space was explored by employing a quadrupole-scan technique for a positive muon beam momentum at \SI{28}{MeV/c} with varying settings of the transfer line magnets, described in detail in Ref.\,\cite{Sakurai2022PhD}. 
The beam profiles under the optimal settings are shown in Fig.~\ref{fig:BeamProperties}a,b, providing a flux of $R = \SI{5.6e6}{\mu^+/s}$ at a proton current of \SI{2}{mA}.

The Twiss parameters of the beam are summarized in Table~\ref{tab:piE1Twiss} and were used to generate an ensemble of Monte Carlo events (see Fig.~\ref{fig:BeamProperties}c,d) used as input for the simulation of the experiment.
Polarization $P_0>0.95$ is known from the ${\rm \mu SR}$ measurements\footnote{\url{https://www.psi.ch/en/smus/dolly}, last visited \today.}. 
\begin{table}%
\centering
\caption{Twiss parameters, $\alpha$, $\beta$, and  $\gamma$, and transverse emittance, $\epsilon$, of the $\pi$E1 beam line at the nominal injection point into the muon EDM experiment~\cite{Sakurai2022PhD}}
\begin{tabular}{|rl|r@{.}l|r@{.}l|}
\hline
	&								&	\multicolumn{2}{l|}{Horizontal (x)}					& \multicolumn{2}{l|}{Vertical (y)}\\ \hline
	$\alpha$& 								&	\qquad\qquad		 -0				&21						& \quad\qquad -0           & 24\\
 $\beta$ & /m							& \qquad\qquad 0								&28 					&	\quad\qquad	0						&90 \\
$\gamma$& /${\rm m^{-1}}$ & \qquad\qquad 3								&69						&	\quad\qquad	1						&17 \\  \hline
 $\epsilon$& /$\pi$ mm\,mrad & \multicolumn{2}{r|}{196}			& \multicolumn{2}{r|}{171} \\ \hline
\end{tabular}
\label{tab:piE1Twiss}
\end{table} 

		\subsection{Injection and magnetic and electric fields}
				\subsubsection{Magnetic field}
					\label{sec:MagneticField}
For the demonstration experiments in Phase~I we will use an existing superconducting solenoid, with mechanical bore dimensions of \SI{200}{mm} diameter and \SI{1000}{mm} length. 
The coils are made of multi-filament Nb-Ti cable and the magnetic field can reach \SI{5}{T}. 
In addition to the principal solenoid coil, a split coil pair also made of Nb-Ti cancels second- and fourth-order axial gradients resulting from the finite length of the solenoid coil.

\begin{figure}%
	\centering
	\subfloat[]{
		\includegraphics[width=0.47\columnwidth]{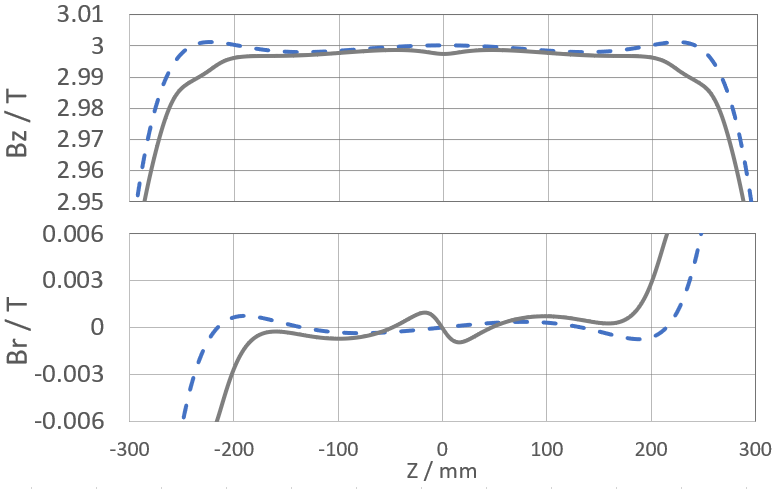}
		\label{subfig:FieldAlongZ}}
	\hfill
	\subfloat[]{
		\includegraphics[width=0.49\columnwidth]{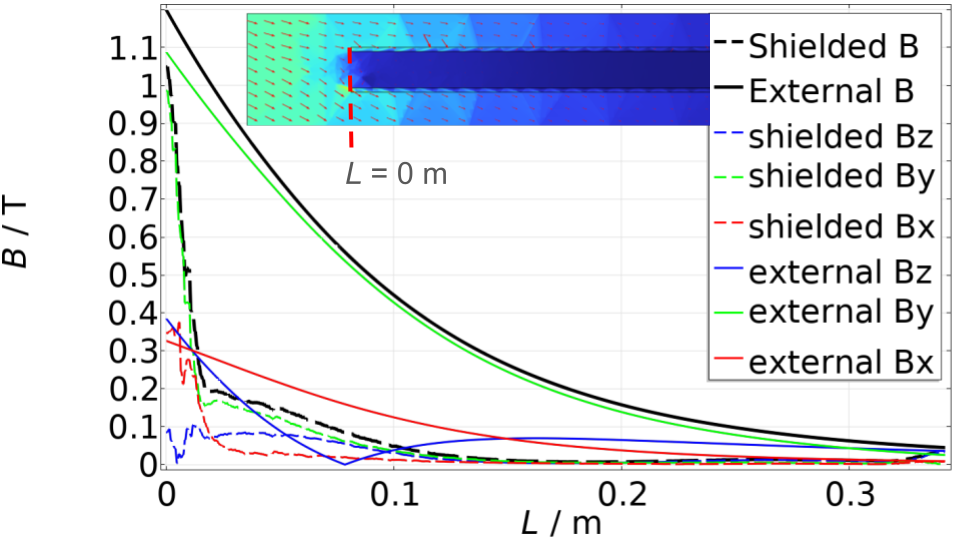}
		\label{subfig:FieldAlongTube}}
	\caption{(a)~Longitudinal~(upper), $B_z$, and radial~(lower), $B_r$, magnetic field  for $r=\SI{30}{mm}$. Continuous gray (dashed blue) line shows magnetic field with(out) weakly focusing and correction coils. (b)~Magnetic field, $\vec{B}$, along the injection tube with (dashed lines) and without (continuous lines) magnetic shield in local right-handed coordinates with $y$ pointing along the axis of the tube. The insert shows the a slice of the \textsc{Comsol}~FEM calculation along the injection tube. All transverse components remain below \SI{0.1}{T}.
    }%
	\label{fig:BFieldProfiles}%
\end{figure}

As no detailed technical documentation of the solenoid exists, the magnetic field was modeled by fitting the measured field data to a calculated field by adapting the parameters of the solenoid coil and the split coil pair in an \textsc{ANSYS} simulation.
The nominal muon injection spiral enters from the ends of the injection channels just within the bore, with a decreasing radius from about \SI{46}{mm} to the final muon storage radius of $\rho=\SI{30}{mm}$.
Figure~\ref{subfig:FieldAlongZ}
shows the magnetic field at the nominal storage radius, $\rho$, inside the bore. 
Without additional coils (blue dashed line), the split coil pair of the SC-solenoid clearly produces two field maxima at $\pm\SI{250}{mm}$. 
For a highly-efficient injection, a monotonically increasing magnetic field towards the center of the magnet is best and can be achieved by a set of correction coils positioned at these local maxima. 
In addition we produce a symmetric weakly-focusing field around $z=0$ generated by a local coil, for the storage of muons.
The effect of the coil is clearly visible in the gradient of $B_r$ in the central region, as shown by the gray continuous line in Fig.~\ref{subfig:FieldAlongZ}).

				\subsubsection{Magnetically shielded injection}
					\label{sec:Injection}
The muon transport into the solenoid bore is shielded from the fringe magnetic field along the two injection tubes to make deflection negligible and to maximize the muon transmission from the exit of the beam line.

Baseline parameters for the magnetically shielded injection tubes are a diameter $d_{\rm inj} = \SI{15}{mm}$ and length $\ell=\SI{800}{mm}$, a mean transverse magnetic field $B_{\perp}$, less than \SI{10}{mT}, and a longitudinal magnetic field along the injection tube $B_{\parallel}$, of less than \SI{1}{T}, resulting in a transmission of about~3\%. 
The simulated phase space at the end of the injection tube was used to generate \num{3e6} events for the next simulation step, the spiral injection.
Exact dimensions are subject of an optimization using numerical methods including measured magnetic shielding factors of prototypes.

%

The magnetic shield up to a magnetic field strength of about \SI{0.2}{T} will be made of cast iron providing a sufficiently large shielding factor~\cite{Noma1981NIM}. 
For fields above \SI{0.2}{T}, we will use a tube made of superconducting material~\cite{Barna2018,Nagasaki2018} to avoid any hysteresis effect close to the magnet bore. 
This SC magnetic shield will be made from any of three materials~\cite{Doinaki2023}: {\it i}\,) low temperature SC multilayer sheets made of NbTi/Nb/Cu, {\it ii}\,) HTS ReBCO tape wrapped around a metal tube, or {\it iii}\,) cast HTS Bi-2223~tubes\footnote{\url{http://www.can-superconductors.com/hts-bulks-and-materials/custom-hts/}, visited 31.01.2025.}, or a combination of them. 

The selection of the optimal design is made by testing the shielding factor, $f_s=B_{\rm sc}/B_{\rm wo}$, where $B_{\rm sc}$ and $ B_{\rm wo}$ are the magnetic field with and without magnetic shields made of different materials, respectively.
These measurements are used to adjust the input parameters of a finite element model implemented in \textsc{Comsol Multiphysics} to simulate the effectiveness and technical design of the magnetic shield. 
For this purpose, we reformulate the Maxwell equations for the electric field, $\vec{E}$, the magnetic-field intensity $\vec{H}$, and the current density $\vec{J}$.
 For the superconducting shield this is done using the ``H'' formulation~\cite{ShenIEEE2020},
\begin{align}
	 \nabla\times\vec{E}(\vec{J}) &= -\mu\Part{\vec{H}}{t} & \quad \vec{J} & = \nabla\times\vec{H}	\nonumber \\
		\vec{E} &= E_c\left(\frac{\vec{J}}{J_c(T,B)}\right)^n & \quad 		q &= \vec{J} \cdot \vec{E},
		\label{eq:H-formulation}
\end{align}
 where $J_c$ and $E_c$ are the critical current density and the critical electric field, respectively, and $n$ characterizes the steepness of the transition from superconducting to the normal state.
All the other domains are modeled with the ``A'' formulation,
\begin{align}
	\vec{B} & = \nabla\times\vec{A}	&\quad \nabla\times\vec{H} &= \vec{J} \nonumber \\
		\vec{J} &= \sigma \vec{E}_c+\vec{J}_e &\quad	\vec{E} &= -\Part{\vec{A}}{t},
		\label{eq:A-formulation}
\end{align}
where $\vec{A}$ is the magnetic potential and $\vec{B}$ the magnetic flux density.

The material properties depend on temperature, so the heat diffusion equation $\rho c_p(T) \partial_t T=\nabla\cdot(k(T)\nabla T)+q$ is coupled with the Maxwell equation. $\rho$ is the density expressed in $\si{kg\per m^3}$, $c_p(T)$ is the specific heat capacity expressed in \si{J\per(kg\,K)} and $k(T)$ is the thermal conductivity expressed in \si{W\per(m\,K)}.

The magnetic field along the last section of the injection channel for a reference trajectory with and without the SC magnetic shield is shown in Figure~\ref{subfig:FieldAlongTube}, for a simulation using very high critical currents.

				\subsubsection{Magnetic field pulse}
					\label{sec:B_FieldPulse}

A gate detector in combination with an active aperture, described in Sec.~\ref{sec:MuInjTrigger}, at the exit of the injection channels generates a trigger signal for every muon within the storage phase space. 
As the muons approach the storage region, their remaining longitudinal momentum, $p_z<\SI{1}{\MeVc}$, is significantly reduced from the initial value of $p_{z_0}\approx \SI{15}{\MeVc}$ at the gate detector, a consequence of the solenoid’s magnetic-field gradient.
To further reduce $p_z$, a radial pulsed magnetic field is employed. This field is generated at the solenoid’s center by a coil array designed to mimic the effect of a split coil pair with a radius of \SI{40}{mm}, separated by \SI{90}{mm}, and supplied with anti-parallel currents by a purpose-built pulse generator.
The static weakly-focusing field, detailed in Sec.~\ref{sec:MagneticField}, will thereafter provide the longitudinal confinement necessary to store the muon and permit spin-phase accumulation under the frozen-spin condition.
By optimizing the currents in the correction, weakly-focusing, and pulse coils, the phase space upon approach to the solenoid center can be matched to the acceptance phase space of the magnetic pulse.
The nominal values for these currents are determined by simulation (see Table~\ref{tab:OptimalParameters}), they will also be tunable and refined empirically during commissioning.

\begin{figure}
    \centering
        \subfloat[]{
    \includegraphics[width=0.45\textwidth]{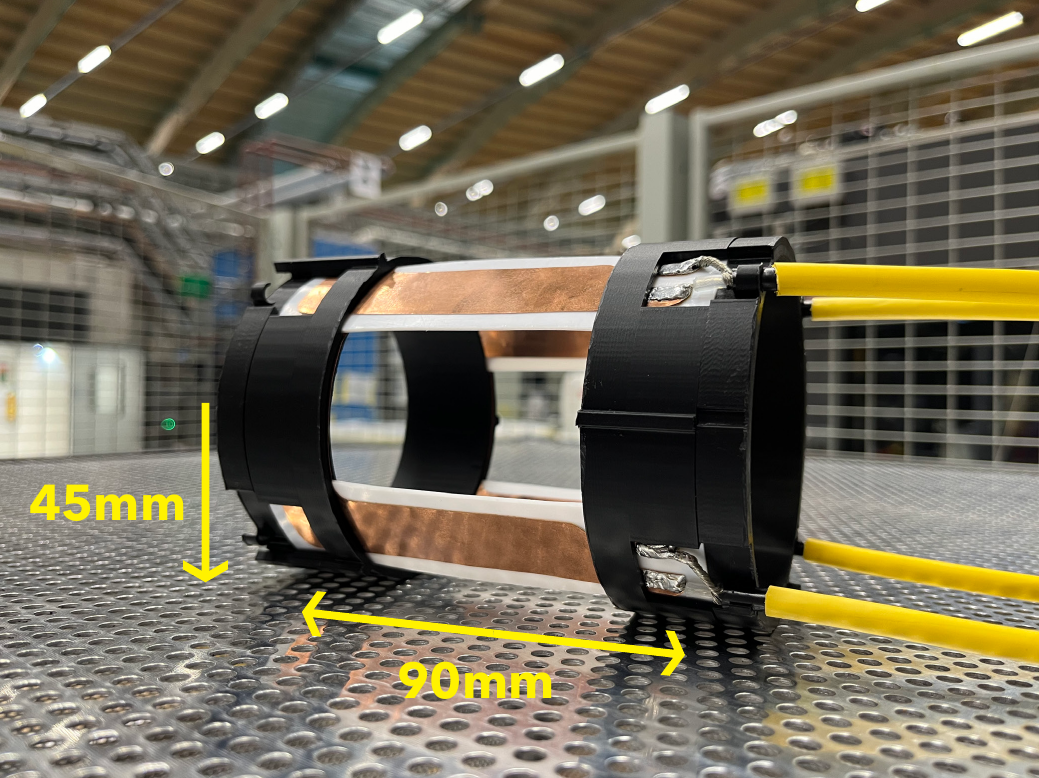}}
		\hfill
		\subfloat[]{
    \includegraphics[width=0.45\textwidth]{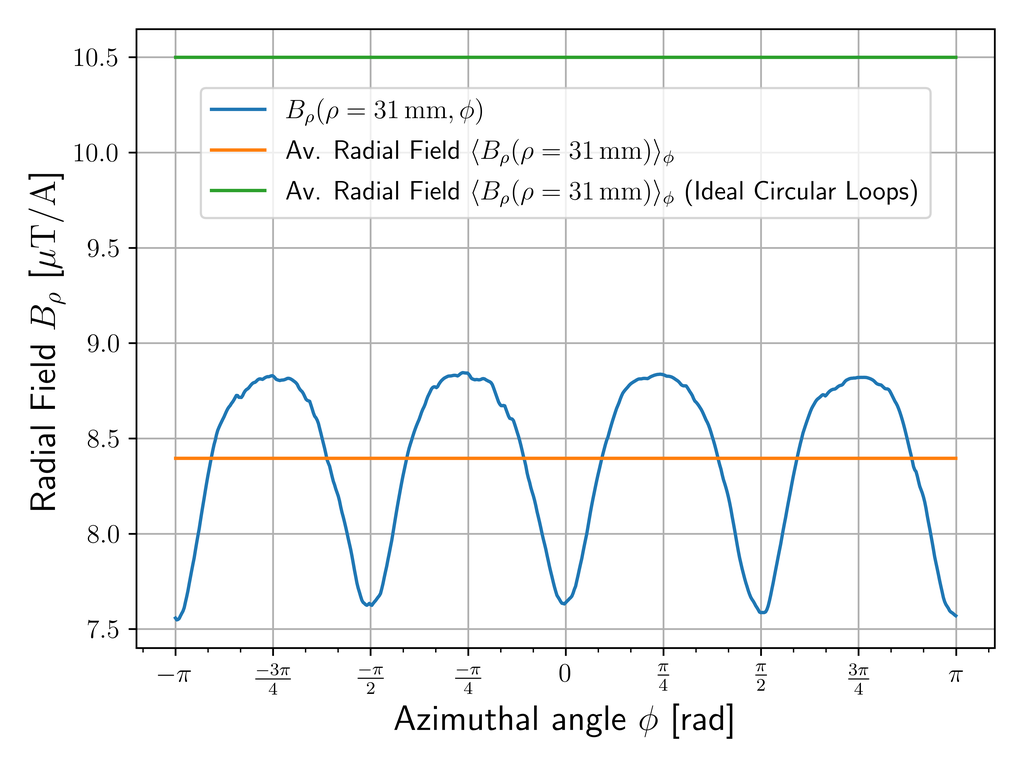}}
    \caption{Pulse coils prototype~(a), made of four quadrant coils. The copper tracks are~\SI{20}{mm} wide and the curved sections of radius~\SI{45}{mm} (to be reduced to \SI{38}{mm} in the final design) are separated by~\SI{90}{mm} from upstream to downstream (center to center of copper tracks). The average field strength over the orbit is \SI{8.4}{\micro\tesla\per\ampere}, as shown by the orange line in~(b), where the field strength is plotted over the orbit parametrized by the azimuthal angle $\phi$.}
    \label{fig:coils}
\end{figure}

A prototype of the pulse coil design is shown in Fig.~\ref{fig:coils}~(a), and was fully simulated
using an \textsc{Ansys} FEM model to characterize coil impedance and produce field maps, from which the field strength over the azimuthal angle of the orbit is plotted in Fig.~\ref{fig:coils}~(b).

By minimizing the inductance of the pulse coil circuit, the current pulse can be delivered at a lower operating voltage. This expands the selection of available MOSFETs that enable fast switching and reduces the required voltage ratings of other circuit components and cables.
Therefore, we effectively divide an anti-Helmoltz coil pair into four quadrants supplied in parallel, each with an inductance of approximately \SI{130}{nH}.
The prototype was made using \SI{100}{\mum} copper sheets, approximately twice the skin depth at the relevant frequency scale of \SI{10}{\mega\hertz}.
Thinner conductors are also favorable for the design to minimize multiple scattering of positrons, with \SI{100}{\mum} copper comparable to that induced by the layers of the scintillating fiber detector.
The plot in~(b) shows the field at each point along the orbit, parametrized by the azimuthal angle $\phi$.
The orange line at \SI{8.4}{\micro\tesla\per\ampere} indicates the average value over the entire orbit, which falls 20\% below the field strength expected for two ideal circular loops of the same separation distance with full $2\pi$ coverage.
Although such a field reduction necessitates a higher current for a given field strength, the expected operating voltage is similar for the ideal loops and the quadrant model because of the larger inductance of the full circular loops.
Therefore, the advantage of the quadrant model lies primarily in the shorter pulse width permitted by low-inductance parallel circuits.

To take full advantage of the quasi-continuous muon source with high flux at PSI, the pulse must be applied on request using the trigger signal. 
The expected transit time from the entrance to the storage is \SI{105}{ns}.
Accounting for trigger signal processing and transmission delays restricts the internal latency of the required pulse generator to less than \SI{60}{ns}.
The principal characteristics of the current pulse with a required maximum current of \SI{200}{A} and a FWHM of approximately \SI{35}{ns}, are defined by the inductance, resistances and stray capacitance of the pulse coil circuit.
A dedicated power supply will produce up to 2000 pulses per second on request, with a minimal inter-pulse spacing of \SI{20}{\micro s}.

\begin{figure}
    \centering

    \includegraphics[width=0.47\textwidth]{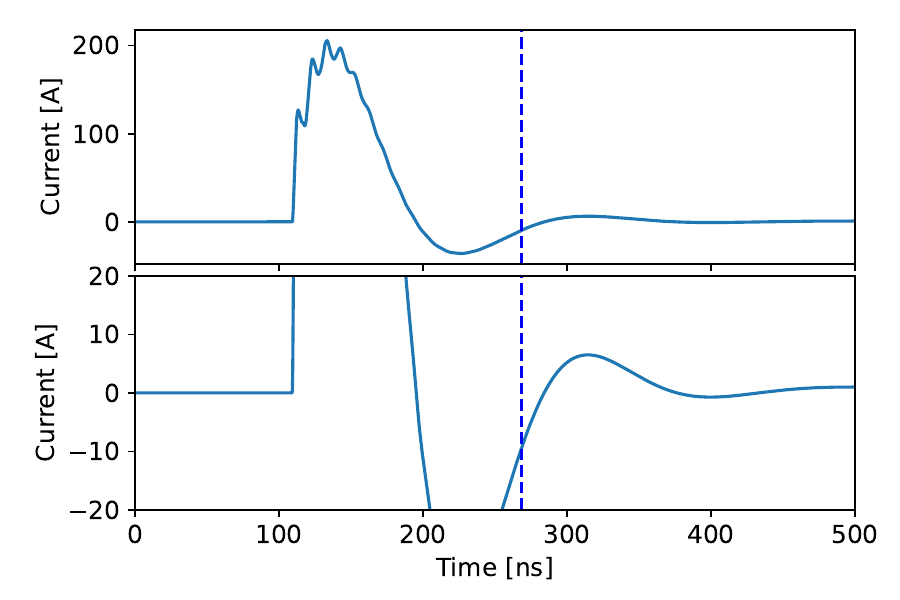}
    \caption{Pulse current profiles simulated using an \textsc{LTspice} circuit model with parameters matching the experimental apparatus. The lower plot shows a zoomed region in the range \SIrange{-20}{20}{A}. The current plotted is for one quadrant only. The dashed vertical line at \SI{260}{ns} indicates the time after which the current remains below $\leq 10\,A$. Note that for a limitation of systematic effects it is sufficient to guarantee a current below $I_{\rm  ripple}\leq\SI{10}{A}$ in a frequency range from approximately \SI{200}{kHz} to \SI{10}{MHz}.
		}
    \label{fig:pulses}
\end{figure}

The load characteristics defined from the described prototype and informed by the \textsc{ANSYS} simulations influence the shape and height of the pulse.
Constraints on pulse characteristics were derived based on injection simulations (see Tab.~\ref{tab:OptimalParameters}) and studies of systematic effect~\cite{Cavoto2024EPJC}, associated with stray and transient fields.
The parameters of the electronic circuit are tuned to critically damp the primary pulse and thus suppress any after-pulse oscillations that could perturb the muon orbit or induce a time-dependent spin precession signal. 
The integral radial magnetic field of the pulse seen by the muon determines the applied momentum kick, providing some freedom in the shape and intensity of the pulse.
The simulated current profiles shown in Fig.~\ref{fig:pulses} show the result for a pulse with optimized damping, such that the peak current is maximized~(a) and the after-pulse oscillations are reduced below \SI{10}{\ampere} within less than \SI{150}{ns}. 
The generator will be designed to achieve a recovery within 1\% of the nominal operating voltage after a \SI{20}{\micro s} delay, the minimum desired interval between muon triggers.

				\subsubsection{Frozen-spin electrodes}
						\label{sec:electrodes}
					The radial electric field will be applied using cylindrical concentric electrodes that surround the muon orbit at $r_{\rm b}=\SI{35}{mm}$ (grounded) and $r_{\rm a}=\SI{25}{mm}$ (high voltage).
A voltage of approximately \SI{1.8}{kV} will be required so that the field strength $E_{\rm f}\approx\SI{1.8}{kV\per cm}$ is applied at the nominal orbit radius, $\rho$, to achieve the frozen-spin condition.

Muons with slightly higher (lower) momentum will orbit with a correspondingly larger (smaller) radius and thus experience a lower (higher) field due to the radial dependence of the electric field.
However, a higher (lower) momentum would actually require a higher (lower) field strength to satisfy the frozen-spin condition, since $E_{\rm f}$ scales linearly with the muon velocity in the relevant range.

Thus, a momentum bite of 0.5\% for the injected muon beam leads to a limit of approximately 1\% for matching the frozen-spin field strength over the momentum distribution, leading to a negligible de-phasing.
Therefore, a relative precision better than 0.5\% on the applied voltage (approximately \SI{10}{V}) is sufficient.
However, controlling systematic effects arising from a longitudinal electric field (see Sec.~\ref{Sec:SystematicEffects} and Ref.~\cite{Cavoto2024EPJC}) requires that the voltage stability be constrained to approximately $10^{-4}$.
A bipolar high-voltage supply voltage with a ripple of less than 0.0125\% and a peak voltage of $\pm\SI{20}{kV}$ enables the calibration of the frozen-spin electric field strength by observing the spin precession due to the AMM, $a$, when the electric field is offset from $E_{\rm f}$, given by 
\begin{equation}
    \omega(E_{\rm offset}) = -\frac{a \, e}{m}B_z\frac{E_\mathrm{offset}}{E_{\rm f}} 
\end{equation}
for the radial field $E_\rho=E_{\rm f}+E_\mathrm{offset}$ and $B_z$ the solenoid field.
By fitting a linear slope to the measured $\omega(E_{\rm offset})$, a calibration offset can be adjusted so as to satisfy the frozen-spin condition for the actual momentum distribution of the stored muons.

The electrodes surround the region where the muons must be exposed to the magnetic pulse to initiate storage. 
However, short magnetic pulses induce eddy currents in metal surfaces that can significantly shield the magnetic pulse strength in this area.  
To study this effect, we tested prototype electrodes made from aluminized Kapton films (\SI{30}{nm} Al on \SI{25}{\mum} Kapton) by measuring the relative shielding with a pick-up coil in comparison to a setup without electrodes.

\begin{figure}
    \centering
    \includegraphics[height=0.5\textwidth]{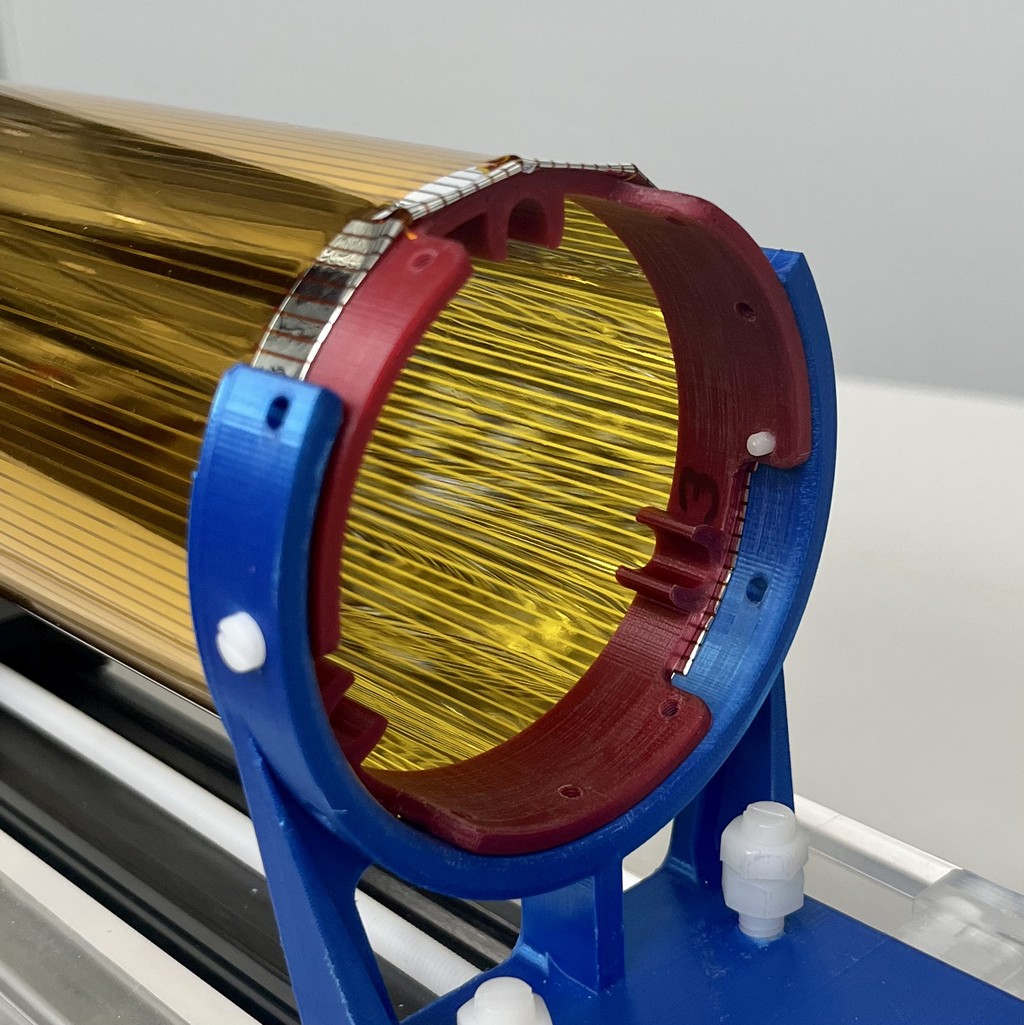}
		\label{fig:stripedElectrode}
    \caption{A segmented electrode prevents eddy current shielding of the magnetic pulse, while preserving the required electric field properties. A prototype made from aluminized Kapton~(a) showed negligible shielding, while simulation studies using a wire array electrode modeled in \textsc{ANSYS}~demonstrated that no additional significant systematic effects are introduced by the azimuthal non-uniformity of the field in the storage region~\cite{Cavoto2024EPJC}.} 
    \label{fig:segmented}
\end{figure}

A ground electrode made of uniformly aluminized Kapton resulted in a maximum shielding factor of $1.9(1)$. This factor increased to $3.0(2)$ when the HV electrode, made of the same material, was added.
In contrast, an electrode with a \SI{2}{mm} striped aluminum coating, with pitch \SI{2.2}{mm}, as shown in Fig.~\ref{fig:stripedElectrode}, exhibited negligible shielding. This indicates that eddy currents are strongly suppressed in the striped configuration compared to the homogeneous metal surface.

A segmented electrode creates periodic field non-uniformity along an orbit, potentially leading to systematic effects from geometric phase accumulation. For instance, splitting the electrodes into 100 longitudinal strands would cause oscillations of the electric field around \SI{40}{\giga\hertz}.
At these high frequencies, geometric phase accumulation remains minimal, even with non-uniformity levels up to approximately 50\% above the nominal field.
Using \textsc{ANSYS}, a segmented HV electrode design was tested with 60 wires of \SI{1}{mm} diameter and 120 ground wires, as shown in Fig.~\ref{fig:segmented}~(b) along with the transverse cross section of the electric field.
The black circle on the field map indicates a \SI{6}{mm} orbit displacement, where non-uniformity-induced oscillations remain below 1\%.
This analysis confirms that a segmented electrode does not introduce additional systematic effects.

		 \subsection{Muon detection and identification} 
				\label{sec:MuDetection}
				Several detectors placed along the muon trajectory, see Fig.~\ref{fig:PhaseISetup}, are used for various tasks, such as beam steering, muon selection, and ToF measurements.

\subsection*{Muon beam monitor}
\begin{figure}
    \centering
    \includegraphics[width=0.5\linewidth]{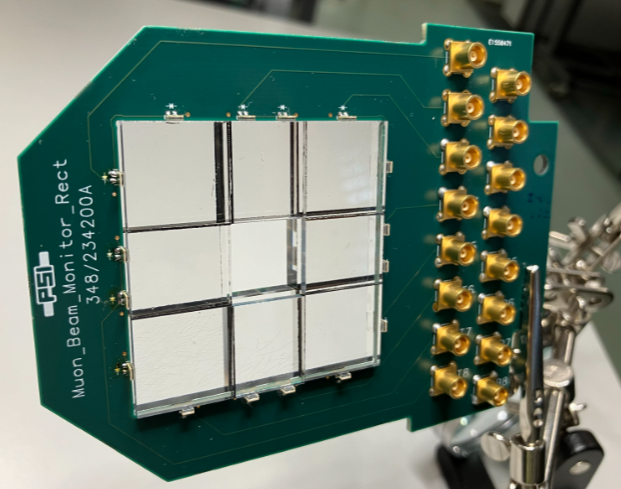}
    \caption{Photographs of the prototype beam monitor with eight rectangular tiles. All tiles are a stack of a \SI{2}{mm} front layer and a \SI{5}{mm} back layer, with approximately \SI{1}{mm} of air in between.}
    \label{fig:EntrMon}
\end{figure}

The first detector mounted on the injection channel, the muon-beam monitor, is used to align the experiment with the beam.
Figure~\ref{fig:EntrMon} shows a prototype, consisting of eight plastic scintillating tiles coupled to silicon photomultiplier~(SiPMs). 
The bulk of the muon beam passes through a central $\SI{15}{mm}\times\SI{15}{mm}$ opening, which corresponds to the diameter of the injection tube. 
The tiles monitor the intensity and spatial distribution of the beam halo, and centering is achieved by balancing the count rates of opposing tiles. 
Differences in detection efficiency are corrected by rotating the detector in steps of $90^{\circ}$.

When the separator in the beam line is turned off, the beam also contains positrons. 
These may not align perfectly with muons. 
To enable particle identification, the monitor features two identically segmented scintillator layers. 
The first, \SI{2}{mm} thick, fully absorbs surface muons, while positrons, as minimum-ionizing particles, deposit only a small fraction of their energy. 
The second, \SI{5}{mm} thick, identifies positrons. 
The design is based on \textsc{G4beamline} simulations evaluating detector performance as a function of the thickness, segmentation, and shape of the scintillator.


The prototype detector has eight independent channels per layer. Each tile in the 8-fold segmented layout is coupled to two SiPMs connected in series.
All surfaces in contact with neighboring tiles are coated with at least \SI{180}{nm} aluminum to suppress crosstalk.

The prototype was tested by measuring the offset between the center of the beam monitor and the beam focus.
It has demonstrated its ability to discriminate positrons in the beam and to align the injection tube with the muon beam to better than \SI{1}{mm}, fulfilling the specifications derived from simulations.

\subsection*{Muon T0 time of flight detector} 
\label{sec:TOFdetector}
\begin{figure}
    \centering
    \includegraphics[width=0.7\linewidth]{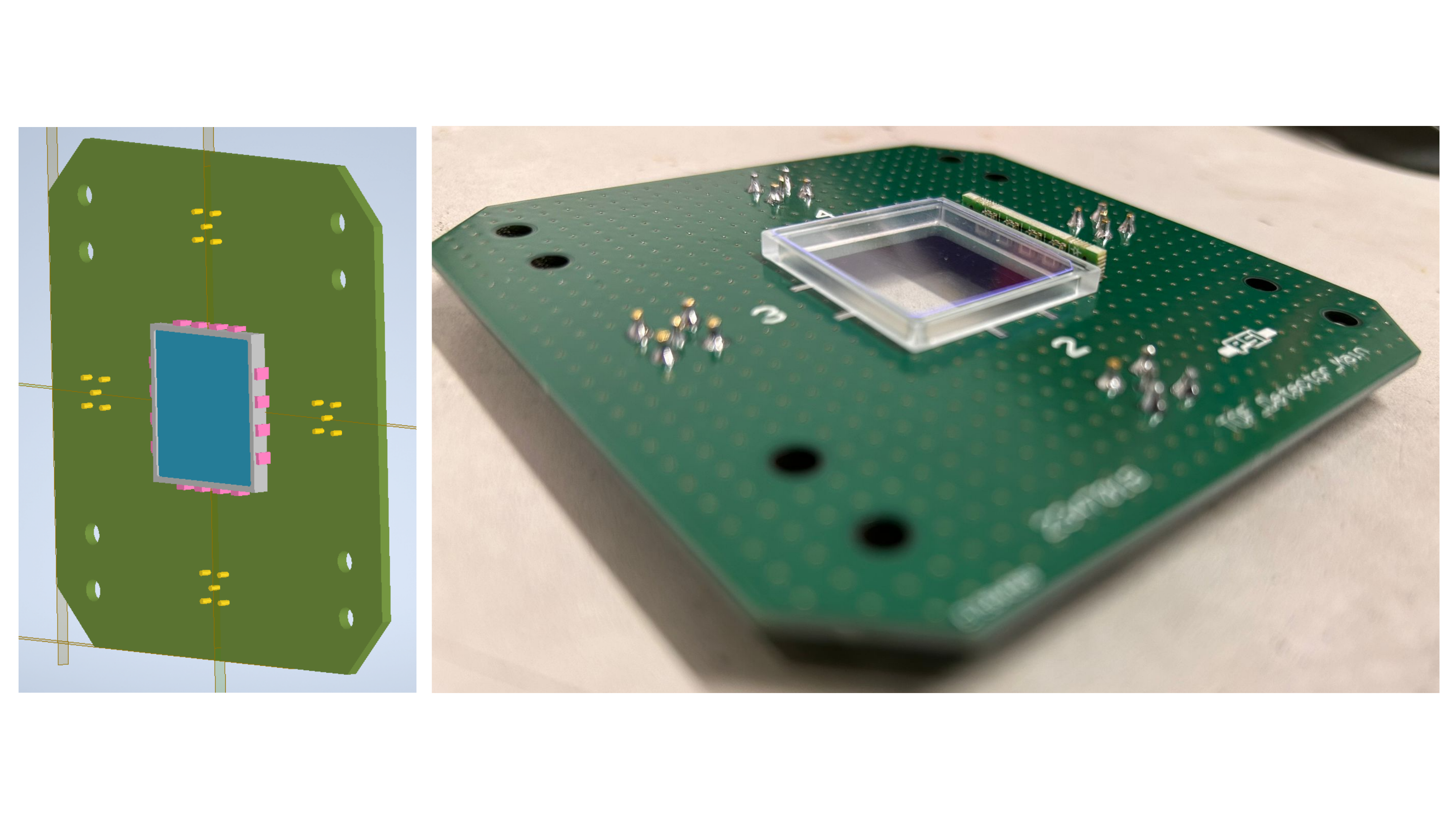}
    \caption{3D view of the CAD design for the TOF detectors (left). The TOF detectors during the assembly procedure. The scintillating windows, the light guide and one board of the MPPC are visible(right}
    \label{fig:muTOF}
\end{figure}

\begin{figure}%
\centering
	\includegraphics[width=0.55\columnwidth]{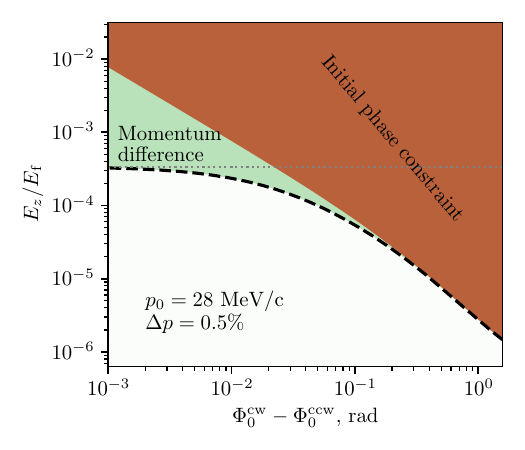}%
\caption{Limit on the longitudinal $E$-field, $E_z$, when considering alternating
		CW/CCW injections, where the
		difference in mean momentum averaged over all injected muons for injections of CW and CCW beams is
		fixed at $\Delta p = 0.5\%$. 
  The limit is shown as a function of the initial phase difference $\Phi_0^\mathrm{cw} - \Phi_0^\mathrm{ccw}$. The horizontal dashed line indicates the limit coming from $\Delta p$, keeping all other parameters equal between the two injection modes. The brown area is the
constraint that results from the difference in the initial phase at $\Delta p =
			0\%$. The green area and its thick dashed line edge show the combined
limits of the two effects.}
	\label{fig:omz_ez_limit}
\end{figure}

A CW and CCW muon storage is crucial to fully utilize statistical sensitivity while mitigating systematic effects. 
The dominant systematic effect arises from misalignment of the electric field, required for the frozen-spin technique, relative to the magnetic field (see Sec.~\ref{Sec:SystematicEffects}). 
This is canceled by CW and CCW storage combined with a reversal of the direction of the magnetic field. 
Figure~\ref{fig:omz_ez_limit} shows how differences in the momentum and initial phase of the spin constrain the angular alignment of the electric field to control systematic effects below statistical sensitivity.
By measuring the time of flight~(ToF) of each muon passing through the injection channel, we can select CW and CCW data with identical momentum distributions and hence the same sensitivity. 
To minimize momentum loss and beam divergence, we use thin scintillators coupled to silicon photomultipliers, which optimizes detection efficiency while reducing multiple scattering. The computer aided design~(CAD) and a picture during the assembly of the TOF detectors are shown in Fig.~\ref{fig:muTOF}

Each injection tube has its own $T_0$-ToF detector, measuring the ToF in combination with the entrance gate of the entrance trigger detector (see Sec.~\ref{sec:MuInjTrigger}). 
The requirements are high detection efficiency (above \SI{95}{\%}) and excellent timing resolution, better than \SI{500}{ps}. 
The T0-ToF detectors cover the injection pipe cross section, with a square scintillator, $23\times 23$~\SI{}{mm^2}, and \SI{50}{\micro\meter} thickness, glued to a square Plexiglass frame as light guide readout by a total of 16 SiPM\@. 
Each side of the Plexiglass frame is coupled to four SiPM connected in series, resulting in a total of four channels per detector. 
To address the low signal-to-noise ratio for the detection of a muon with a momentum of about \SI{27}{\MeVc}, we combine the signal of four SiPMs in one channel and use the coincidence of at least three channels to discriminate against ubiquitous dark noise.
This permits us to trigger on only a few photon-electrons while maintaining a highest possible detection efficiency and timing resolution.

\begin{figure}
    \centering
		\subfloat[]{
    \includegraphics[width=0.475\linewidth]{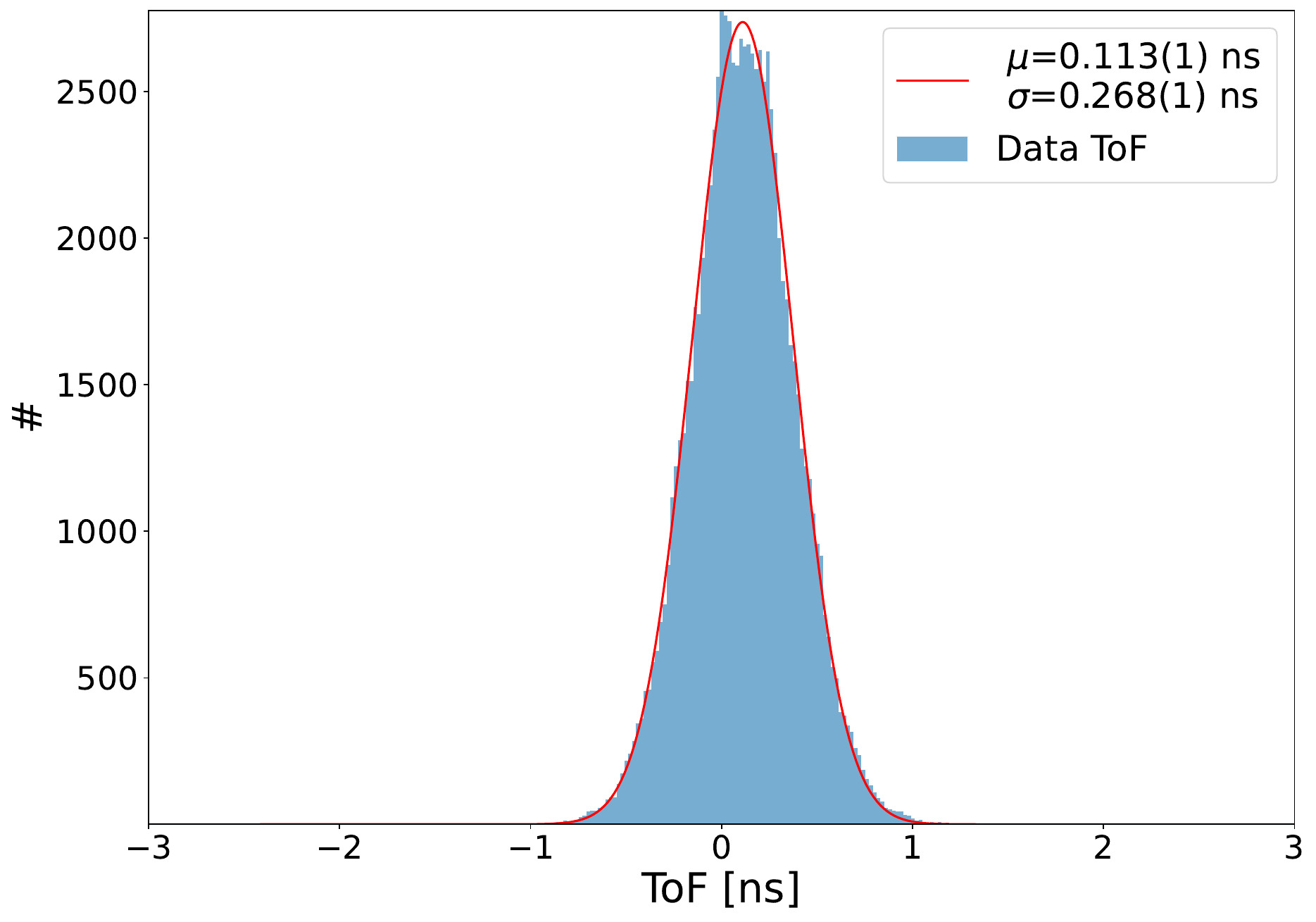}}
		\hfill
		\subfloat[]{
    \includegraphics[width=0.475\linewidth]{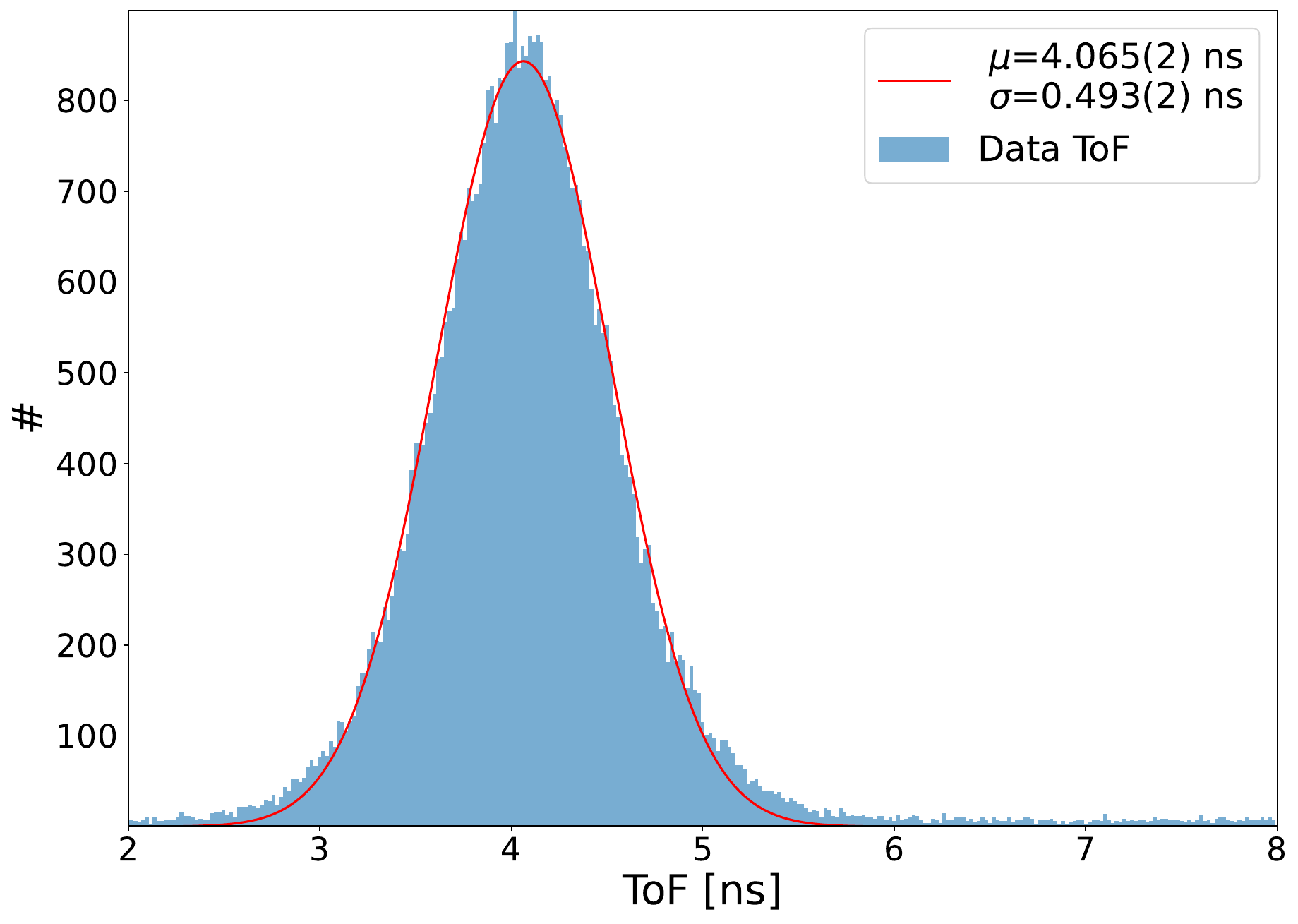}}
    \caption{
    (a)~Timing resolution of one \SI{100}{\micro m} and one \SI{50}{\micro m}  detector pair, mounted back-to-back with a distance of a few millimeters, and a (b)~time of flight measurement of positrons using a \SI{50}{\micro\meter} thick scintillator upstream and a \SI{50}{\micro\meter}~scintillator thick scintillator downstream.}
    \label{fig:TOF_TOFresolutionpair}
\end{figure}

Figure~\ref{fig:TOF_TOFresolutionpair}\,a) shows the intrinsic time resolution for a \SI{50}{\micro m} / \SI{100}{\micro m} detector pair using a constant fraction method to extract the time of each channel. 
A detection efficiency >98\% was achieved with an intrinsic time resolution better than \SI{300}{ps}. 
Figure~\ref{fig:TOF_TOFresolutionpair}\,b) shows a ToF measurement using a pair of \SI{50}{\micro m} thick detectors. 

%
%

\subsection*{Muon trajectory characterization}
The muon beam monitor and ToF detector enable alignment monitoring and CW/CCW symmetry checks during data collection. 
However, during beam commissioning, a detailed characterization of muon trajectories is essential to ensure proper alignment with the magnet and validate trajectory distributions against expectations. 
This is critical to optimize muon storage efficiency and confirm that magnet support and movement systems maintain trajectory symmetry between CW and CCW injections, minimizing systematic uncertainties.

To achieve this, we are developing a precision muon tracking detector for beam commissioning, temporarily placed at the muon entrance trigger detector's location. The detector aims for a momentum resolution of 0.5\% and angular/position resolutions within a few mrad, respectively mm, to resolve anomalies and biases of the phase space.
Achieving such precision at \SI{28}{\mega\electronvolt/c} initial muon momentum, requires an ultra-light detector, to minimize tracking deterioration from the multiple Coulomb scattering.

The proposed solution is a Time Projection Chamber (TPC) using an ultra-light helium-based gas mixture, potentially at sub-atmospheric pressure(\SI{0.4}{\bar}
), separated from the vacuum in the magnet bore by a vacuum-tight \SI{300}{\nano\meter} Silicon Nitride window. 
The GridPix detector is used as TPC readout structur, a gaseous detector made of a conductive mesh implanted \SI{50}{\micro\meter} above a Timepix chip~\cite{Llopart2007NIMA}. 
A voltage difference between mesh and chip creates avalanche when drift electrons reach the mesh, functioning similarly as a sort of microscopic Micromegas~\cite{Giomataris1996NIMA}.

\begin{figure}
    \centering
    \subfloat[]{
    \includegraphics[width=0.37\linewidth]{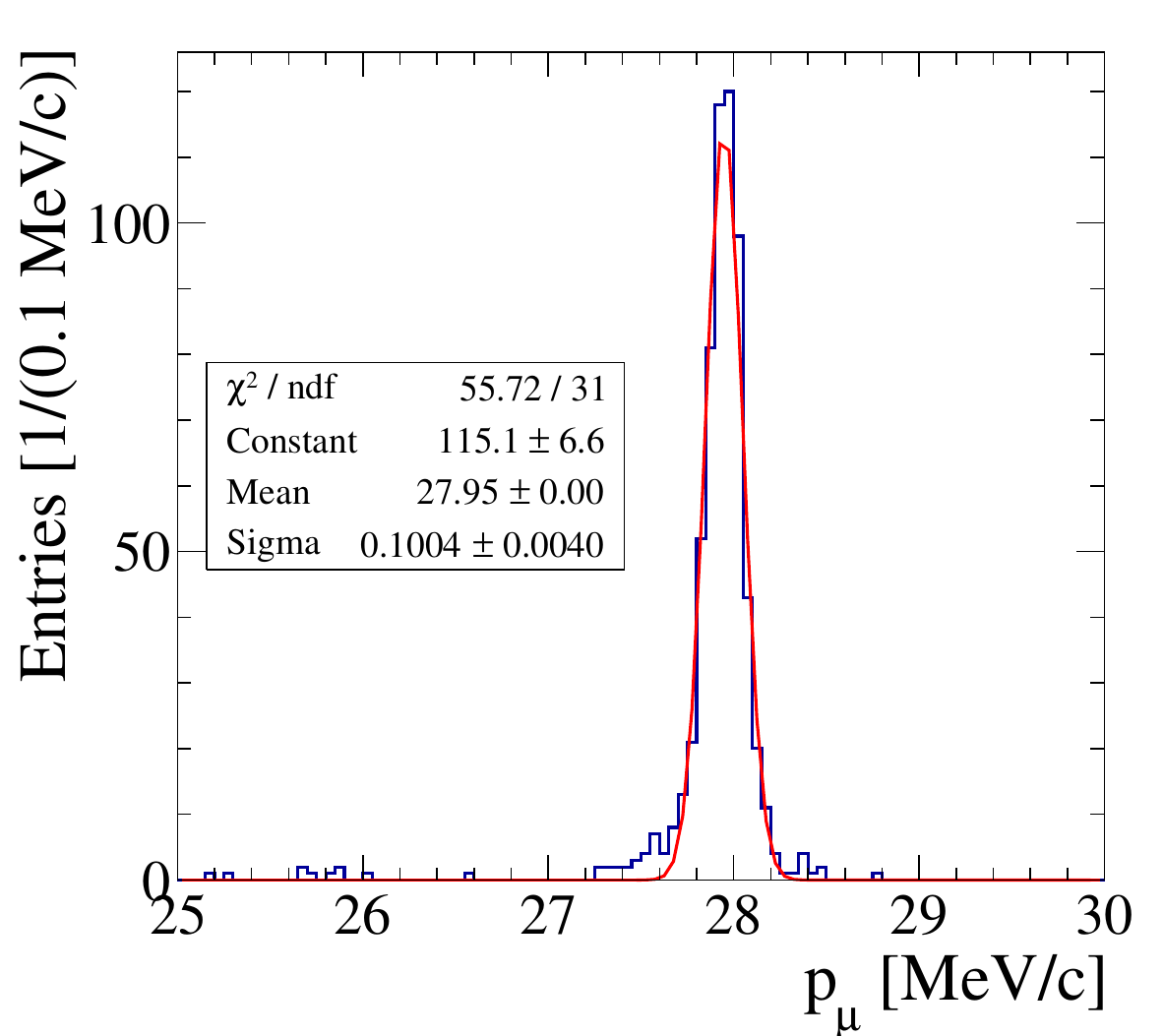}
    }
    \subfloat[]{
    \includegraphics[width=0.6\linewidth]{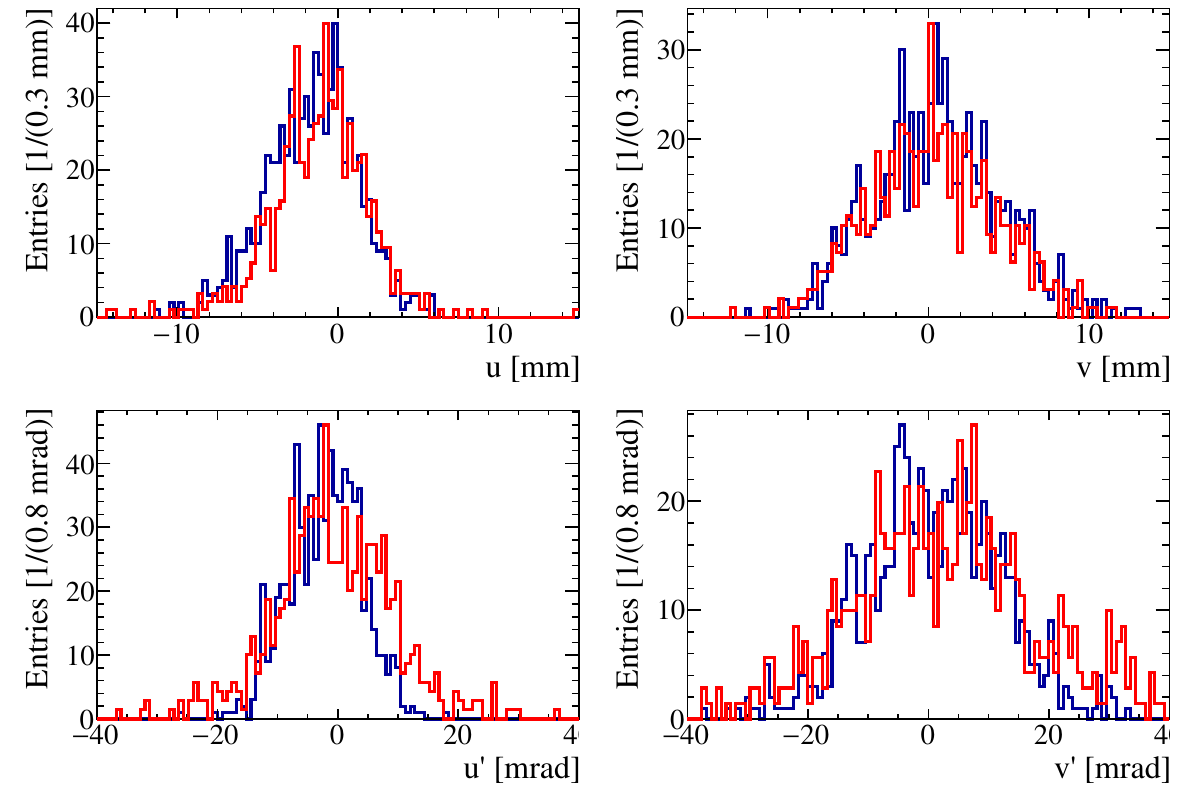}
    }
    \caption{Performance of a longitudinal TPC for the reconstruction of injected muons. (a)~Reconstructed momentum (assuming \SI{28}{\MeVc} initial momentum). (b)~True (blue) and reconstructed (red) phase space coordinates and angles (u: horizontal; v: vertical).}
    \label{fig:tpc_reso}
\end{figure}

In Fig.~\ref{fig:tpc_reso} we show the expected momentum resolution and the comparison of true and reconstructed phase-space coordinates at the exit of the injection channel. A momentum resolution better than \SI{0.4}{\percent} may be obtained, and the phase space resolutions look sufficient for alignment purposes.

				\subsection*{Muon entrance trigger}
						\label{sec:MuInjTrigger}

In Phase~I, only about \num{4e-3} of the muons that have passed through the magnetically shielded injection channel (see Section~\ref{sec:Injection}) are on or close to a spiral trajectory that may result in storage. 
Any other muon needs to be vetoed to avoid triggering magnetic kicks that, in any event, would not store muons. 

\REVIEW{Angela I would show a picture of the detector prototype, instead of the sketch}The muon entrance trigger generates a signal triggering the magnetic kick at the right moment when the muon passes the weakly-focusing field region (approximately \SI{100}{\nano\second} after having left the injection channel). 
\begin{figure}
    \centering
    \subfloat[]{
    \includegraphics[width=0.39\textwidth]{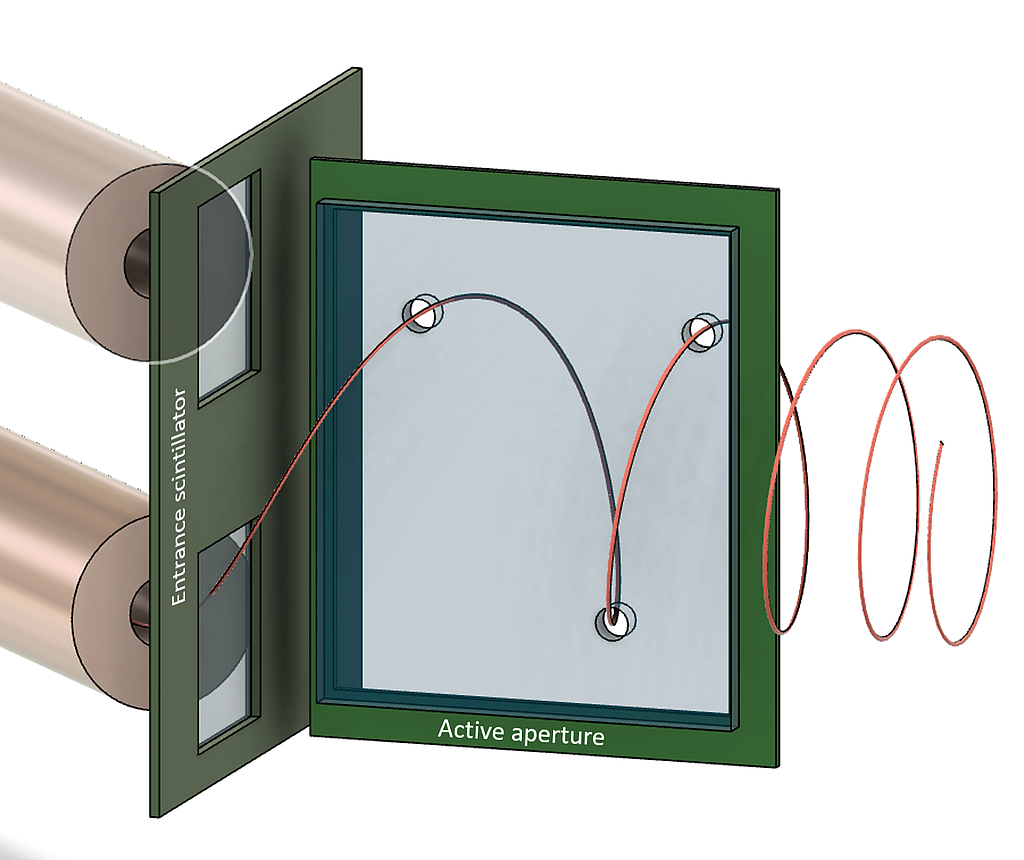}}
    \hfill
    \subfloat[]{
    \includegraphics[width=0.59\textwidth]{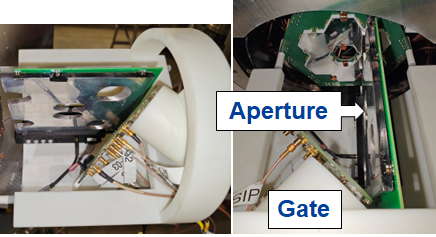}}
    \caption{(a)~CAD sketch of the entrance trigger. On the left, the muons exit the collimation tube shielded by a super-conducting shield. The strong magnetic field immediately bends the muons onto a spiral trajectory. First, they pass through a thin \SI{100}{\micro\meter} gate scintillator. A thick second scintillator with holes at positions along the reference muon trajectory forms an active aperture, stopping and detects muons that are outside the acceptance phase space. A trigger for the magnetic pulse (see Sec.~\ref{sec:B_FieldPulse}) will only be generated in the case of an anti-coincidence between the gate scintillator and the active-aperture scintillator. (b)~Photographs of the fabricated trigger detector.}
    \label{fig:MuonEntranceTriggerCAD}
\end{figure}
It is made of two thin scintillator tiles, \SI{100}{\micro\meter}-thick gate scintillators, one each at the exit of the CW and CCW injection tube, in combination with an active aperture, the anti-coincidence scintillator~(ACS), made of a scintillator approximately \SI{5}{mm} thick, which has holes around the reference muon trajectory. The anti-coincidence between the thin gate scintillator and thick active aperture is used to create a TTL-signal to trigger a pulsed magnetic kick to deflect the trajectory of the muon onto a storage orbit.
Due to this, the pulse rate of the kicker power supply reduces from about \SI{12e4}{\per\second} to \SI{500}{\per\second}.  Figure~\ref{fig:MuonEntranceTriggerCAD} shows the CAD of the entrance detector with the injection trajectory. 

Simulation studies emphasize the need for pulse latency within a narrow window of \SI{120}{ns} to \SI{150}{ns}, between muon detection and the rising edge of the magnetic pulse. 
By designing a high-speed SiPM readout electronic circuit to generate the TTL signal, we could demonstrate a propagation delay of less than \SI{5}{ns}. 
This alleviates the demand for the internal delay of the HV kicker current supply, as various elements, i.e.\ TTL and pulse transmission lines, contribute.

In addition to a minimal propagation delay for the TTL signal, we also want to be able to record the signal waveforms with the DAQ\@.
For this purpose, the SiPM signals are pre-amplified and divided into two by a passive divider.  
One of the signals is digitized using the DAQ for monitoring purposes, while the other will be sent to a custom discriminator generating digital signals that are used for the anticoincidence logic to create the trigger TTL-signal.

The first custom component prototypes were built and tested. The measured delay through the preamplifier plus the splitter was below \SI{1}{\nano\second}, while the delay through the discriminator plus the Low-Voltage Differential Signal~(VDS) to TTL converter was below \SI{3}{\nano\second}, both are well within the required performance. The first prototype muon trigger detector for in-field measurement was constructed and tested in a beam test at PSI in November 2024. The preliminary results indicated that the prototype detector performed as expected. 

				\subsection*{Muon Exit Detector}
					\label{sec:MuExit}
\begin{figure}
	\centering
		\includegraphics[width=0.48\textwidth]{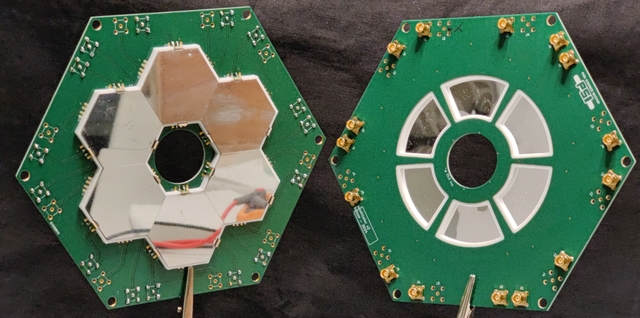}
		\caption{Prototype of the muon exit detector. Six hexagonal scintillating tiles cover the radial range $\SI{15}{mm}<r<\SI{45}{mm}$ detecting muons passing the central area. Each tile is coated with a \SI{100}{nm}-thick aluminum layer, and read-out with two SiPM.} 
		\label{fig:muonExitDetector}
\end{figure}

Most muons passing through the entrance detector will be kicked onto a stable orbit and stored in the center of the trap until they decay. 
However, a small fraction of muons will not be kicked sufficiently and will pass through the central area. 
These events will be registered by the muon exit detector. A prototype is shown in Fig.~\ref{fig:muonExitDetector}, to erase the veto and reset the trigger of the entrance detector.
The detector comprises six hexagonal plastic scintillator tiles, each coated with an \SI{100}{nm}-thick aluminum layer. 
The tiles and PCB are designed such that each tile can be connected to up to four SiPMs in parallel, permitting a coincidence readout to reduce dark noise.

During commissioning, the muon exit detector is used to fine tune the timing and amplitude of the magnetic kick by minimizing the coincidence rate between the entrance trigger and a detection in the exit detector. 

				\subsection{Positron detection}
				\label{sec:PositronTracker}

\begin{figure}
   \begin{center}
    \includegraphics[width=0.88\textwidth]{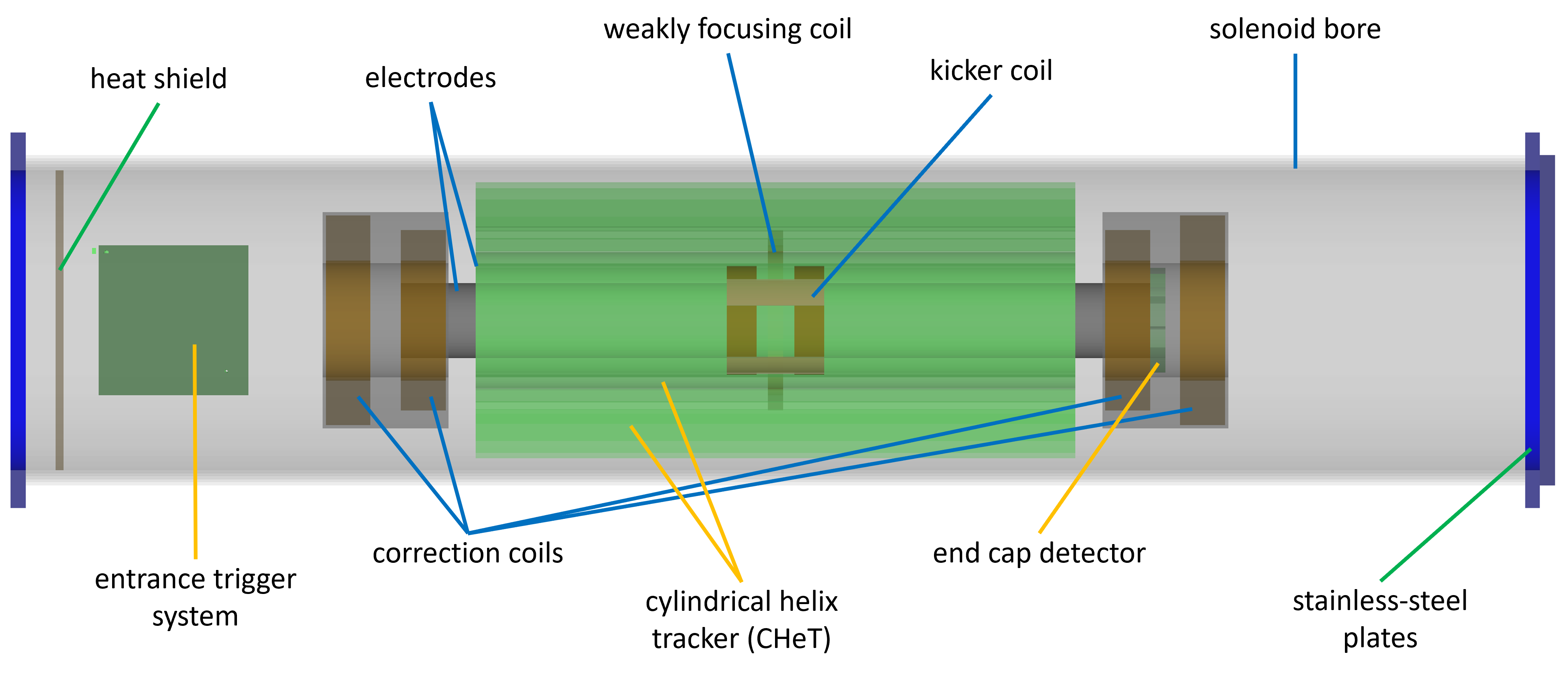}
    \caption{Geometry implemented in the \textsc{Geant4} simulation. The muons enter from the left through the injection channel (not shown here).}
    \label{fig:geant4_geometry}
   \end{center}
\end{figure}

A detailed model of the experiment in \textsc{Geant4}, shown in Fig.~\ref{fig:geant4_geometry}, permits the study of positron decays from muons stored in the trap but also from muons stopped at the entrance trigger detector.

The critical aspect of the muon EDM search is the ability to measure the direction of the outgoing positron.
This holds true for the three sequential steps of the experiment: demonstrating muon storage, measuring $g-2$ frequency as a function of the electric field, and acquiring EDM data with an electric field matched to the frozen-spin condition.
The kinematics of an $e^+$ coming from decay is described in Sec.~\ref{sec:kinematics}. 
A significant positron background is generated by stopping muons in the ACS\@. \REVIEW{LM: ACS?}As the expected injection efficiency is approximately 0.45\%, 99.55\% of the muons are being stopped and decay in the ACS\@, emitting positrons. 
A fraction of these will produce a non-correlated background in the positron detector. 
Simulations of the three distinct measurements and the positron background provide the specifications for the positron tracker:
\begin{description}
    \item[Momentum resolution] around a few $\MeVc$, necessary for selection cuts on positron momentum to increase the sensitivity to the change in asymmetry in a momentum binned analysis, see also Sec.~\ref{sec:kinematics};
    \item[Single-hit resolution] around \SI{1}{mm}, necessary to reconstruct tracks with the required angular resolution for a higher sensitivity to a change in asymmetry.
    \item[Timing resolution] less than \SI{1}{ns}, necessary to separate different hits produced by the same positron when it curls through the detector. Positrons from muon decays travel approximately at \textit{c}, meaning that in the expected magnetic field a complete circle in the transverse plane takes $\gtrsim$ \SI{0.6}{ns}, imposing a limit on the time resolution. 
\end{description}

\begin{figure}
    \centering
    \subfloat[]{
    \includegraphics[width=.6\textwidth]{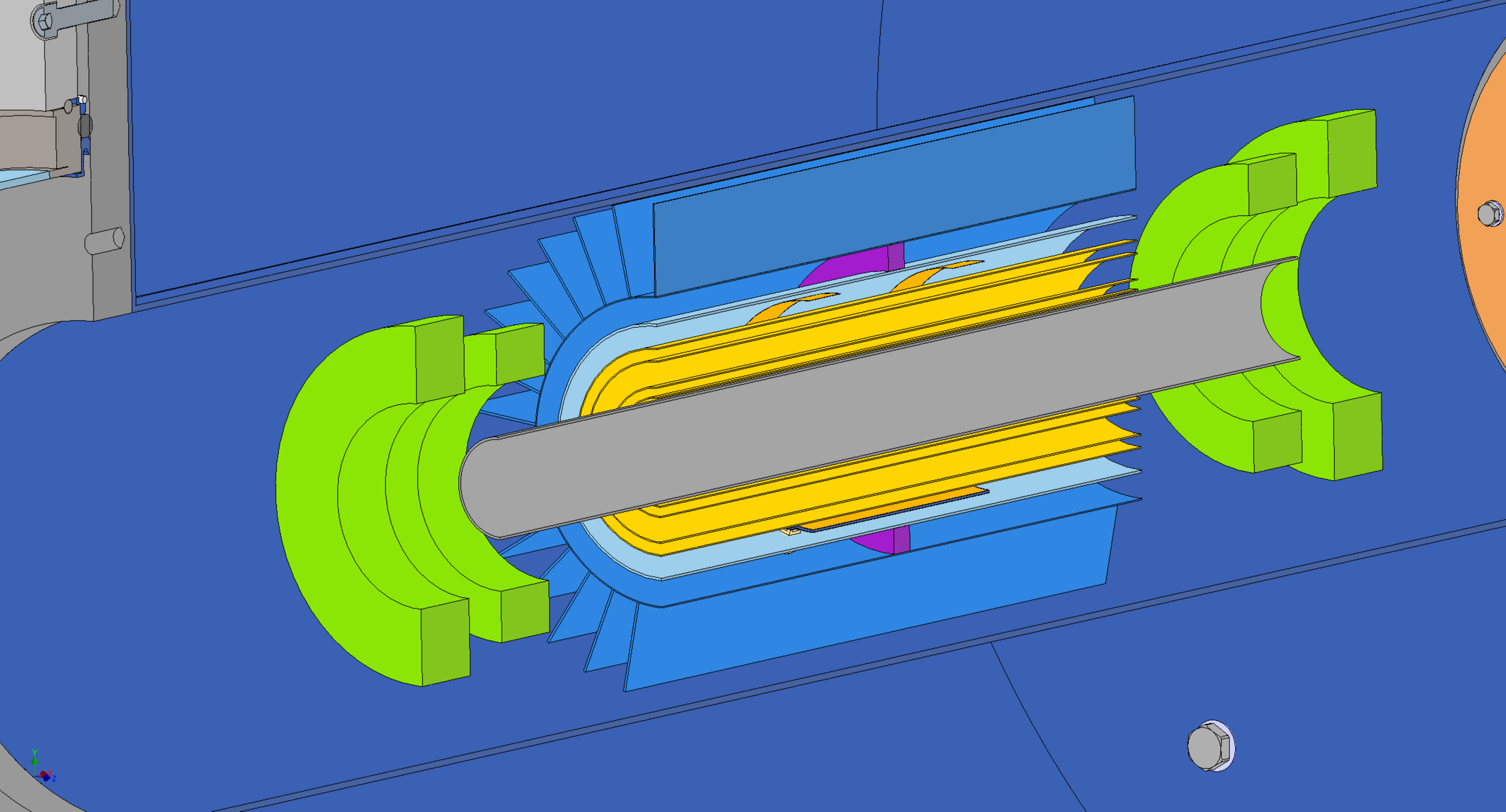}

    }
    \hfill
    \subfloat[]{
     \includegraphics[width=.35\textwidth]{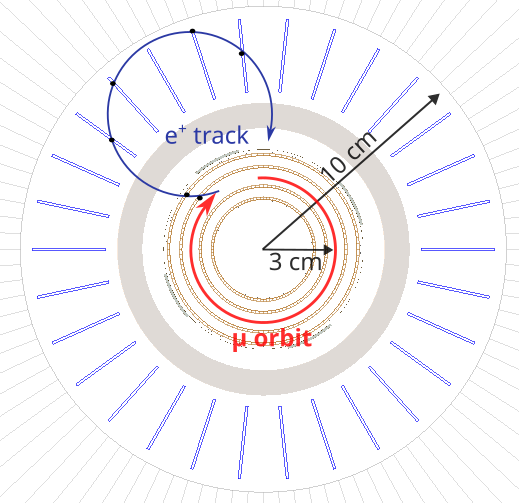}
    }
    \caption{(a)~A 3D view of the CHeT detector inside the magnet.
    (b)~Frontal projection of the CHeT with highlighted sizes and a simulated event shown for reference. The longitudinal size of the detector extends for \SI{400}{mm}.}
    \label{fig:CHeT_sketch}
\end{figure}


To meet these specifications we will deploy the Cylindrical Helix Tracker~(CHeT) in Phase~I\@.
Figure~\ref{fig:CHeT_sketch}\,a) shows a computer-aided detector design. 
It consists of two concentric cylinders inside the high voltage electrode, $r=\SI{20}{mm}$ and \SI{24}{mm}, and two concentric cylinders outside the ground electrode, $r=\SI{37}{mm}$ and \SI{40}{mm}, made of scintillating fibers of \SI{500}{\micro\meter} diameter, coupled to SiPMs\@. 
Each cylinder consists of two layers of scintillating fibers, tilted under a small angle with respect to each other, to precisely measure the position of the positron hit on the detector.
These scintillating fiber cylinders are complemented by 30 tiles, depicted in blue in Fig.~\ref{fig:CHeT_sketch}\,b), each comprising two layers of orthogonal scintillating fibers for the precise measurement of the radial and longitudinal coordinates of the hits.

The tiles are important in Phase~I of the experiment when the $(g - 2)$ precession will be measured, since they allow the measurement of the decay angle of the positrons in the muon orbit plane with a resolution of a few milliradians.
The cylinders are deployed as near to the muon orbit as possible, which is essential to detect decay positrons emitted almost parallel to the magnetic field with a small curvature radius, which are most sensitive to a muEDM\@.   

This design allows for fast, versatile, modular, and low-cost detector technology to be operational in large magnetic fields and in vacuum and adheres to the geometrical constraints imposed by the frozen-spin electrodes, the kicker, the weakly focusing coils, and their support structures.
With \SI{500}{\micro\meter} fibers, this detector design requires approximately 200 meters of fibers and 2000 read-out channels, assuming that 4 fibers are coupled per SiPM\@. This reduction in the number of readout channels ensures a spatial resolution better than \SI{1}{mm} in the $x-y$ coordinates.


\begin{figure}
    \centering
    \includegraphics[width=.9\linewidth]{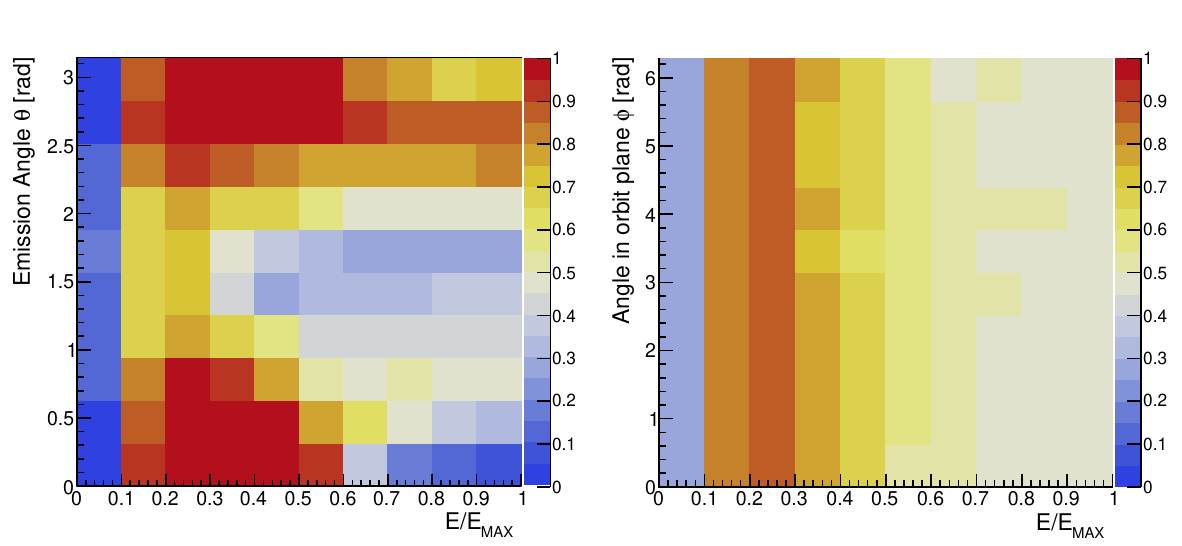}
    \caption{CHeT acceptance as a function of momentum and $\theta$ and momentum and $\phi$. While the acceptance in $\phi$ is flat because of the detector cylindrical symmetry, in $\theta$ the acceptance is maximum for $\theta \sim 0$ and $\pi$: this correspond to particles emitted perpendicular to the magnetic field.}
    \label{fig:CHeT_acceptance}
\end{figure}

The detector's acceptance and resolutions of the kinematical variables were evaluated through a \textsc{Geant4} simulation.
The bi-dimensional distributions of the acceptance are shown in Fig.~\ref{fig:CHeT_acceptance}, while the average resolutions are summarized in Tab.~\ref{tab:CHeT_resolutions}.

\begin{table}
    \centering
    \caption{Track resolution extracted from simulations of the CHeT detector.}
    \begin{tabular}{|r| l|} \hline
        Variable & Resolution  \\ \midrule
        $\sigma_p / p$ & 0.05 \\
        $\sigma_{\theta}$ & \SIrange{10}{30}{mrad} \\
        $\sigma_\phi$ &\SI{3}{mrad}\\ \hline
    \end{tabular}
    \label{tab:CHeT_resolutions}
\end{table}



    \subsection{Data acquisition and experimental control}
		The data acquisition of the muEDM Phase~I experiment records data from all particle detectors, including the positron tracker, for event reconstruction and system stability assessment. 
The system processes approximately 2000 read-out channels.

The trigger signal, generated by the muon entrance trigger, opens a gate during which hits in the positron tracker and muon scintillation detectors are registered to trace positron tracks from muon decay. 
Scintillators in the entrance trigger are also crucial for identifying potential muon pile-up, which can create positron background. 
The gate is closed and the trigger is reset upon detecting a muon hit on the end detector or after a \SI{10}{\micro\second} timeout.
%

Given the positron tracker’s expected detection rate of about \SI{1}{kHz}, the likelihood of multiple positron tracks within a gate is significant, requiring the DAQ system to register all tracks. The data acquisition system must therefore meet the following prerequisites, considering the expected resolution of the tracker: 

\begin{itemize}
    \item Measurement of the hit time with a resolution better than \SI{1}{ns};
    \item Record more than one hit during each gate from the same channel;
    \item Measurement of the hit rate for any connected input;
    \item Time synchronization over all channels;
    \item DAQ rate up to approximately \SI{1}{kHz} to avoid read out dead time, the requirement for the Phase~II is larger by at least a factor of 100.
\end{itemize}

Furthermore, we plan to use SiPMs for nearly all channels, requiring a device capable of providing high-voltage and integrated, programmable front-end electronics.

The selected baseline solution is the CAEN FERS A5202\footnote{\url{https://www.caen.it/products/a5202/}, last visited \today{}.}, equipped with two Weeroc Citiroc-1A ASICs for high-performance data acquisition. 
Its on-board A7585D SiPM power supply ensures precise sensor biasing for optimal operation. 
The A5202 supports multiple acquisition modes, including charge integration (energy), counting (rate), and timing, with \SI{250}{ps} resolution for hit time-to-digital-conversion~(TDC) and time-over-threshold~(ToT) capabilities.
With a maximum read-out rate exceeding \SI{10}{kHz}, it meets all experimental requirements, including those for Phase~II.

Scalability is a key feature of A5202, allowing the management and synchronization of up to 128 cards, i.e.\ 8192 channels, by a single DT5215 concentrator board. 
Optical TDlink technology ensures efficient data transfer and synchronization across channels. 
The A5202 readout software will integrate into ``Maximum Integrated Data Acquisition System''~(MIDAS\footnote{\url{https://daq00.triumf.ca/MidasWiki/}, last visited \today{}.}), merging seamlessly with the experimental data stream.

The muon monitors at the injection tube entrance will be read out by independently running WaveDAQ cards, reporting scintillating tile count rates to MIDAS for real-time beam alignment monitoring.

\section{Conclusions}
At PSI, a first application of the frozen-spin technique in a compact storage trap to search for the electric dipole moment of the muon is being developed. 

In this paper, we describe the derivation of the specification deduced in the simulations and the experimental design to achieve the predicted statistical sensitivity of $\sigma\left(d\right)\leq\SI{4E-21}{\ecm}$, for a data accumulation period of 200 days, meeting the requirements to control systematic effects accordingly. In a separate paper~\cite{Cavoto2024EPJC} we have assessed the most important systematic effects and deduced limitations on experimental design. 

In this demonstration experiment, we are using muons with a momentum of \SI{28}{\MeVc} from a surface beam line injected through a superconducting magnetically shielded path into a \SI{2.5}{T} solenoid field.

The muon loses momentum when passing through a set of scintillating detectors and air and is finally stored in a circular orbit with \SI{23}{\MeVc}. 
By applying a radial electric field, matching the frozen-spin condition, we nullify the $(g-2)$-precession and the measurement becomes highly sensitive to the EDM coupling of the muon with the electric field in its rest frame.

In the future experiment in Phase~II, we will use a dedicated superconducting magnet with a magnetic-field gradient of about 0.1\% between the injection area and the weakly-fcousing field in the storage region.
Coupled to a muon beam delivering more than $\SI{1E8}{\mu/s}$ at a momentum of $p=\SI{125}{\MeVc}$ results in a sensitivity of better than $\sigma\left(d\right)\leq\SI{6E-23}{\ecm}$ being sensitive to theoretical predictions~\cite{Hiller2010PRD,Hiller2020PRD,Bigaran2022PhysRevD}.
\REVIEW{LM: sensitivity being sensitive to theo predicitons?}

\section{Acknowledgments}
\label{sec:acknowledgments}
The collaboration is grateful for the excellent technical support and advice of F.~Barchetti, M.~Gantert, U.~Greuter, A.~Hofer, L.~Kuenzi, M.~Meier and R.~Senn from LTP, T.~Rauber and P.~Simon from CAS supporting the test beam, and T.~Höwler and his colleagues from the PSI survey group for aligning the solenoid. We thank C.~Klauser for supporting us in the coating of the scintillators with aluminum. We also acknowledge the great help by A.~Knecht and A.~Antognini before and during our test beam times on the piE1 beam line. Finally, we express our special thanks to all the colleagues from the UCN group at PSI for generously lending electronic devices to us. 

This work is partially funded by the Swiss National Science Foundation grants \textnumero~\textbf{204118}, \textbf{220487}, \textbf{PP00P21\_76884}, \textbf{TMCG-2\_213690} and receives funding from the Swiss State Secretariat for Education, Research, and Innovation~(SERI) under grant number \textnumero~\textbf{MB22.00040}. 
It is supported by the Instituto Nazionale di Fisica Nucleare~(INFN) and the Ministry of University and Research~(MUR) in Italy, grant: PRIN~\textnumero~\textbf{2022ENJMRS}.
China supports the project through the Science Foundation of China under Grant \textnumero~\textbf{12050410233}. 
In addition, it has received funding from the European Union Horizon 2020
research and innovation program under the Marie Sk\l{}odowska-Curie grant
agreement \textnumero~\textbf{884104}~(PSI-FELLOW-III-3i).  
The numerical simulations for design optimization were performed on the PSI Local High Performance Computing cluster, Merlin6, the Siyuan-1 cluster supported by the Center for High Performance Computing at Shanghai Jiao Tong University, and the Euler cluster operated by the High Performance Computing group at ETH Zürich.


\clearpage
\appendix
\section{Derivation of the statistical sensitivity}
\label{app:StatSens}
The measured asymmetry is defined as
\begin{equation}
	A_m = \frac{N_a - N_c}{N_a + N_c},
	\label{eq:measured_asymmetry0}
\end{equation}
where $N_a$ and $N_c$ are the number of decay positrons emitted
in the direction of the magnetic field and opposite to the magnetic field, respectively.
We denote $N = N_a + N_c$ the total number of decay positrons and $\delta = N_a - N_c$ their difference.

The relative statistical uncertainty on the measured asymmetry is
\begin{equation}
	\left( \frac{\sigma_{A_m}}{A_m}\right)^2 = \left( \frac{\sigma_{\delta}}{\delta}
	\right)^2 + \left( \frac{\sigma_{N}}{N} \right)^2.
	\label{eq:delta_am}
\end{equation}
The relative variance of $N$ is simply $1/N$ and the variance of $\delta$ is
\begin{equation}
	\sigma_{\delta}^2 = \sigma_{N_\uparrow}^2 + \sigma_{N_\downarrow}^2 - 2\rho
	\sigma_{N_\uparrow} \sigma_{N_\downarrow},
	\label{eq:sigma_delta}
\end{equation}
where $\rho$ is the correlation coefficient between $N_a$ and $N_c$.
Considering that $N_a \sim N_c \sim N/2$ and assuming $\rho = 0$,
\begin{equation}
	\sigma_{\delta}^2 = \frac{N}{2} + \frac{N}{2} = N.
	\label{eq:sigma_delta_2}
\end{equation}
It should be noted that the correlation coefficient $\rho$ is not necessarily
zero, but it is expected to be small in the case of a small EDM.

Combining Eqs.~\eqref{eq:measured_asymmetry0}, \eqref{eq:delta_am}
and~\eqref{eq:sigma_delta_2}, the variance of the measured asymmetry is
\begin{equation}
	\sigma_{A_m}^2 = \left(\frac{N}{\delta^2} +
	\frac{1}{N}\right) \left(\frac{\delta}{N}\right)^{-2}
	\approx \frac{1}{N},
	\label{eq:delta_am_2}
\end{equation}
where for $N \gg \delta$ the $1/N$ term in brackets can be neglected.

What we measure in the experiment is the time derivative of the asymmetry $\dot
	A_m \sim \sin(\Omega_\EDM t)$, where $\Omega_\EDM$ is the angular frequency
of the spin precession caused by an EDM. As the value of $\Omega_\EDM$ is expected to be very small, the rate
of change of the measured asymmetry will be proportional to $\Omega_\EDM$.
Thus, the measured asymmetry as a function of time, in the absence of systematic
effects, will be described by a linear function $y = a + b \, t$.

In each time bin $i$, the measured asymmetry is given by
Eq.~\eqref{eq:measured_asymmetry0} and the variance by
$\sigma_i^2 = 1 / N_i$, where
\begin{equation}
	N_i = \frac{N}{\gamma\tau}e^{-t_i / \gamma\tau},
	\label{eq:ni}
\end{equation}
is the total number of decay positrons in the time bin $i$, which reduces
exponentially with the time dilated muon lifetime $\gamma\tau$.
The statistical uncertainty on the measured slope $\sigma_b$ is
\begin{equation}
	\sigma_b^2 = \frac{\sigma^2_{y_w}}{\mathrm{var}(t)},
	\label{eq:sigma_b}
\end{equation}
where $\mathrm{var}(t) = (\gamma\tau)^2$ is the variance of the time distribution of the
decay positrons, and
\begin{equation}
	\sigma^2_{y_w} = 1/\sum_{i=1}^n \frac{1}{\sigma^2_i},
	\label{eq:sigma_y_w}
\end{equation}
is the variance of the weighted mean of $y$.
Substituting Eq.~\eqref{eq:ni} into Eq.~\eqref{eq:sigma_y_w} and taking the limit of infinitesimally small time bins
\begin{equation}
	\sigma^2_{y_w} = 1/\int_0^\infty \frac{1}{\sigma_i^2(t)}dt = \gamma \tau / N
	\int_0^\infty e^{-t/\gamma\tau} = \frac{1}{N},
	\label{eq:sigma_y_w_2}
\end{equation}
which leads to the statistical uncertainty on the slope
\begin{equation}
	\sigma_b = \frac{1}{\gamma\tau\sqrt{N}}.
	\label{eq:sigma_b_2}
\end{equation}

The slope $b$ is related to the EDM by
\begin{equation}
	b = \dot A_m = \frac{2c}{\hbar}\beta B \tilde\alpha P_0 d_\mu,
	\label{eq:uncertainty_slope}
\end{equation}
where the parameter $\tilde\alpha$ is the weighted average of the parity violating
decay asymmetry and $P_0$ is the initial muon polarization. The calculation of
$\alpha$ is discussed in detail in Section~\autoref{sec:kinematics}.
Thus, the statistical uncertainty on the EDM is
\begin{equation}
	\sigma_{d_\mu} = \frac{\hbar}{2c\beta B \tilde\alpha P} \frac{1}{\gamma\tau\sqrt{N}}.
	\label{eq:uncertainty_edm}
\end{equation}

\end{document}